\documentclass[a4paper,12pt]{report}
\usepackage[english]{babel}

\usepackage{graphicx,color}
\usepackage{cite}

\newcommand{\be}{\begin{equation}}
\newcommand{\ee}{\end{equation}}

\newcommand{\tit}{\textit}
\newcommand{\ba}{\begin{eqnarray}}
\newcommand{\ea}{\end{eqnarray}}

\newcommand{\ket}{\rangle}

\begin{document}

\begin{titlepage}
\begin{center}
\Large
{Universit\`a degli Studi dell'Insubria}\\
\large{ DIPARTIMENTO DI SCIENZA E ALTA TECNOLOGIA} \\
\rule{14.5cm}{0.5mm}\\
{\vspace{0.2cm}
\large{Ph.D. Thesis / Dissertation }}\\
\vspace{2.0cm} \huge {\textbf{The Renormalization Scale Setting Problem in QCD}}\\
\vspace{3.cm}
\large{Author:} \\
\large{\bf{Leonardo Di Giustino}}\\
\vspace{1cm} \large
\begin{tabular}{p{9cm}r}
Advisor:\\
\bf{Prof. Philip G. Ratcliffe} \\
 Chair of the Doctoral Programme:\\
 \bf{Prof. Giuliano Benenti} \\
\end{tabular}
\newline
\newline
\rule{14.5cm}{0.5mm} XXXIV CICLO - January 2022

\end{center}
\end{titlepage}
\newpage
\thispagestyle{empty}
 \hspace{30cm} \nonumber


%
%
%
\newpage
 \setcounter{page}{0}

 \pagenumbering{roman}
\tableofcontents

\newpage
\thispagestyle{empty}
 \nonumber

\section*{Glossary}
\begin{description}
    \item[AF] asymptotic freedom.
    \item[BLM] Brodsky-Lepage-Mackenzie.
    \item[BSM] beyond standard model.
    \item[CSR] commensurate scale relation.
    \item[CSS] conventional scale setting.
    \item[EW] electroweak.
    \item[FAC] fastest apparent convergence principle.
    \item[GCR] generalized Crewther relation.
    \item[GM--L] Gell-Mann--Low.
    \item[GUT] grand unified theory.
    \item[HQ] heavy quark.
    \item[iCF] intrinsic conformality.
    \item[IR] infrared.
    \item[mMOM] minimal momentum-subtraction scheme.
    \item[MOM] momentum-subtraction scheme.
    \item[MS] minimal-subtraction scheme.
    \item[\boldmath$\overline{\rm MS}$] modified minimal-subtraction scheme.
    \item[NP] new physics.
    \item[OPT] optimized perturbation theory.
    \newpage
\thispagestyle{empty}
 \nonumber
    \item[pQCD] perturbative quantum chromodynamics.
    \item[PMC] principle of maximum conformality.
    \item[PMCa] alternative principle of maximum conformality scale-setting.
    \item[PMCm] multi-scale principle of maximum conformality scale-setting.
    \item[PMCs] single-scale principle of maximum conformality scale-setting.
    \item[PMC$_\infty$] infinite-order scale-setting using the principle of maximum conformality.
    \item[PMS] principle of minimal sensitivity.
    \item[RG] renormalization group.
    \item[RGE] renormalization group equations.
    \item[RS] renormalization scheme.
    \item[RSS] renormalization scale setting.
    \item[SM] standard model.
    \item[UV] ultraviolet.
    \item[V-scheme] scheme of the static potential between two heavy quarks.
    \item[xRGE] extended renormalization group equations.
\end{description}


\newpage
\thispagestyle{empty}
 \nonumber

\vspace*{6cm}

{\it ``In the perspective of a theory unifying all the
interactions, electromagnetic, weak and strong nuclear, such as a
so-called grand unified theory or GUT, we are constrained to apply
the same scale-setting procedure in all sectors of the theory."}



\newpage

 \setcounter{page}{1}
 \pagenumbering{arabic}

\chapter*{Introduction}
\addcontentsline{toc}{chapter}{\numberline{}Introduction}

A key issue in making precise predictions in QCD is the
uncertainty in setting the renormalization scale $\mu_R$ in order
to determine the correct running coupling $\alpha_s(\mu_R^2)$ in
the perturbative expansion of a scale invariant quantity. Other
quantities such as the top and Higgs mass uncertainty, the strong
coupling uncertainty also affect the theoretical predictions for
perturbative QCD, but one of the most important sources of error
is given by the renormalization scale and scheme ambiguities.
These scale and scheme ambiguities are an important source of
uncertainties for predictions in many fundamental processes in
perturbative QCD, such as the gluon fusion in Higgs
production~\cite{Anastasiou:2016cez}, or bottom-quark
production~\cite{Catani:2020kkl}, which are essential for the
physics investigated at the Large Hadron Collider (LHC) and at
future colliders.

Precise predictions are crucial for both standard model (SM) and
beyond the standard model (BSM) physics, in order to enhance
sensitivity to new physics (NP) at colliders.

In principle, an infinite perturbative series is void of this
issue, given the scheme and scale invariance of the entire
physical
quantities~\cite{StueckelbergdeBreidenbach:1952pwl,GellMann:1954fq,Peterman:1978tb,Callan:1970yg,Symanzik:1971vw},
in practice perturbative corrections are known up to a certain
order of accuracy and scale invariance is only approximated in
truncated series leading to so-called scheme and scale
ambiguities~\cite{Celmaster:1979km,Buras:1979yt,Stevenson:1980du,Stevenson:1981vj,Stevenson:1982qw,Stevenson:1982wn,Grunberg:1980ja,Grunberg:1982fw,Grunberg:1989xf,Brodsky:1982gc,Abbott:1980hwa}.
On one hand, according to the conventional practice or
conventional scale setting (CSS), this problem cannot be avoided
and is responsible for part of the theoretical errors. Using the
CSS approach the renormalization scale $\mu_R$ is set to the
typical scale of a process Q, and errors are estimated by varying
the scale over a range of two $[Q/2;2Q]$. This method gives
predictions with large theoretical uncertainties comparable with
the calculated order correction. Moreover, this procedure leads to
results that depend on the renormalization scheme and on the
initial scale choice, to a perturbative series that diverges as
$\sim \alpha_s^{n+1} \beta_0^{n} n!$~\cite{thooftscheme}, and does
not agree with QED. Moreover, in several processes there is more
than one typical scale and results can be significantly different
for perturbative calculations according to the different choices
of scale~\cite{Berger:2009ep}.

On the other hand, some strategies for the optimization of the
truncated expansion have been proposed, such as the Principle of
Minimal Sensitivity (PMS) proposed by
Stevenson~\cite{Stevenson:1982wn,Stevenson:1982qw,Stevenson:1981vj,Stevenson:1980du},
the Fastest Apparent Convergence (FAC) criterion introduced by
Grunberg~\cite{Grunberg:1980ja,Grunberg:1982fw,Grunberg:1989xf}.
These are procedures commonly in use for scale setting in
perturbative QCD together with the CSS and an introduction to
these methods can also be found in
Refs.~\cite{Wu:2013ei,Deur:2016tte}. However, as shown in
Refs.~\cite{Wu:2013ei}, these methods not only have the same
difficulties of CSS, but they also lead to incorrect and
unphysical results~\cite{Kramer:1990zt} in particular kinematic
regions.

In general a scale-setting procedure is considered reliable if it
preserves important self-consistency requirements. All
Renormalization Group properties such as: {\it uniqueness,
reflexivity, symmetry,} and {\it transitivity} should be preserved
also by the scale-setting procedure in order to be generally
applied~\cite{Brodsky:2012ms}. Other requirements are also
suggested by known tested theories, by the convergence behavior of
the series and also for phenomenological reasons or scheme
independence.

It has been shown how all the theoretical requirements for a
reliable scale-setting procedure can be satisfied at once, leading
to accurate results by using the PMC~(the Principle of Maximum
Conformality~\cite{Brodsky:2011ig,Brodsky:2011ta,
Brodsky:2012rj}). This method is the generalization and extension
of the original Brodsky-Lepage-Mackenzie (BLM)
method~\cite{Brodsky:1982gc} to all orders and to all observables.
The primary purpose of the PMC method is to solve the
scale-setting ambiguity, it has been extended to all
orders~\cite{Mojaza:2012mf,Brodsky:2013vpa} and it determines the
correct running coupling and the correct momentum flow according
to RGE invariance~\cite{Wu:2014iba,Wu:2019mky}. This leads to
results that are invariant respect to the initial renormalization
scale in agreement with the requirement of scale invariance of an
observable in pQCD~\cite{Wu:2013ei}. This method provides a
systematic way to eliminate renormalization scheme and scale
ambiguities from first principles by absorbing the $\beta$ terms
that govern the behavior of the running coupling via the
renormalization group equation. Thus, the divergent renormalon
terms cancel, which improves convergence of the perturbative QCD
series. Furthermore, the resulting PMC predictions do not depend
on the particular scheme used, thereby preserving the principles
of renormalization group invariance~\cite{Brodsky:2012ms,
Wu:2014iba}. The PMC procedure is also consistent with the
standard Gell-Mann--Low method in the Abelian limit,
$N_c\rightarrow0$~\cite{Brodsky:1997jk}. Moreover, in a theory of
unification of all forces, electromagnetic, weak and strong
interactions, such as Grand Unified theories, one cannot simply
apply a different scale-setting or analytic procedure to different
sectors of the theory. The PMC offers the possibility to apply the
same method in all sectors of a theory, starting from first
principles, eliminating the renormalon growth, the scheme
dependence, the scale ambiguity, and satisfying the QED
Gell-Mann-Low scheme in the zero-color limit $N_c\to 0$.

We show in this thesis, how the implementation at all orders of
the PMC simplifies in many cases by identifying only the
$\beta_0$-terms at each order of accuracy due to the presence of a
new property of the perturbative corrections: the {\it intrinsic
conformality} (iCF). First we recall that there is no ambiguity in
setting the renormalization scale in QED. The standard
Gell-Mann-Low scheme determines the correct renormalization scale
identifying the scale with the virtuality of the exchanged
photon~\cite{GellMann:1954fq}. For example, in electron-muon
elastic scattering, the renormalization scale is the virtuality of
the exchanged photon, i.e. the spacelike momentum transfer squared
$\mu_R^2 = q^2 = t$. Thus
\begin{equation} \alpha(t) = {\alpha(t_0) \over 1 - \Pi(t,t_0)}
\label{qed1}
\end{equation} where
$$ \Pi(t,t_0) = {\Pi(t) -\Pi(t_0)\over 1-\Pi(t_0) } $$
From Eq.~\ref{qed1} it follows that the renormalization scale
$\mu^2_R=t$ can be determined by the $\beta_0$-term at the lowest
order. This scale is sufficient to sum all the vacuum polarization
contributions into the dressed photon propagator, both proper and
improper, at all orders. Again in QED, considering the case of
two-photon exchange, a new different scale is introduced in order
to absorb all the $\beta$-terms related to the new subprocess into
the scale. Also in this case the scale can be determined by
identifying the lowest-order $\beta_0$-term alone. This term
identifies the virtuality of the exchanged momenta causing the
running of the scale in that subprocess. This scale again would
sum all the contributions related to the $\beta$-function into the
renormalization scale and no further corrections need to be
introduced into the scale at higher orders. Given that the pQCD
and pQED predictions match analytically in the $N_C\to 0 $ limit
where $C_F\alpha_{QCD} \to \alpha_{QED}$ (see
Ref.~\cite{Brodsky:1997jk}), we extend the same procedure to pQCD.
In fact, in many cases in QCD the $\beta_0$ terms alone can
determine the pQCD renormalization scale at all
orders~\cite{Brodsky:1997vq} eliminating the renormalon
contributions $\alpha_s^{n+1} \beta_0^n n!$. Although in
non-Abelian theories other diagrams related to the three- and
four-gluon vertices arise, these terms do not necessarily spoil
this procedure. In fact, in QCD, the $\beta_0$ terms arising from
the renormalization of the three-gluon or four-gluon vertices as
well as from gluon wavefunction renormalization determine the
renormalization scales of the respective diagrams and no further
corrections to the scales need to be introduced at higher orders.
In conclusion, if we focus on a particular class of diagrams we
can fix the PMC scale by determining the $\beta_0$-term alone and
we show this to be connected to the general scheme and scale
invariance of an observable in a gauge theory.




This thesis is organized as follows: in {\bf section I} we
introduce QCD as a gauge theory based on the $\rm SU(3)_c$ local
symmetry; in {\bf section II} we give a description of the
renormalization group equations and their extended version,
starting from the renormalization procedure of the strong coupling
$\alpha_s(\mu)$ and its renormalization scale dependence; in
\textbf{section III} we present the state of the art for the
renormalization scale setting problem in QCD, introducing the
basic concepts of scale setting procedures in use at present:
conventional scale setting (CSS), FAC-scale setting, PMS-scale
setting and BLM scale setting, highlighting their basic features
and showing some important physical results; in \textbf{section IV
} we describe the PMC approach; in \textbf{section V} we introduce
the newly developed method PMC$_\infty$ and its new features; we
show results for the application of the PMC$_\infty$ to the
event-shape variables: thrust and C-parameter comparing the
results under the CSS and the PMC$_\infty$ methods; in
\textbf{section VI} we apply the PMC$_\infty$ to thrust in the
perturbative conformal window of QCD and we investigate the QED
limit of the QCD thrust calculations; in {\bf section VII} we show
a novel method to determine the strong coupling and its behavior
$\alpha_s(Q)$ over a wide range of scales, from a single
experiment at a single scale, using the event shape variable
results; in {\bf section VIII} we apply the PMC/PMC$_\infty$ to a
multi-renormalization scale process: heavy-quark pair production
at threshold, comparing results in different schemes and taking
the QED limit of the QCD calculations.

\color{black}

\newpage
\thispagestyle{empty}

\begin{center}
 \vspace*{0.0cm}

This thesis mostly summarizes results published in the following
articles or conference proceedings:
\begin{itemize}

\item L. Di Giustino, Stanley J. Brodsky, Xing-Gang Wu and
Sheng-Quan Wang, PMC$_\infty$: Infinite-Order Scale-Setting method
using the Principle of Maximum Conformality and preserving the
Intrinsic Conformality; Proceedings of the conference RADCOR 2021,
arXiv:2110.05171 [hep-ph].

\item L. Di Giustino, Francesco Sannino, Sheng-Quan Wang and
Xing-Gang Wu, Thrust Distribution for 3-Jet Production from
$e^+e^-$ Annihilation within the QCD Conformal Window and in QED,
Physics Letters B 823C (2021) 136728; arXiv:2104.12132[hep-ph];

\item  L. Di Giustino, Stanley J. Brodsky, Sheng-Quan Wang and
Xing-Gang Wu, Infinite-order scale-setting using the principle of
maximum conformality: A remarkably efficient method for
eliminating renormalization scale ambiguities for perturbative
QCD, Phys.Rev.D 102 (2020) 1, 014015: arXiv: 2002.01789[hep-ph];

\item  Sheng-Quan Wang, Stanley J. Brodsky, Xing Gang Wu, L. Di
Giustino and Jian-Ming Shen, Renormalization scale setting for
heavy quark pair production in $e^+e^-$ annihilation near the
threshold region, Phys.Rev.D 102 (2020) 1, 014005: arXiv:
2002.10993[hep-ph];

\item Sheng-Quan Wang, Stanley J. Brodsky, Jian-Ming Shen,
Xing-Gang Wu and L. Di Giustino A Novel Method for the Precise
Determination of the QCD Running Coupling from Event Shapes
Distributions in Electron-Positron Annihilation, Phys. Rev. D 100
(2019) 9, 094010 :   arXiv:1908.00060[hep-ph];

\newpage
\thispagestyle{empty}

 \item  Sheng-Quan Wang, Stanley J. Brodsky,
Xing-Gang Wu, L. Di Giustino Thrust Distribution in
Electron-Positron Annihilation using the Principle of Maximum
Conformality, Phys. Rev. D 99 (2019) 11, 114020: arXiv:
1902.01984[hep-ph];

\item  X. D. Huang, J. Yan, H. H. Ma, L. Di Giustino, J. M. Shen,
X. G. Wu and S. J. Brodsky, Detailed Comparison of Renormalization
Scale-Setting Procedures based on the Principle of Maximum
Conformality, arXiv:2109.12356 [hep-ph];

\item Sheng-Quan Wang,  Chao-Qin Luo, Xing-Gang Wu, Jian-Ming
Shen, Leonardo Di Giustino, New analyses of event shape
observables in electron-positron annihilation and the
determination of $\alpha_s$ running behavior in perturbative
domain, arXiv:2112.06212 [hep-ph];

\end{itemize}

\end{center}

\newpage

\chapter{Quantum Chromodynamics}
\label{sez:teorie di gauge}
\section{Introduction}

Quantum Chromodynamics is the theory of the strong interaction,
which describes how elementary constituents such as quarks and
gluons interact in order to bind together and form hadrons. In
particular, the theory exhibits the symmetries observed
experimentally and it explains why quarks interact strongly at low
energies and weakly in processes at large momentum transfer, where
quarks appear to be almost free.

Despite the many successful achievements, there are still several
problems that need to be solved in order to test the theory to the
highest precision. In particular, the confinement mechanism still
remains unexplained, though some evidence of such phenomenon has
been obtained in recent years by investigating gauge theories on
the lattice and by studying the topology of the non-Abelian gauge
theories. In order to overcome this obstacle, several alternative
approaches have also been proposed, which so far unfortunately are
not completely void of ambiguities and lead to results that are
rather model dependent. The non-perturbative region is in fact
affected by the peculiar infrared (IR) dynamics, which involves
mechanisms like hadronization, soft radiation and confinement. In
particular, in the IR region the behavior of the strong coupling
$\alpha_s$ is not yet predictable or explained in QCD. Recent
studies of light front holographic QCD (LFHQCD) (i.e. AdS/CFT
theory) suggest the presence of a finite limit of the coupling at
zero momentum, which means the presence of an interacting fixed
point in the $\beta$-function. This hypothesis has not been
rejected by comparison with experimental data and the behavior of
the coupling seems to be in agreement. At the moment the AdS/CFT
correspondence with QCD is still under investigation and the
LFHQCD can be considered only as a model or as an ``ansatz" for
the non-perturbative region of QCD. Despite the lack of a
perturbative expansion at low energies, there are still several
strong theorems that hold and that can be used to relate processes
with analogous characteristics in the non-perturbative region and
model parameters can be constrained by fits to data. Thus, if on
one hand the dynamics at long distances is not yet totally
revealed, on the other hand the dynamics at high energies reached
its apex with the discovery of {\it asymptotic freedom}~(AF),
which perfectly explains experimental evidence of the weakly
interacting partons inside hadrons at high momentum transfer. The
perturbative nature of QCD at short distances is still affected by
errors due to uncertainties, which in several cases can spoil the
theoretical predictions and which are related to both the
confining nature of the strong forces and the renormalization
scale and scheme ambiguities. The latter are the main sources of
errors in many processes where QCD corrections can be calculated
perturbatively. In order to make reliable theoretical predictions
and improve the precision, it is crucial to eliminate the scale
and scheme ambiguities in order to improve the theoretical
calculations in QCD enhancing the sensitivity to any possible
signal of new physics (NP) at the large hadron collider (LHC).
Furthermore, QCD is the most complex gauge theory we have ever
dealt with, it represents the most difficult sector of the
Standard Model (SM) for quantitatively predictions. In fact, if in
the electroweak (EW) sector, perturbation theory is always
applicable, given the particular weak nature of the interaction at
energies accessible in accelerators or in weak decays involving
leptons and vector bosons, this is not always possible in QCD
since the coupling $\alpha_s(Q) \gg \alpha_{QED}$ in a wide range
of scales $Q$ and quarks cannot be observed as real states. Thus
the complex nature of QCD can only be investigated by comparing
its predictions with all possible results from different
experiments in different kinematic regions. Given the vast variety
of the phenomena and phenomenological applications, QCD is
currently still a field of great interest and under continuous and
active investigation.


\section{The flavor symmetry: ${\rm SU(3)_f}$ }

Neutron, proton, pions and the other particles listed in the
Particle Data Group (PDG)~\cite{ParticleDataGroup:2020ssz} tables
are classified as \tit{hadrons} and have in common the particular
property of interacting via the strong interaction. All these
particles can be simply classified using global symmetries, in
particular the ${\rm SU(3)}$ symmetry. This was originally
discovered by Gell-Mann and Neeman with the quarks \tit{up, down,
strange} (u, d, s) and was later expanded to include the quarks
\tit{charm}~(c), \tit{bottom}~(b) and \tit{top}~(t). These
symmetries apply, following the Gell-Mann and Zweig approach,
starting from the assumption that light hadrons are formed by
quarks, respectively three for baryons and two (a quark and an
antiquark) for mesons. Quarks are fermions of spin 1/2 divided
into six flavors. Three quarks, u, c, t, have charge $ Q = 2/3 $,
while d, s, b have a charge of $ Q = -1/3 $. Following the
original idea of the isospin ${\rm SU(2)_I}$ symmetry of
Heisenberg for the states of the nucleon, Gell-Mann introduced the
group ${\rm SU(3)_f}$ for the quarks: u, d, s. The ${\rm SU(3)_f}$
symmetry applies straightforwardly assuming first that the three
light quarks u, d, s transform as a fundamental representation $3$
of the flavor symmetry group ${\rm SU(3)_f}$. All other particles
formed by the u, d, s quarks can be classified as higher
representations of the group, which can be generated by taking
multiple products of the fundamental representation: \be 3 = q_i =
\left(
\begin{array}{c} u \\ d \\ s  \end{array} \right). \ee

The ${\rm SU(3)_f}$ group is a rank 2 Lie group and its
representations are identified with two quantum numbers: by
convention we take the third component of isospin $T_3$ and the
hypercharge $Y$. The two quantum numbers strangeness and
hypercharge are not independent, but are related by the Gell-Mann
and Nishijma formula:$$Q =T_3+Y/2,$$ where $Y=B+S$, with $B$ the
baryon number, which for all quarks has the value $B=1/3$, $S$ is
the strangeness and $Q$ the electric charge. In this model, we
expect mesons and light baryons to be classified according to the
following representations:

\be \mbox{Mesons}=3\otimes \bar{3}=8\oplus 1,\nonumber \ee \be
\mbox{Baryons}=3\otimes 3 \otimes 3 =10 \oplus 8 \oplus 8 \oplus
1. \ee

The symmetry of Gell-Mann and Neeman predicts that mesons are
arranged in terms of octets and singlets, with baryons in octets,
decuplets and singlets. The fact that this simple representation
can in fact classify the lightest mesons and baryons present in
nature, was a remarkable achievement.

At present six different flavors of quarks are known, and in
principle, one could think of generalizing ${\rm SU(3)_f}$ to
${\rm  SU(6)_f}$. Unfortunately, the large differences among the
heavier quark states, break the global ${\rm SU(6)}$ symmetry
strongly. In contrast, the original ${\rm SU(3)_f}$ symmetry of
the light quarks is nearly conserved, given that the u, d, s
quarks have approximately the same masses. This justifies the
approximation of the quasi-conserved chiral symmetries, under
which the left-handed and right-handed fields transform
independently of each other. Such chiral symmetries are given by
the invariance of the Lagrangian under the group ${\rm SU(3)_L
\times SU(3)_R}$ or considering a better approximation, by the
group ${\rm SU(2)_L \times SU(2)_R}$, involving only the up and
down quarks, which can be actually considered massless ($m_u,m_d
\ll\Lambda_{QCD} \simeq 300{\rm  MeV}$).


\section{Hints for color}

The discovery of the $\Delta^{++}(J=3/2)$ baryon in
1951~\cite{Brueckner:1952zz} provided a first indication of the
existence of a new degree of freedom necessary to antisymmetrize
its wavefunction. In fact, the particle is in a fundamental $L=0$
state with all spins aligned along the same direction
$u^{\uparrow}u^{\uparrow}u^{\uparrow}$, which leads to a symmetric
wavefunction. This is in contrast with the Pauli exclusion
principle and Fermi-Dirac statistics, given that baryons have
semi-integer spin. In order to overcome this inconsistency, a new
quantum number was introduced: \tit{color}, with at least $N_c=3$
number of colors and with each quark $q^\alpha,$ having a color
index $\alpha=1,2,3$. Thus, according to this assumption the
wavefunction can be antisymmetric:
\be \Delta^{++}=\frac{1}{\sqrt{6}}\epsilon^{\alpha \beta
\gamma}|u^{\uparrow}_\alpha u^{\uparrow}_\beta
u^{\uparrow}_\gamma\ket. \ee Other experimental evidence for color
was provided by measurements of the decay width of the neutral
pion, $\Gamma(\pi\rightarrow \gamma \gamma) \propto N_c^2$ and by
the ratio $R_{e^+ e^-}=N_c\sum^{N_f}_{f=1}Q_f^2,$ that showed a
neat $N_c$ proportionality. These experimental hints were also
supported by other theoretical indications like the anomaly
cancellation and the ${\rm U(1)}$ problem of the $\eta'$ meson
mass.

\section{Quarks and gluons}

The ``quark model" introduced by Gell-Mann, comprehensively
explains the spectrum of hadronic states and in particular the
spectra of states containing heavy quarks such as {\it charm} and
{\it bottom}: charmonium, bottomonium respectively and also D and
B mesons. The charmonium, bottomonium states appear as resonances
of $c\bar{c}$ and $b\bar{b}$ quark states. Taking into account the
reduced mass difference, these are in fact similar to the states
of positronium ($e^-e^+$) and can be obtained assuming that the
heavy quarks interact via a non-relativistic potential that at
short-distances has the characteristics of an effective
Coulomb-like static potential, while at large-distances diverges
linearly according to the confinement mechanism. For the sake of
completeness, in order to complement the description of the
interaction of heavy quarks, relativistic and pQCD corrections
should be also taken into account.
It was also found that, in processes with large momentum transfer,
hadrons appear to be made up of almost free elementary
constituents (i.e. weakly interacting with each other) called
\tit{partons} by Feynman.

After careful analysis and measurements of charges and spins,
Feynman's partons were identified with Gell-Mann's quarks. It was
deduced that these values are in agreement with those of the quark
model. In particular, at the moment we know of 6 quark flavors:
namely, $ u, d, s, c, b $ and $ t,$ where each flavor also
corresponds to three different states of color $(r,g,b)$. Quarks
are then elementary particles that interact via strong forces
exchanging other massless particles, i.e. \tit{gluons}. However,
quarks have never been observed as free particles in laboratories,
so they cannot be represented as asymptotic states. They only
exist as elementary constituents of hadrons, wherein they are
confined. Moreover, also real states of single gluons cannot be
directly observed. Gluons are the bosons that convey the strong
forces inside hadrons analogously to photons in QED, but unlike
photons they carry the color charge. In fact, they exist in the
${\rm SU(3)}$ octet configuration or { \it adjoint}
representation. In this picture, quarks and gluons are carriers of
the color quantum number, while baryons and mesons are colorless
particles. This means that all hadrons are singlets under
rotations in the color space and cannot exchange long range
gluons. In order to avoid the existence of extra unobserved
hadronic states, to therefore satisfy hadron spectroscopy, one
must postulate furthermore that all asymptotic states are
colorless. This assumption is known as the \tit{hypothesis of
confinement}; in fact, it implies the non-observability of free
quarks and gluons, which excludes the color singlet gluon state.

\section{Gauge theories}

QCD is a non-Abelian {\it gauge theory}, i.e. a Lagrangian quantum
field theory with the gauge symmetry given by the invariance of
the Lagrangian under the local special unitary Lie group ${\rm
SU(3)_c}$ transformations. In general, gauge theories satisfy two
fundamental requirements: invariance under the particular gauge
group, which for the Standard Model is ${\rm SU(3)_c \times
SU(2)_L\times U(1)_Y}$ and {\it renormalizability}. These two
requirements lead to a theory that is symmetric under the local
transformation of the particular special unitary group and under
the {\it renormalization group equations} (RGE). The groups ${\rm
SU(3)_c}$ and ${\rm SU(2)_L}$ are non-Abelian special unitary
groups related to the strong and weak forces. The ${\rm U(1)_Y}$
is the Abelian group of hypercharge, which leads to QED after the
{\it spontaneous symmetry breaking} (SSB) mechanism. Lie groups
have the following Standard representations:
$$ U={\rm exp}\{i\sum^{n^2-1}_{i=1}\epsilon_i T^i\},$$
where $\epsilon_i$ are real parameters and $T^i$ Hermitian
traceless matrices defined as {\it group generators}. Special
unitary groups have a total number of $2n^2-n^2-1=n^2-1$
generators , where ``$n$" is the dimension of the fundamental
representation of the group, and $n-1$ generators are diagonal.
The algebra of Lie groups is set by the commutation relations of
the generators: \be
\left[\frac{\sigma_i}{2},\frac{\sigma_j}{2}\right]=i\epsilon_{ijk}\frac{\sigma_k}{2}
\ee  and
\begin{equation}
\left[\frac{\lambda_i}{2},\frac{\lambda_j}{2}\right]=if_{ijk}\frac{\lambda_k}{2},
\end{equation}
where $\epsilon_{ijk}$ and $f_{ijk}$ are total antisymmetric
tensors and are the {\it structure constants} of the group, while
the $\sigma_i$ with $i=1,2,3$ and the $\lambda_j$ with
$j=1,2,..,8$ are the adjoint representations of the $SU(2)$ and
$SU(3)$ group given by the Pauli and the Gell-Mann matrices
respectively.

\section{The QCD Lagrangian}
\label{sez:lagrangian}

The QCD Lagrangian can be derived from the \tit{Yang-Mills}
Lagrangian by adding all flavor contributions. The Yang-Mills
Lagrangian is itself a gauge invariant Lagrangian under local
special-unitary transformations:

\be U_xU_x^\dag=U_x^\dag U_x=I\mbox{,}\quad \mbox{det}U_x=1. \ee
Local gauge ${\rm SU(3)}$ matrices, $U_x$, can be parametrized
using the form: \be U_x={\rm exp}[-ig_s\sum_A \theta^A(x)
\lambda^A/2], \ee where $g_s$ is the coupling constant, which is
related to $\alpha_s$ by: $g_s^2=4 \pi \alpha_s $ and $\lambda^A$
$(A=1,2,...,8)$ are the group generators of ${\rm SU(3)_c}$, which
are given by the Gell-Mann matrices:
$$ \lambda^{1}=\left(\begin{array}{ccc}
0 & 1 & 0 \\
1 & 0 & 0 \\
0 & 0 & 0
\end{array}\right), \lambda^{2}=\left(\begin{array}{ccc}
0 & -i & 0 \\
i & 0 & 0 \\
0 & 0 & 0
\end{array}\right), \lambda^{3}=\left(\begin{array}{ccc}
1 & 0 & 0 \\
0 & -1 & 0 \\
0 & 0 & 0
\end{array}\right),$$

\begin{equation}
\lambda^{4}=\left(\begin{array}{ccc}
0 & 0 & 1 \\
0 & 0 & 0 \\
1 & 0 & 0
\end{array}\right), \lambda^{5}=\left(\begin{array}{ccc}
0 & 0 & -i \\
0 & 0 & 0 \\
i & 0 & 0
\end{array}\right), \lambda^{6}=\left(\begin{array}{ccc}
0 & 0 & 0 \\
0 & 0 & 1 \\
0 & 1 & 0
\end{array}\right)
\end{equation}
$$ \lambda^{7}=\left(\begin{array}{ccc}
0 & 0 & 0 \\
0 & 0 & -i \\
0 & i & 0
\end{array}\right),
\lambda^{8}=\frac{1}{\sqrt{3}}\left(\begin{array}{ccc}
1 & 0 & 0 \\
0 & 1 & 0 \\
0 & 0 & -2
\end{array}\right).$$
The $\theta^A(x)$ are arbitrary parameters depending on space-time
coordinates. In order to preserve the gauge invariance locally, a
new definition of the derivative occurs, the \tit{covariant
derivative}: \be (D^\mu q_f)^\alpha=\left\{\partial^\mu
\delta^{\alpha \beta}-ig_s G_{\alpha \beta}^\mu(x)
\right\}q_f^\beta,\ee where $q_f^{\beta}$ is the fermionic field
in its fundamental representation of the group ${\rm SU(3)_c}$,
with $f=1,...,6$ and $\beta=1,2,3,$ the flavor and color indices
respectively, and:
 \be G_{\alpha \beta}^\mu(x)=\sum_A
\frac{\lambda^A_{\alpha \beta}}{2} G^{A\mu}(x)\ee the eight gauge
fields relative to the eight gluons with $A=1,...,8$. Following
these considerations we can write the QCD Lagrangian as:
 \be
 {\mathcal L}_{QCD}=-\frac{1}{4}
\mbox{tr}(F_{\mu\nu}F^{\mu\nu})+\sum_{f}
 \bar{q}^\beta_f(i\gamma^\mu
(D_\mu)_{\beta\alpha}-m_f \delta_{\beta\alpha})q^\alpha_f,
 \label{eqn:lqcd}
 \ee
where the gluon field strength tensor $F^A_{\mu\nu}$ is given by:

\begin{eqnarray}
F^{\mu \nu}(x) &\equiv &
\partial^\mu G^\nu-\partial^\nu G^\mu-ig_s[G^\mu\mbox{,}G^\nu],\nonumber\\
&\equiv & (\lambda^A/2)F^{A\mu \nu}(x),
\end{eqnarray}
with \be F^{A\mu \nu}(x)=\partial^\mu G^{A\nu}-\partial^\nu
G^{A\mu}+g_sf^{ABC}G^{B\mu}G^{C\nu}. \ee

Due to the non-commutative nature of ${\rm SU(3)}$, the trace of
the field strength tensor introduces gluon cubic and quartic self
interactions, which is the main difference between QCD and QED.
These interactions are particularly important in the calculation
of the $\beta$-function and lead to a fundamental property of QCD:
asymptotic freedom. Eq.~\ref{eqn:lqcd} is still missing two terms
${\mathcal L}_{GF}$ and ${\mathcal L}_{Ghosts}$, respectively the
Lagrangians for the gauge fixing and ghost terms, which must be
introduced for consistency (for a review see Ref.~\cite{peskin}).
However, these terms do not introduce extra real final or initial
particle states, but fictitious particles; namely ghosts, which
occur only in loop integration.

\section{Quark states and ${\rm SU(3)_c}$}

According to the Lagrangian in Eq.~\ref{eqn:lqcd}, the strong
force preserves the flavor quantum number. The interactions that
change flavor stem from weak interactions. In fact,
flavor-changing transitions among different quarks have a lower
intensity with respect to flavor conserving interactions.
Moreover, intermediate bosons of the electroweak interactions
($\gamma$, $Z^0$, $W^\pm$), do not couple via strong forces or to
color charges. Thus, the ${\rm SU(3)_c}$ color symmetry is well
preserved by the strong forces and it can be considered the
perfect candidate for the QCD symmetry group. Moreover, QCD
includes the following features:
 \begin{itemize}
 \item color ${\rm SU(3)_c}$ is an exact symmetry;
 \item $N_c=3$;
 \item quarks and antiquarks have different fundamental representations: $\bar{3}\neq 3$,
 this leads to a complex representation;
 \item hadrons are colorless particles, so they must be represented as color singlets;
 \item asymptotic freedom.
 \end{itemize}

These conditions exclude other groups, but lead straightforwardly
to the ${\rm SU(3)}$ group symmetry. Assuming the fundamental
representations for quarks and antiquarks, $3$ and $\bar{3}$
respectively, we can determine the other possible ${\rm SU(3)_c}$
states considering multiple combinations of quarks and/or
antiquarks:
 \begin{eqnarray}
 q\bar{q} &:&
3\otimes \bar{3}=1\oplus 8,\nonumber\\
 qqq &:& 3\otimes 3\otimes 3=1\oplus 8
\oplus 8 \oplus 10,\nonumber\\
 qq &:&  3\otimes 3=\bar{3} \oplus
6.
 \end{eqnarray}

Simple color singlets result from $q\bar{q}$ and  $qqq$ and lead
to mesons and baryons respectively. Recently other unstable
particles assumed to be tetraquarks (e.g. X(3872))  and
pentaquarks (e.g. $\Theta^+$ (1540))~\cite{Liu:2019zoy} have been
observed by Belle, D$\emptyset$, BaBar and LHCb, while there is no
evidence for a six quark state. A first six quark state, namely
hexaquark, was introduced  by Jaffe
(H-dibaryon~\cite{Jaffe:1976yi}) and recently by Farrar
(sexaquark~\cite{Farrar:2017eqq}). Actually color representations
for mesons and baryons are not the only color singlets possible
that can be realized with multiple combinations of quarks and
antiquarks. In fact, color singlets can also be formed considering
the combinations: $qq\bar{q}\bar{q}, qqqq\bar{q}, qqqqqq$.
Tetraquark and pentaquark states are both unstable, while the
sexaquark is claimed to be stable. However, it is still under
investigation if these novel quark states are simple bound states
or molecules.



\chapter{The Renormalization Group \label{RGroup}}


\section{Renormalization}

Renormalization is a procedure that applied quantum field theories
(QFT) in order to cancel an infinite number of ultraviolet (UV)
singularities that arise in loop integration absorbing them into a
finite number of parameters entering the Lagrangian, like the mass
or coupling constant and fields. This procedure starts from the
assumption that the variables entering the Lagrangian are not the
effective quantities measured in experiments, but are unknown
functions affected by singularities. The origin of the ultraviolet
singularities is often interpreted as a manifestation that a QFT
is a low-energy effective theory of a more fundamental yet unknown
theory.

The use of regularization UV cut-offs shields the very short
distance domain, where the perturbative approach to QFT ceases to
be valid.

Once the coupling has been renormalized to a measured value and at
a given energy scale, the effective coupling is no longer
sensitive to the ultraviolet (UV) cut-off nor to any unknown
phenomena arising beyond this scale. Thus, the scale dependence of
the coupling can be well understood formally and
phenomenologically. Actually gauge theories are affected not only
by UV, but also by infrared (IR) divergencies. The cancellation of
the latter is guaranteed by the Kinoshita - Lee - Nauenberg (KLN)
theorem~\cite{Kinoshita:1962ur,Lee:1964is}.

Considering first the Lagrangian of a massless theory, which is
free from any particular scale parameter, in order to deal with
these divergences a regularization procedure is introduced.
Referring to the dimensional regularization
procedure~\cite{tHooft:1972tcz,Cicuta:1972jf,Bollini:1972ui}, one
varies the dimension of the loop integration, $D=4-2\varepsilon$
and introduces a scale $\mu$ in order to restore the correct
dimension of the coupling.


In order to determine the renormalized gauge coupling, we consider
the quark-quark-gluon vertex and its loop corrections.
UV-divergences arise from loop integration for higher order
contributions for both the external fields and the vertex. The
renormalization constants are related by: \be
Z^{-1}_{\alpha_s}=(\sqrt{Z_3} Z_2 /Z_1)^2, \label{eqn:zalphas} \ee
where $Z_{\alpha_s}$ is the coupling renormalization factor,
$Z_1\left(Q^{2}\right)$ is the vertex renormalization constant and
$Z_{3}\left(Q^{2}\right)$ and $Z_{2}\left(Q^{2}\right)$ are:
$$
A_{a, \mu}=Z_{3}^{1 / 2}\left(Q^{2}\right) A_{a,
\mu}^{R}\left(Q^{2}\right), \quad \psi=Z_{2}^{1 /
2}\left(Q^{2}\right) \psi^{R}\left(Q^{2}\right),
$$
the renormalization constants for gluon and quark fields
respectively. The superscript $R$ indicates the renormalized
field. The renormalization factors in dimensional regularization
are given by: \ba Z_1(Q^2)&=&
1-\frac{\alpha_s(Q^2)}{4\pi} (N_c+C_F) \frac{1}{\varepsilon} \\
Z_2(Q^2)&=& 1-\frac{\alpha_s(Q^2)}{4\pi} C_F  \frac{1}{\varepsilon} \\
Z_3(Q^2) &=& 1+\frac{\alpha_s(Q^2)}{4\pi}
(\frac{5}{3}N_c-\frac{2}{3}N_f) \frac{1}{\varepsilon}
 \ea
where ${\rm \varepsilon}$ is the regulator parameter for the
UV-ultraviolet singularities. By substitution, we have that the UV
divergence : \be Z_{\alpha}\left(Q^{2}\right)=1-\frac{\beta_{0}
\alpha_{s}\left(Q^{2}\right)}{4 \pi \varepsilon}, \ee with \be
\beta_{0}=11-\frac{2 N_f}{3}. \ee

The singularities related to the UV poles are subtracted out by a
redefinition of the coupling. In the $\rm MS$ scheme, the
renormalized strong coupling $\alpha_s(Q^2)$ is related to the
bare coupling $\overline{\alpha_{s}}$ by:
 \be
\overline{\alpha_{s}}=Q^{2 \varepsilon}
Z_{\alpha}\left(Q^{2}\right) \alpha_{s}\left(Q^{2}\right).
 \label{eqn:alphasren}\ee

In the minimal subtraction scheme ($\rm MS$) only the pole
$1/\varepsilon$ related to the UV singularity is subtracted out. A
more suitable scheme is $\overline{\rm
MS}$~\cite{Bardeen:1978yd,tHooft:1973mfk,Weinberg:1973xwm}, where
also the constant term $\ln(4 \pi)-\gamma_E$ is subtracted out.
Different schemes can also be related by scale redefinition, e.g.
$\mu^2 \rightarrow 4 \pi \mu^2 e^{-\gamma_E}$. Thus, the
renormalization procedure depends both on the particular choice of
the scheme and on the subtraction point $\mu$.


Hence, even though there are no dimensionful parameters in the
initial bare Lagrangian, a mass scale $\mu$ is acquired during the
renormalization procedure. The emergence of $\mu$ from a
Lagrangian without any explicit scale is called dimensional
transmutation~\cite{Coleman:1973sx}. The value of $\mu$ is
arbitrary and is the momentum at which the UV divergences are
subtracted out. Hence $\mu$ is called the subtraction point or
renormalization scale. Thus, the definition of the renormalized
coupling $\alpha_s^{\overline{\rm MS}}(\mu)$ depends at the same
time on the chosen scheme, $\overline{\rm MS}$ and on the
renormalization scale $\mu$.

However, different schemes and scales can be related according to
the so-called extended renormalization group equations, which we
will introduce in the following sections. Given the perturbative
nature of the theory, these relations are known up to a certain
level of accuracy and the truncated formulas are responsible for
an important source of uncertainties: {\it the scheme and scale
ambiguities}.


\section{The running coupling constant $\alpha_s(\mu)$  \label{sec:alphas}}

In this section we will discuss the dependence of the strong
coupling on the renormalization scheme (RS) and scale ($\mu$) and
we will show how this dependence can be controlled by means of the
{renormalization group equations} (RGE). The strong coupling
$\alpha_s$, is a fundamental parameter of the SM theory and
determines the strength of the interactions among quarks and
gluons in quantum chromodynamics (QCD). As shown in the previous
section, its value depends on the renormalization scale $\mu$
(i.e. the subtraction point). In order to understand hadronic
interactions, it is necessary to determine the magnitude of the
coupling and its behavior over a wide range of values, from low to
high energy scales. Long and short distances are related to low
and high energies respectively. In the high energy region the
strong coupling has an {\it asymptotic behavior} and QCD becomes
perturbative, while in the region of low energies, e.g. at the
proton mass scale, the dynamics of QCD is affected by processes
such as quark confinement, soft radiation and hadronization. In
the first case experimental results can be matched with
theoretical calculations and a precise determination of the
$\alpha_s$ depends both on experimental accuracy and on
theoretical errors. In the latter case experimental results are
difficult to achieve and theoretical predictions are affected by
the confinement and hadronization mechanisms, which are rather
model dependent. Various processes also involve a precise
knowledge of the coupling in both the high and low momentum
transfer regions and in some cases calculations must be improved
with electroweak (EW) corrections. Thus, the determination of the
QCD coupling over a wide range of energy scales is a crucial task
in order to achieve results and to test QCD to the highest
precision. Theoretical uncertainties in the value of
$\alpha_s(Q^2)$ contribute to the total theoretical uncertainty in
the physics investigated at the Large Hadron Collider (LHC), such
as the Higgs sector, e.g. gluon fusion Higgs
production~\cite{Anastasiou:2016cez}. The behavior of the
perturbative coupling at low-momentum transfer is also fundamental
for the scale of the proton mass, in order to understand hadronic
structure, quark confinement and hadronization processes. IR
effects such as soft radiation and renormalon factorial growth
spoil the perturbative nature of the QCD in the low-energy domain.
Higher-twist effects can also play an important role. Processes
involving the production of heavy quarks near threshold require
the knowledge of the QCD coupling at very low momentum scales.
Even reactions at high energies may involve the integration of the
behavior of the strong coupling over a large domain of momentum
scales including IR regions. Precision tests of the coupling are
crucial also for other aspects of QCD that are still under
continuous investigation, such as the hadron masses and their
internal structure. In fact, the strong interaction is responsible
for the mass of hadrons in the zero-mass limit of the u, d quarks.

The origin and phenomenology of the behavior of $\alpha_s(\mu)$ at
small distances, where asymptotic freedom appears, is well
understood and explained in many textbooks on Quantum Field Theory
and Particle Physics.

 Numerous reviews exist; see e.g. Refs.~\cite{Prosperi:2006hx,Altarelli:2013bpa}.
 However, standard explanations often create an
apparent puzzle, as will be addressed in this thesis. Other
questions remain even in this well understood regime: a
significant issue is how to identify the scale Q that controls a
given hadronic process, especially if it depends on many physical
scales. A fundamental requirement, called ``renormalization group
invariance", is that physical observables cannot depend on the
choice of the renormalization scale and scheme.

In the perturbative regime theoretical predictions are affected by
several sources of errors, e.g. the top and Higgs mass
uncertainty, the strong coupling uncertainty and the main source
of errors is given by the renormalization scale and scheme
ambiguity. In this section we discuss the scale and scheme
dependence of the effective coupling $\alpha_s(\mu)$ in QCD.


\section{The evolution of $\alpha_s(\mu)$ in perturbative QCD}

As shown in the previous section the renormalization procedure is
not void of ambiguities. The subtraction of the singularities
depends on the subtraction point or renormalization scale $\mu$
and on the renormalization scheme (RS). Observables in physics
cannot depend on the particular scheme or scale, given that the
theory stems from a conformal Lagrangian. This implies that scale
invariance must be recovered imposing the invariance of the
renormalized theory under the renormalization group equation
(RGE). We discuss in this section the dependence of the
renormalized coupling $\alpha_s(Q^2)$ on the scale $Q^2$. As shown
in QED by Gell-Mann and Low, this dependence can be described
introducing the $\beta$-function given by:
 \be
\frac{1}{4 \pi}\frac{d\alpha_s(Q^2)}{d \log Q^2}=\beta(\alpha_s),
\label{betafun1}\ee and \be
\beta\left(\alpha_{s}\right)=-\left(\frac{\alpha_{s}}{4
\pi}\right)^{2} \sum_{n=0} \left(\frac{\alpha_{s}}{4
\pi}\right)^{n} \beta_{n}. \label{betafun10}\ee

Neglecting quark masses, the first two $\beta$-terms are RS
independent and they have been calculated in
Refs.~\cite{Gross:1973id,Politzer:1973fx,Caswell:1974gg,Jones:1974mm,Egorian:1978zx}
for the $\overline{\rm MS}$ scheme:
$$\beta_{0}=\!\frac{11}{3}C_{A}\!-\!\frac{4}{3}T_{R}N_{f},$$
$$\beta_{1}=\!
\frac{34}{3}C_{A}^{2}\!-\!4\left(\frac{5}{3}C_{A}\!+\!C_{F}\right)T_{R}N_{f}$$
where $C_F=\frac{\left(N_{c}^{2}-1\right)}{2 N_{c}}$, $C_A=N_c$
and $T_R=1/2$ are the color factors for the ${\rm SU(3)}$ gauge group~\cite{Mojaza:2010cm}. \\
At higher loops we have that $\beta_i,$ $i\geq 2$ are scheme
dependent and results for $\overline{\rm MS}$ are calculated in
Ref.~\cite{Larin:1993tp}:
$$
 \beta_{2}=\frac{2857}{2}-\frac{5033}{18}
N_{f}+\frac{325}{54} N_{f}^{2};
$$
in Ref.~\cite{vanRitbergen:1997va}:
\begin{eqnarray}
\beta_{3}&=& \left(\frac{149753}{6}+3564  \zeta_{3} \right)- \left(\frac{1078361}{162}+\frac{6508}{27}  \zeta_{3} \right) N_{f} \nonumber \\
&+& \left(\frac{50065}{162}+\frac{6472}{81} \zeta_{3} \right)
N_{f}^{2}+\frac{1093}{729} N_{f}^{3}; \end{eqnarray} and in
Ref.~\cite{Baikov:2016tgj}:
\begin{eqnarray}
\beta_{4}&=& \left\{\frac{8157455}{16}+\frac{621885}{2} \zeta_{3}-\frac{88209}{2} \zeta_{4}-288090 \zeta_{5}\right. \nonumber \\
&+&N_{f}\left[-\frac{336460813}{1944}-\frac{4811164}{81} \zeta_{3}+\frac{33935}{6}\zeta_{4}+\frac{1358995}{27}\zeta_5 \right] \nonumber \\
&+&N_{f}^{2}\left[\frac{25960913}{1944} +\frac{698531}{81} \zeta_{3}-\frac{10526}{9} \zeta_{4}-\frac{381760}{81}\zeta_{5}\right] \nonumber \\
&+&N_{f}^{3}\left[-\frac{630559}{5832}-\frac{48722}{243}
\zeta_{3}+\frac{1618}{27} \zeta_{4}+\frac{460}{9} \zeta_{5}\right]
\nonumber \\
 &+& \left.
N_{f}^{4}\left[\frac{1205}{2916}-\frac{152}{81}\zeta_{3}\right]\right\},
\end{eqnarray}
 with $\zeta_{3} \simeq 1.20206$, $ \zeta_{4}\simeq
1.08232 $ and $ \zeta_{5} \simeq 1.03693$, the Riemann zeta
function. Given the renormalizability of QCD, new UV singularities
arising at higher orders can be cancelled by redefinition of the
same parameter, i.e. the strong coupling. This procedure leads to
 the renormalization factor:
 \begin{eqnarray}
Z_{a}(\mu)&=& 1-\frac{\beta_{0}}{\epsilon} a+\left(\frac{\beta_{0}^{2}}{\epsilon^{2}}-\frac{\beta_{1}}{2 \epsilon}\right) a^{2} \nonumber \\
& &-\left(\frac{\beta_{0}^{3}}{\epsilon^{3}}-\frac{7}{6} \frac{\beta_{0} \beta_{1}}{\epsilon^{2}}+\frac{\beta_{2}}{3 \epsilon}\right) a^{3} \nonumber \\
& &+\left(\frac{\beta_{0}^{4}}{\epsilon^{4}}-\frac{23 \beta_{1}
\beta_{0}^{2}}{12 \epsilon^{3}}+\frac{5 \beta_{2} \beta_{0}}{6
\epsilon^{2}}+\frac{3 \beta_{1}^{2}}{8
\epsilon^{2}}-\frac{\beta_{3}}{4 \epsilon}\right) a^{4}+\cdots,
\label{eqn:zexp}
\end{eqnarray}
where $a=\alpha_s(\mu)/(4 \pi)$ in the $\rm MS$ scheme. Given the
arbitrariness of the subtraction procedure of including also part
of the finite contributions (e.g. the constant $[\ln(4
\pi)-\gamma_E]$ for the $\overline{\rm MS}$), there is an inherent
ambiguity for these terms which translates into the RS dependence.
In order to solve any truncated Eq.~\ref{betafun1}, this being a
first order differential equation, we need an initial value of
$\alpha_s$ at a given energy scale $\alpha_s(\mu_0)$. For this
purpose we set the initial scale $\mu_0=M_{Z}$ the $Z^0$ mass and
the value $\alpha_s(M_Z)$ is determined phenomenologically. In QCD
the number of colors $N_c$ is set to 3, while $N_f$, i.e. the
number of active flavors, varies with energy scale across quark
thresholds.


\section{One-loop result and asymptotic freedom \label{oneloop}}

When all quark masses are set to zero two physical parameters
dictate the dynamics of the theory and these are the numbers of
flavors $N_f$ and colors $N_c$. We determine in this section the
exact analytical solution to the truncated Eq.~\ref{betafun1}.
Considering the formula : \be
\int_{\alpha_s(\mu_0^2)}^{\alpha_s(\mu^2)} \frac{1}{4\pi}
\frac{d\alpha_{s}}{\beta(\alpha_{s}) }=-\int_{\mu_0^2}^{\mu^2}
\frac{d Q^{2}}{Q^{2}}, \ee
 and retaining only the first term:
\be \frac{Q^{2}}{\alpha_{s}^{2}} \frac{\partial
\alpha_{s}}{\partial Q^{2}}=-\frac{1}{4 \pi} \beta_{0} \ee we
achieve the solution for the coupling:
 \be \frac{4
\pi}{\alpha_{s}\left(\mu_{0}^{2}\right)}-\frac{4
\pi}{\alpha_{s}\left(Q^{2}\right)}=\beta_{0} \ln
\left(\frac{\mu_{0}^{2}}{Q^{2}}\right). \label{1loopalphas} \ee
This solution can be given in the explicit form: \be
\alpha_{s}(Q^{2}) = \frac{\alpha_s(\mu^2_0)}{1+\beta_0 \frac{
\alpha_s(\mu_0^2)}{4 \pi} \ln(Q^2/\mu_0^2)}. \label{1loopalphas2}
\ee This solution relates one known (measured value) of the
coupling at a given scale $\mu_0$ with an unknown value
$\alpha_s(Q^2)$. More conveniently, the solution can be given
introducing the QCD scale parameter $\Lambda$. At $\beta_{0}$
order, this is defined as: \be \Lambda^{2} \equiv \mu^{2}
e^{-\frac{4 \pi}{\beta_{0} \alpha_{x}\left(\mu^{2}\right)}}
\label{lambda1loop}\ee which yields the familiar one-loop
solution:
$$
\alpha_{s}\left(Q^{2}\right)=\frac{4 \pi}{\beta_{0} \ln
\left(Q^{2} / \Lambda^{2}\right)}.
$$

Already at the one loop level one can distinguish two regimes of
the theory. For the number of flavors larger than $11N_c/2$ (i.e.
the zero of the $\beta_0$ coefficient) the theory possesses an
infrared non-interacting fixed point and at low energies the
theory is known as non-abelian quantum electrodynamics
(non-abelian QED). The high energy behavior of the theory is
uncertain, it depends on the number of active flavors and there is
the possibility that it could develop a critical number of flavors
above which the theory reaches an UV fixed
point~\cite{Antipin:2017ebo} and therefore becomes safe. When the
number of flavors is below $11 N_c /2$ the non-interacting fixed
point becomes UV in nature and then we say that the theory is {\it
asymptotically free}.

It is straightforward to check the asymptotic limit of the
coupling in the deep UV region: \be \lim_{s \rightarrow \infty}
\alpha_s(s)=0. \ee This result is known as {\it asymptotic
freedom} and it is the outstanding result that has justified QCD
as the most accredited candidate for the theory of strong
interactions. On the other hand, we have that the perturbative
coupling diverges at the $\Lambda \sim (200-300){\rm MeV}$ scale.
This is sometimes referred to as the {\it Landau ghost pole} to
indicate the presence of a singularity in the coupling that is
actually unphysical and indicates the breakdown of the
perturbative regime. This itself is not an explanation for
confinement, though it might indicate its presence. When the
coupling becomes too large the use of a nonperturbative approach
to QCD is mandatory in order to obtain reliable results. We remark
that the scale parameter $\Lambda$ is RS dependent and its
definition depends on the order of accuracy of the coupling
$\alpha_s(Q^2)$. Considering that the solution $\alpha_{s}$ at
order $\beta_{0}$ or $\beta_{1}$ is universal, the definition of
$\Lambda$ at the first two orders is usually preferred, i.e. the
$\Lambda$ given at 1-loop by Eq.~\ref{lambda1loop} or at 2-loops
(see later) by Eq.~\ref{landaupole} .




\section{Two-loop solution and the perturbative conformal window \label{twoloops}}


In order to determine the solution for the strong coupling
$\alpha_s$ at NNLO, it is convenient to introduce the following
notation: $x(\mu)\equiv \frac{\alpha_s(\mu)}{2 \pi}$,
$t=\log(\mu^2/\mu_0^2)$, $B=\frac{1}{2}\beta_0$ and
$C=\frac{1}{2}\frac{\beta_1}{\beta_0}$, $x^*\equiv -\frac{1}{C}$.
The truncated NNLO approximation of the Eq.~\ref{betafun1} leads
to the differential equation:
\begin{equation}
\frac{dx}{dt}=-B x^2(1+C x) \label{lambert1}
\end{equation}
An implicit solution of Eq.~\ref{lambert1} is given by the Lambert
$W(z)$ function:
\begin{equation}
W e^W = z \label{W}
\end{equation}
with: $ W=\left(\frac{x^*}{ x}-1\right)$. The general solution for
the coupling is:
\begin{eqnarray}
x &=& \frac{x^*}{1+W} , \\
 z &=& e^{\frac{x^*}{x_0}-1} \left(\frac{x^*}{x_0}-1 \right) \left( \frac{\mu^2}{\mu_0^2}
\right)^{x^* B}. \label{xz}
\end{eqnarray}
We will discuss here the solutions to Eq.~\ref{lambert1} with
respect to the particular initial phenomenological value
$x_0\equiv \alpha_s(M_Z) /(2\pi)= 0.01876 \pm 0.00016$ given by
the coupling determined at the $Z^0$ mass
scale~\cite{ParticleDataGroup:2020ssz}.

The signs of $\beta_0,\beta_1$ and consequently of $B,x^*$,
depends on the values of the $N_c,N_f$, since the number $N_c$ is
set by the theory $\rm SU(N_c)$, we discuss the possible regions
varying only the number of flavors $N_f$. We point out that
different regions are defined by the signs of the
$\beta_0,\beta_1$, that have zeros in
$\bar{N_f^0}=\frac{11}{2}N_c$, $\bar{N_f^1}=\frac{34 N_c^3}{13
N_c^2-3}$ respectively with $\bar{N_f^0}> \bar{N_f^1}$.

In the range $N_f<\bar{N_f^1}$ and $N_f>\bar{N_f^0}$ we have
$B>0$, $C>0$ and the physical solution is given by the $W_{-1}$
branch, while for $\bar{N_f^1}< N_f < \bar{N_f^0}$ the solution
for the strong coupling is given by the $W_{0}$ branch. By
introducing the phenomenological value $x_0$, we define a
restricted range for the IR fixed point discussed by Banks and
Zaks~\cite{Banks:1981nn}. Given the value $\bar{N}_f =x^{*-1}(x_0)
= 15.222 \pm 0.009$, we have that in the range $\frac{34 N_c^3}{13
N_c^2-3}< N_f<\bar{N}_f$ the $\beta$-function has both a UV and an
IR fixed point, while for $N_f> \bar{N}_f$ we no longer have the
asymptotically free UV behavior. The two-dimensional region in the
number of flavors and colors where asymptotically free QCD
develops an IR interacting fixed point is colloquially known as
the {\it conformal window of pQCD}.

\begin{figure}[htb]
\centering
\includegraphics[width=8cm]{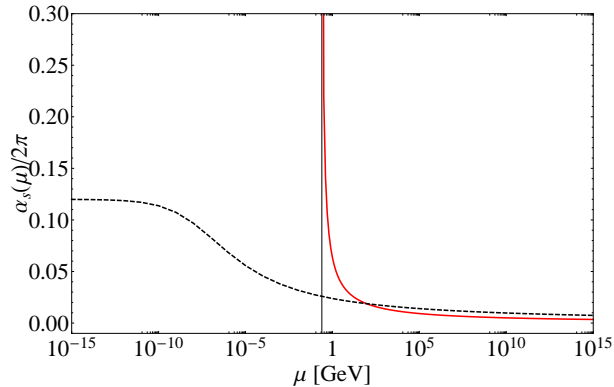}
\caption{The strong running coupling $\alpha_s(\mu)$ for $N_f=12$
(black dashed) and for $N_f=5$ (solid
red).~\cite{DiGiustino:2021nep} } \label{Lambert}
\end{figure}

Thus, the actual physical range of a conformal window for pQCD is
given by $\frac{34 N_c^3}{13 N_c^2-3}< N_f<\bar{N}_f$. The
behavior of the coupling is shown in Fig.~\ref{Lambert}. In the IR
region the strong coupling approaches the IR finite limit, $x^*$,
in the case of values of $N_f$ within the conformal window (e.g.
black dashed curve of Fig.~\ref{Lambert}), while it diverges at
\begin{equation} \Lambda= \mu_0 \left(1+ \frac{|x^*|}{x_0}
\right)^{\frac{1}{2 B |x^*|}} e^{-\frac{1}{2 B x_0}}
\label{landaupole}\end{equation} outside the conformal window
given the solution for the coupling with $W_{-1}$ (e.g. the solid
red curve of Fig.~\ref{Lambert}). The solution of the NNLO
equation for the case $B>0, C>0$, i.e. $N_f<\frac{34 N_c^3}{13
N_c^2-3}$ , can also be given using the standard QCD scale
parameter $\Lambda$ of Eq.~\ref{landaupole},
\begin{eqnarray}
x &=& \frac{x^*}{1+W_{-1}} , \\
 z &=& -\frac{1}{e}  \left( \frac{\mu^2}{\Lambda^2}
\right)^{x^* B}. \label{xz2}
\end{eqnarray}
Different solutions can be achieved using different schemes, i.e.
different definitions of the $\Lambda$ scale
parameter~\cite{Gardi:1998qr}. We underline that the presence of a
Landau ``ghost" pole in the strong coupling is only an effect of
the breaking of the perturbative regime, including
non-perturbative contributions, or using non-perturbative QCD, a
finite limit is expected at any $N_f$~\cite{Deur:2016tte}. Both
solutions have the correct UV asymptotic free behavior. In
particular, for the case $\bar{N}_f<N_f<\frac{11}{2}N_c$, we have
a negative $z$, a negative $C$ and a multi-valued solution, one
real and the other imaginary, actually only one (the real) is
acceptable given the initial conditions, but this solution is not
asymptotically free. Thus we restrict our analysis to the range
$N_f<\bar{N}_f$ where we have the correct UV behavior. In general
IR and UV fixed points of the $\beta$-function can also be
determined at different values of the number of colors $N_c$
(different gauge group $SU(N)$) and $N_f$ extending this analysis
also to other gauge theories~\cite{Ryttov:2017khg}.

\section{$\alpha_s$ at higher loops}

The 3-loop truncated RG equation~\ref{betafun1}, written using the
same normalization of Eq.~\ref{lambert1} is given by: \be
\beta(x)=\frac{d x}{d t}=-B x^{2}\left(1+C x+C_{2} x^{2}\right)
\label{3loopbeta}\ee
 with  $C_2=\frac{\beta_2}{4\beta_0}$.

A straightforward integration of this equation would be hard to
invert, as shown in Ref.~\cite{Gardi:1998qr} it is more convenient
to extend the approach of the previous section
 by using the Pad\'e Approximant (PA).
The Pad\'e Approximant of a given quantity calculated
perturbatively in QCD up to the order $n$, i.e. of the series: \be
S(x)=x\left(1+r_{1} x+r_{2} x^{2}+\cdots+r_{n} x^{n}\right)
\label{pade0} \ee is defined as the rational function:
 \be x[N / M]=x \frac{1+a_{1}
x+\ldots+a_{N} x^{N}}{1+b_{1} x+\ldots+b_{M} x^{M}}:\quad x[N /
M]=S+x \mathcal{O}\left(x^{N+M+1}\right) \label{padeapp} \ee whose
Taylor expansion up to the $N+M=n$ order is identical to the
original truncated series. The use of the PA makes the integration
of Eq.~\ref{3loopbeta} straightforward. PA's may be also used
either to predict the next term of a given perturbative expansion,
called a Pad\'e Approximant prediction (PAP), or to estimate the
sum of the entire series, called Pad\'e Summation. Features of the
PA are shown in Ref.~\cite{Gardi:1996iq}.


The  Pad\'e Approximant ($x^{2}[1 / 1]$) of the 3-loop $\beta$ is
given by the: \be \beta_{\rm PA}(x)=-B x^{2}
\frac{1+\left[C-\left(C_{2} / C\right)\right] x}{1-\left(C_{2} /
C\right) x} \label{padebeta} \ee that leads to the solution: \be B
\ln \left(Q^{2} / \Lambda^{2}\right)=\frac{1}{x}-C \ln
\left[\frac{1}{x}+C-\frac{C_{2}}{C}\right] \ee and finally,

\ba
x\left(Q^{2}\right)&=&-\frac{1}{C} \frac{1}{1-\left(C_{2} / C^{2}\right)+W(z)} \\
z&=&-\frac{1}{C} \exp \left[-1+\left(C_{2} / C^{2}\right)-B t /
C\right], \ea the sign of $C$ determines the sign of $z$ and also
the physically relevant branches of the Lambert function $W(z)$ :
for $C>0, z<0$ and the physical branch is $W_{-1}(z)$, taking real
values in the range $(-\infty,-1)$, while for $C<0, z>0$ and the
physical branch is given by the $W_{0}(z)$, taking real values in
the range $(0, \infty)$.

We notice that the only significant difference between the 3-loop
solution and the 2-loop solution \ref{xz} is in the solution
 $x\left(Q^{2}\right)$. This is because the
difference in the definition of $z$ can be reabsorbed into an
appropriate redefinition of the scale parameter:
$$
\Lambda^{2} \longrightarrow \tilde{\Lambda}^{2}=\Lambda^{2}
e^{C_{2} /\left(B C\right)}.$$


%
%
%
%


For orders up to $\beta_{4}$, an approximate analytical solution
is obtained integrating Eq.~\ref{betafun1} :\begin{eqnarray}
\ln \frac{\mu^{2}}{\Lambda^{2}} &=& \int \frac{d a}{\beta(a)}=\frac{1}{\beta_{0}}\left[\frac{1}{a}+b_{1} \ln a+a\left(-b_{1}^{2}+b_{2}\right)\right.\nonumber \\
&&+a^{2}\left(\frac{b_{1}^{3}}{2}-b_{1} b_{2}+\frac{b_{3}}{2}\right)+a^{3}\left(-\frac{b_{1}^{4}}{3}+b_{1}^{2} b_{2}-\frac{b_{2}^{2}}{3}\right. \nonumber \\
&&-\left.\left.\frac{2}{3} b_{1}
b_{3}+\frac{b_{4}}{3}\right)+O\left(a^{4}\right)\right]+C
\end{eqnarray}
where $a=\alpha_s(\mu)/(4 \pi)$ and $b_N\equiv \beta_N/\beta_0$,
$(N=1,..,4)$ and performing the inversion of the last formula by
iteration as shown in Ref.~\cite{Kniehl:2006bg}, achieving the
final result of the coupling at five-loop accuracy :
\begin{eqnarray}
a &=& \frac{1}{\beta_{0} L}-\frac{b_{1} \ln L}{\left(\beta_{0} L\right)^{2}}+\frac{1}{\left(\beta_{0} L\right)^{3}}\left[b_{1}^{2}\left(\ln ^{2} L-\ln L-1\right)+b_{2}\right] \nonumber \\
&&+\frac{1}{\left(\beta_{0} L\right)^{4}}\left[b_{1}^{3}\left(-\ln ^{3} L+\frac{5}{2} \ln ^{2} L+2 \ln L-\frac{1}{2}\right)\right. \nonumber\\
&&-\left. 3 b_{1} b_{2} \ln L+\frac{b_{3}}{2}\right]+\frac{1}{\left(\beta_{0} L\right)^{5}}\left[b _ { 1 } ^ { 4 } \left(\ln ^{4} L-\frac{13}{3} \ln ^{3} L\right.\right. \nonumber \\
&&-\left. \frac{3}{2} \ln ^{2} L+4 \ln L+\frac{7}{6}\right)+3 b_{1}^{2} b_{2}\left(2 \ln ^{2} L-\ln L-1\right) \nonumber \\
&&-\left. b_{1} b_{3}\left(2 \ln L+\frac{1}{6}\right)+\frac{5}{3}
b_{2}^{2}+\frac{b_{4}}{3}\right]+O\left(\frac{1}{L^{6}}\right) .
\end{eqnarray}
where $L=\ln(\mu^2/\Lambda^2).$ The same definition of $\Lambda$
scale given in Eq.~\ref{landaupole} has been used for the
$\overline{\rm MS}$ scheme which leads to set the constant $C=
\left(b_{1} / \beta_{0}\right) \ln (\beta_{0})$.


\section{The $\beta$ coefficients in different schemes}

The $\beta_{i}$ are the coefficients of the $\beta$-function
arising in the loop expansion, i.e. in orders of $\hbar$. Although
the first two coefficients $\beta_0,\beta_1$ are universal scheme
independent coefficients depending only on the number of colors
$N_c$ and flavors $N_f$, the higher-order terms are in contrast
scheme dependent. In particular, for the 't Hooft
scheme~\cite{thooftscheme} the higher $\beta_i, i\geq 2$ terms are
set to zero, leading to the solution of Eq.~\ref{twoloops} for the
$\beta$-function valid at all orders. Moreover, in all $\rm
MS$-like schemes all the $\beta_{i}$ coefficients are gauge
independent, while other schemes like the momentum space
subtraction (MOM) scheme~\cite{Celmaster:1979km} depend on the
particular gauge. Using the Landau gauge, the $\beta$ terms for
the MOM scheme are given by~\cite{Boucaud:2005rm}:
$$
\beta_{2}=3040.48-625.387 N_f +19.3833 N_f^{2}
$$
and
$$
\beta_{3}=100541-24423.3 N_f+1625.4 N_f^{2}-27.493 N_f^{3} .
$$
Results for the minimal MOM scheme and Landau gauge are shown in
Ref.~\cite{Chetyrkin:2000dq}. The renormalization condition for
the MOM scheme sets the virtual quark propagator to the same form
as a free massless propagator. Different MOM schemes exist and the
above values of $\beta_{2}$ and $\beta_{3}$ are determined with
the MOM scheme defined by subtracting the 3-gluon vertex to a
point with one null external momentum. This leads to a coupling
which is not only RS dependent but also gauge-dependent. The
values of $\beta_{2}$ and $\beta_{3}$ given here are only valid in
the Landau gauge. Values in the $\mathrm{V}$-scheme defined by the
static heavy quark potential~\cite{Appelquist:1977tw,
Fischler:1977yf, Peter:1996ig, Schroder:1998vy, Smirnov:2008pn,
Smirnov:2009fh, Anzai:2009tm} can be found in
Ref.~\cite{Kataev:2015yha}. They result in:
$\beta_{2}=4224.18-746.01 N_f+20.88 N_f^{2}$ and
$\beta_{3}=43175-12952 N_f+707.0 N_f^{2}$ respectively. We recall
that the sign of the $\beta_{i}$ controls the running of
$\alpha_{s}$. We have $\beta_{0}>0$ for $N_f \leq 16, \beta_{1}>0$
for $N_f \leq 8, \beta_{2}>0$ for $N_f \leq 5$, and $\beta_{3}$ is
always positive. Consequently, $\alpha_{s}$ decreases at high
momentum transfer, leading to the asymptotic freedom of pQCD. Note
that, $\beta_{i}$ are sometimes defined with an additional
multiplying factor $1 /(4 \pi)^{i+1}$.
 Different schemes are characterized by different $\beta_i, i\geq 2
 $ and lead to different definitions for the effective coupling.


\section{The $\Lambda$ parameter and quark thresholds \label{sec:lambdapar}}

The $\Lambda$ parameter represents the Landau ghost pole in the
perturbative coupling in QCD. We recall that the Landau pole was
initially identified in the context of Abelian QED. However, the
presence of this pole does not affect QED. Given its value,
$\Lambda \sim 10^{30-40}{\rm GeV}$, above the Planck
scale~\cite{Gockeler:1997dn}, at which new physics is expected to
occur in order to suppress the unphysical divergence. The QCD
$\Lambda$ parameter in contrast is at low energies, its value
depends on the RS, on the order of the $\beta$-series,
$\beta_{i}$, on the approximation of the coupling $\alpha_s(\mu)$
at orders higher than $\beta_{1}$ and on the number of flavors
$N_f$. Although mass corrections due to light quarks at higher
order in perturbative calculations introduce negligible terms,
they actually indirectly affect $\alpha_s$ through $N_f$. In fact,
the number of active quark flavors runs with the scale $Q^2$ and a
quark $q$ is considered active in loop integration if the scale
$Q^2\geq m^2_q$. Thus, in general, light quarks can be considered
massless regardless of whether they are active or not, while
$\alpha_s$ varies smoothly when passing a quark threshold, rather
than in discrete steps. The matching of the values of $\alpha_s$
below and above a quark threshold makes $\Lambda$ depend on $N_f$.
Matching requirements at leading order $\beta_0$, imply that:
$$
\alpha_{s}^{N_f-1}\left(Q^{2}=m_{q}^{2}\right)=\alpha_{s}^{N_f}\left(Q^{2}=m_{q}^{2}\right)
$$
and therefore that:
$$
\Lambda^{N_f}=\Lambda^{N_f-1}\left(\frac{\Lambda^{N_f-1}}{m_{q}}\right)^{2
/\left(33-2 N_f\right)}
$$
The formula with $\beta_{1}$, can be found in~\cite{Larin:1994va}
and the four-loop matching in the $\overline{\rm MS}$ RS is given
in~\cite{Chetyrkin:1997sg}.

As shown in the previous section at the lowest order $\beta_{0}$,
the Landau singularity is a simple pole on the positive real axis
of the $Q^2$-plane, whereas at higher order it acquires a more
complicated structure. This pole is unphysical and is located on
the positive real axis of the complex $Q^{2}$-plane. This
singularity of the coupling indicates that the perturbative regime
of QCD breaks down and it may also suggest that a new mechanism
takes over, such as the confinement. Thus, the value of $\Lambda$
is often associated with the confinement scale, or equivalently to
the hadronic mass scale. An explicit relation between hadron
masses and the $\Lambda$ scale has been obtained in the framework
of holographic QCD~\cite{Brodsky:2014yha}. Landau poles on the
other hand, usually do not appear in nonperturbative approaches
such as AdS/QCD. Approximate values of $\Lambda$ in different
schemes are given in Table \ref{tablambda}:
\begin{table}
\begin{center}
\begin{tabular}{|c|c|c|c|c|}
\hline \multicolumn{5}{|c|}{ The numerical values of $\Lambda$ in
different schemes, ${\rm MeV}$}\\ \hline $N_{f}$ & the order of
approximation $\nu$ & $\Lambda_{\overline{\rm
MS}}^{\left(N_{f}\right)}$ & $\Lambda_{\rm
V}^{\left(N_{f}\right)}$ &
$\Lambda_{\rm mMOM}^{\left(N_{f}\right)}$ \\
\hline 4 & 2 & 350 & 500 & 625 \\
\hline 4 & 3 & 335 & 475 & 600 \\
\hline 4 & 4 & 330 & 470 & 590 \\
\hline 5 & 2 & 250 & 340 & 435 \\
\hline 5 & 3 & 245 & 335 & 430 \\
\hline 5 & 4 & 240 & 330 & 420 \\
\hline
\end{tabular}
\caption{Results for the $\Lambda$ parameter in different schemes,
at different values of the number of active flavor, $N_{f}$, and
at different orders of accuracy $\nu$~\cite{Kataev:2015yha}.}
\label{tablambda}
\end{center}
\end{table}

Different schemes are related perturbatively by: \be
\alpha_{s}^{(2)}\left(Q^{2}\right)=\alpha_{s}^{(1)}\left(Q^{2}\right)\left[1+v_{1}
\alpha_{s}^{(1)}\left(Q^{2}\right) /(4 \pi)\right]
+\mathcal{O}(\alpha_s^2)\ee where $v_{1}$ is the leading order
difference between $\alpha_{s}\left(Q^{2}\right)$ in the two
schemes. In the case of the V-scheme and $\overline{\rm MS}$
scheme we have : $v_1^{\overline{\mathrm{MS}}}=\frac{31}{9}
C_{A}-\frac{20}{9} T_{F} n_{l}$.

Thus, the relation between $\Lambda_{1}$ in a scheme 1 and
$\Lambda_{2}$ in a scheme 2 is, at the one-loop order, given by:
$$
\Lambda_{2}=\Lambda_{1} e^{\frac{ v_{1}}{2 \beta_{0}}}.
$$
For example, the $\overline{\rm MS}$ and V-scheme scale parameters
are related by:
$$
\Lambda_{\rm V}=\Lambda_{\overline{\rm MS}} e^{\frac{93-10 N_f}{2(
99-6 N_f)}}
$$
The relation is valid at each threshold translating all values for
the scale from one scheme to the other.

\begin{figure}
\begin{center}
  \includegraphics[width=6cm]{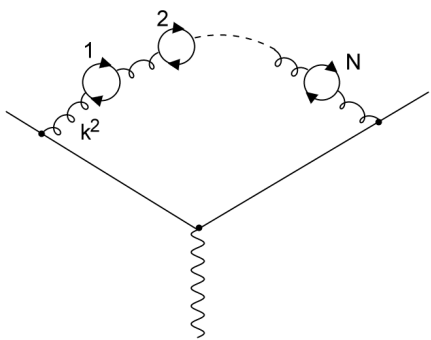}\\
  \caption{Example of a diagram with the ``bubble-chain" insertion.}\label{bubbles}
  \end{center}
\end{figure}

In general, one may think that lower values of the scale parameter
lead to slower increasing couplings in the IR. Unfortunately other
effects can occur spoiling this criterion. In fact the nature of
the perturbative expansion is affected by the renormalon
growth~\cite{thooftscheme} of the coefficients. The renormalons
affect renormalizable gauge theories only, they stem from
particular diagrams known as ``bubble-chain" diagrams and shown in
Fig.~\ref{bubbles}.

In loop integration these terms introduce a factorial growth,
using Eq.~\ref{1loopalphas} and considering $k^2 \geq \mu^2$ we
obtain:
$$\!\int \! d^4k k^{-2n}\alpha_s(k^2)\!=\! \alpha_s(\mu) \sum_N \int d^4k
k^{-2n} (\beta_0 \alpha_s(\mu) \ln(k^2/\mu^2))^N \!\sim\! \sum_N
N! (\frac{\beta_0}{n-2})^N \alpha_s(\mu)^{N+1},$$ with $n \geq 3$.
Performing the Borel transform of the last series one obtains a
geometric series, \be B(z)=\sum_N
\left(\frac{\beta_0}{n-2}\right)^N z^N
=\frac{1}{1-\left(\frac{\beta_0}{n-2}\right)z}\ee which has poles
on the real positive axis: \be z_n=\frac{n-2}{\beta_0}, \quad
n=3,4,5,... \ee These singularities introduce an ambiguity in the
inverse Borel transform: \be F(\alpha_s(\mu))=\int^\infty_0 dt
e^{-t/\alpha_s(\mu)} B(t), \label{borel2}\ee since they lie on the
path of integration and lead to non-zero residue contributions of
the type: \be \Delta=\left(\frac{\Lambda}{\mu}\right)^{2 \beta_0
z_n }. \ee IR and UV renormalons arise as singularities on the
real negative and positive axis of the complex $z$ plane of the
Borel transform, (analogously to instanton poles, this explains
the name given by 't Hooft), and are related to the $\Lambda$
scale and thus to the Landau pole of the strong coupling. These
terms affect the coefficients of the perturbative QCD series and
its convergence. It has been shown in several applications of
resummed quantities (i.e. applying the technique of resummation of
large IR logarithms) renormalons do not affect the final result if
one uses an appropriate prescription (e.g. the minimal
prescription (MP) formula~\cite{Catani:1989ne,DiGiustino:2011jn}).
Reviews on renormalons exist in the literature, e.g.
Refs.~\cite{Altarelli:1995kz,Beneke:1998ui}.

Thus, different growths of the coefficients in different
renormalization schemes can balance the differences in values of
the corresponding scales $\Lambda_{1,2}$. However, the growth of
the coefficients is not only due to renormalons, in some cases the
coefficients have inherently fast rising behavior as shown in
Ref.~\cite{Brodsky:2000cr}.
For further insights, relations among different RS and their associated $\Lambda$ are discussed in Ref.~\cite{Celmaster:1979km}.\\

\section{The Renormalization Group Equations}

The scale dependence of the coupling can be determined considering
that the bare coupling $\overline{\alpha_{s}}$ and renormalized
couplings, $\alpha_s$, at different scales are related by:
 \be
\overline{\alpha_{s}}=Q^{2 \varepsilon}
Z_{\alpha}\left(Q^{2}\right) \alpha_{s}\left(Q^{2}\right)=\mu^{2
\varepsilon} Z_{\alpha}\left(\mu^{2}\right)
\alpha_{s}\left(\mu^{2}\right),
 \label{eqn:alphascales}\ee
where $\varepsilon$ is the regularization parameter, integrals are
carried out in $4-2 \varepsilon$ dimensions and the UV divergences
are regularized to $1 / \varepsilon$ poles. The $Z_{i}$ are
constructed as functions of $1 / \varepsilon$, such that they
cancel all $1 / \varepsilon$ poles. From Eq.~\ref{eqn:alphascales}
we can obtain the relation from two different couplings at two
different scales: \be
\alpha_{s}\left(Q^{2}\right)=\mathcal{Z}_{\alpha}\left(Q^{2},
\mu^{2}\right) \alpha_{s}\left(\mu^{2}\right), \ee with
$\mathcal{Z}_{\alpha}\left(Q^{2}, \mu^{2}\right)
\equiv\left(\mu^{2 \varepsilon} / Q^{2
\varepsilon}\right)\left[Z_{\alpha}\left(Q^{2}\right) /
Z_{\alpha}\left(\mu^{2}\right)\right]$.

The $\mathcal{Z}_{\alpha}$ form a group with a composition law:
\be \mathcal{Z}_{\alpha}\left(Q^{2},
\mu^{2}\right)=\mathcal{Z}_{\alpha}\left(Q^{2}, \mu_{0}^{2}\right)
\mathcal{Z}_{\alpha}\left(\mu_{0}^{2}, \mu^{2}\right),\ee a unity
element: $\mathcal{Z}_{\alpha}\left(Q^{2}, Q^{2}\right)=1$ and an
inversion law: $\mathcal{Z}_{\alpha}\left(Q^{2},
\mu^{2}\right)=\mathcal{Z}_{\alpha}^{-1}\left(\mu^{2},
Q^{2}\right)$. Fundamental properties of the Renormalization group
are:{\it reflexivity, symmetry and transitivity}. Thus the scale
invariance of a given perturbatively calculated quantity is
recovered by the invariance of the theory under the
Renormalization Group Equations (RGE). As previously discussed,
the QCD Lagrangian has only one dimensionless coupling constant
$g_s$ and in the case of a massless theory the Lagrangian has no
particular mass scale, but is completely conformal. The only
energy scale, $\mu$, is introduced by the subtraction point with
the renormalization procedure. Thus, all phenomena involving
strong interactions can be described by only one parameter, the
strong coupling $\alpha_s(Q^2)$, and the same parameter is
responsible for the interaction at low and high energy scales. As
shown in the previous section, the strong coupling has a running
behavior, it evolves and increases with the energy of the probe,
$Q^2$, from high to low values of the momentum transfer. Thus if
we consider the case of the physical observable at an energy
$s\geq m_f^2$, in order to neglect masses, for example the ratio:
 \be
 \label{eqn:R}
 R_{e^+
e^-}
(s)=\frac{\sigma(e^+e^-\rightarrow\mbox{hadrons})}{\sigma(e^+e^-\rightarrow
\mu^+ \mu^-)}
 \ee
where the cross sections are given at the lowest order by:
  \be
\sigma(e^+e^-\rightarrow \mu^+ \mu^-)=\frac{4\pi\alpha^2}{3s},
 \ee
 and
  \be
 \sigma(e^+e^-\rightarrow
{\rm hadrons })=\frac{4\pi\alpha^2}{3s}N_c\sum_f Q_f^2,
 \ee
with $\alpha\equiv e^2/4\pi$ the fine structure constant of QED.
As shown in the formula the single cross sections depend on the
center-of-mass energy $s=Q^2$, but this dependence cancels in the
$R_{e^+ e^-}$ at lowest order. Also by dimensional analysis, a
constant value of the observable $R_{e^+ e^-}$ would be predicted
independently of any energy scale $s$, the ratio being a
dimensionless quantity. However, higher order loop integrations
and the renormalization procedure of the coupling constant
introduce a scale dependence. Since $R$ is dimensionless and since
there is no mass scale in the QCD Lagrangian, scale dependence can
be introduced only via the $\mu$ scale dependence of the coupling
$\alpha_{s}(\mu)$ and via a ratio
$\left(Q^{2}/\mu^{2}\right)$-like dependence of the perturbative
coefficients. In fact, except for the first two terms
$r_{0},r_{1},$ which are scale independent, the coefficients
$r_{n}$ are polynomials of $\ln\left( Q^{2}/\mu^{2}\right)$ with
highest power $n-1$. By means of the RGE all these logarithms can
be reabsorbed into the running coupling. The purpose of taking
$\alpha_{s}$ scale-dependent is to transfer to $\alpha_{s}$ all
terms involving $\mu$ in the perturbative series of
$R\left(s\right)$.
 The independence of $R$ with respect to $\mu$ is given by the Callan-Symanzik
relation for QCD~\cite{Callan:1970yg,Symanzik:1971vw}:

%

 \be
 R_{e^+ e^-}(s;\mu)=N_c\sum_f Q_f^2\left[1+R(s;\mu)\right].
 \ee
where $Q_f$ are quark charges summed over the flavor index $f$ and
\be R(s;\mu)=\sum_{n=1}^{\tilde{n}} r_n(s;\mu) \left(
\alpha_s(\mu)/\pi\right)^n\ee.

\be \mu^2 \frac{d}{d\mu^2}R_{e^+ e^-}(s,\mu)=0, \ee equivalently:
\be \label{eqn:Rcallan} \left[ \mu^2\frac{\partial}{\partial
\mu^2}+\beta(\alpha_s)\frac{\partial}{\partial \alpha_s}\right]
R_{e^+ e^-}(s;\mu)=0, \ee where \be \label{eqn:betag}
\beta(\alpha_s)=\mu^2 \frac{\partial \alpha_s}{\partial \mu^2},
\ee setting the renormalization scale $\mu$ equal to the physical
scale $Q$ would remove the $\ln(Q^{2}/\mu^{2})$ in the
coefficients $r_{n}$ and fold the $\mu$-dependence into
$\alpha_{s}\left(\mu^{2}=s\right)$. Thus the option of choosing
$\mu^{2}=s$ yields the simplest form for the perturbative
expansions of given observable.




\section{The Extended Renormalization Group Equations}

Given that physical predictions cannot depend on the choice of the
renormalization scale nor on the scheme, the same approach used
for the renormalization scale based on the invariance under RGE is
extended to scheme transformations. This approach leads to the
Extended Renormalization Group Equations, which were introduced
first by St\"uckelberg and
Peterman~\cite{StueckelbergdeBreidenbach:1952pwl}, then discussed
by
Stevenson~\cite{Stevenson:1980du,Stevenson:1981vj,Stevenson:1982wn,Stevenson:1982qw}
and also improved by Lu and Brodsky~\cite{Lu:1992nt}. A physical
quantity, $R$, calculated at the $N$-th order of accuracy is
expressed as a truncated expansion in terms of a coupling constant
$\alpha_{S}(\mu)$ defined in the scheme $\rm S$ and at the scale
$\mu$, such as: \be R_{N}=r_{0} \alpha_{S}^{p}(\mu)+r_{1}(\mu)
\alpha_{S}^{p+1}(\mu)+\cdots+r_{N}(\mu) \alpha_{S}^{p+N}(\mu).
\label{eqn:truncated}\ee At any finite order, the scale and scheme
dependencies of the coupling constant $\alpha_{S}(\mu)$ and of the
coefficient functions $r_{i}(\mu)$ do not totally cancel, this
leads to a residual dependence in the finite series and to the
scale and scheme ambiguities.

In order to generalize the RGE approach it is convenient to
improve the notation by introducing the universal coupling
function as the extension of an ordinary coupling constant to
include the dependence on the scheme parameters
$\left\{c_{i}\right\}:$

\be \alpha=\alpha\left(\mu / \Lambda,\left\{c_{i}\right\}\right) .
\ee where $\Lambda$ is the standard two-loop $\overline{\rm MS}$
scale parameter. The subtraction prescription is now characterized
by an infinite set of continuous {\it scheme parameters}
$\left\{c_{i}\right\}$ and by the renormalization scale $\mu$.
Stevenson~\cite{Stevenson:1981vj} has shown that one can identify
the beta-function coefficients of a given renormalization scheme
with the scheme parameters.  Considering that the first two
coefficients of the $\beta$-function are scheme independent, each
scheme is identified by its $\left\{\beta_{i}, \quad i=2,3,
\ldots\right\}$ parameters.

More conveniently, let us define the rescaled coupling constant
and the rescaled scale parameter as \be
a=\frac{\beta_{1}}{\beta_{0}} \frac{\alpha}{4 \pi}, \quad
\tau=\frac{2 \beta_{0}^{2}}{\beta_{1}} \log (\mu / \Lambda) . \ee
Then, the rescaled $\beta$-function takes the canonical form: \be
\beta(a)=\frac{d a}{d \tau}=-a^{2}\left(1+a+c_{2} a^{2}+c_{3}
a^{3}+\cdots\right) \ee with $c_{n}=\beta_{n} \beta_{0}^{n-1} /
\beta_{1}^{n}$ for $n=2,3, \cdots$.


The scheme and scale invariance of a given observable $R$, can be
 expressed as: \ba \frac{\delta R}{\delta \tau} &=& \beta
\frac{\partial
R}{\partial a}+\frac{\partial R}{\partial \tau}=0 \nonumber \\
\frac{\delta R}{\delta c_{n}}&=&\beta_{(n)} \frac{\partial
R}{\partial a}+\frac{\partial R}{\partial c_{n}}=0.
\label{extendedrge} \ea



The fundamental beta function that appears in
Eqs.~\ref{extendedrge} reads: \be
\beta\left(a,\left\{c_{i}\right\}\right) \equiv \frac{\delta
a}{\delta \tau}=-a^{2}\left(1+a+c_{2} a^{2}+c_{3}
a^{3}+\cdots\right) \ee and the extended or scheme-parameter beta
functions are defined as: \be
\beta_{(n)}\left(a,\left\{c_{i}\right\}\right) \equiv \frac{\delta
a}{\delta c_{n}} . \ee The extended beta functions can be
expressed in terms of the fundamental beta function. Since the
$(\tau,\{c_i\})$ are independent variables, second partial
derivatives respect the commutativity relation: \be
\frac{\delta^{2} a}{\delta \tau \delta c_{n}}=\frac{\delta^{2}
a}{\delta c_{n} \delta \tau}, \ee which implies \be \frac{\delta
\beta_{(n)}}{\delta \tau}=\frac{\delta \beta}{\delta c_{n}},\ee
\be \beta \beta_{(n)}^{\prime}=\beta_{(n)} \beta^{\prime}-a^{n+2},
\ee where $\beta_{(n)}^{\prime}=\partial \beta_{(n)} / \partial a$
and $\beta^{\prime}=\partial \beta / \partial a$. From here \be
\beta^{-2}\left(\frac{\beta_{(n)}}{\beta}\right)^{\prime}=-a^{n+2},
\ee \be
\beta_{(n)}\left(a,\left\{c_{i}\right\}\right)=-\beta\left(a,\left\{c_{i}\right\}\right)
\int_{0}^{a} d x
\frac{x^{n+2}}{\beta^{2}\left(x,\left\{c_{i}\right\}\right)}, \ee
where the lower limit of the integral has been set to satisfy the
boundary condition
$$
\beta_{(n)} \sim O\left(a^{n+1}\right) .
$$
That is, a change in the scheme parameter $c_{n}$ can only affect
terms of order $a^{n+1}$ or higher in the evolution of the
universal coupling function.

The extended renormalization group equations
Eqs.~\ref{extendedrge} can be written in the form: \ba
\frac{\partial R}{\partial \tau}&=& -\beta \frac{\partial R}{\partial a} \nonumber \\
\frac{\partial R}{\partial c_{n}}&=&-\beta_{(n)} \frac{\partial
R}{\partial a}. \label{eqn:pms}\ea

Thus, provided we know the extended beta functions, we can
determine any variation of the expansion coefficients of $R$ under
scale-scheme transformations. In particular, we can evolve a given
perturbative series into another determining the expansion
coefficients of the latter and vice versa.


\chapter{Renormalization Scale Setting in QCD \label{RSSQCD}}

The scale-scheme ambiguities are an important source of
uncertainties in many processes in perturbative QCD preventing
precise theoretical predictions for both SM and BSM physics. In
principle, an infinite perturbative series is void of this issue,
given the scheme and scale invariance of the entire physical
quantities~\cite{StueckelbergdeBreidenbach:1952pwl,GellMann:1954fq,Peterman:1978tb,Callan:1970yg,Symanzik:1971vw},
in practice perturbative corrections are known up to a certain
order of accuracy and scale invariance is only approximated in
truncated series, leading to the scheme and scale
ambiguities~\cite{Celmaster:1979km,Buras:1979yt,Stevenson:1980du,Stevenson:1981vj,Stevenson:1982qw,Stevenson:1982wn,Grunberg:1980ja,Grunberg:1982fw,Grunberg:1989xf,Brodsky:1982gc,Chishtie:2015lwk,Chishtie:2016wob,Abbott:1980hwa}.
If on one hand, according to the conventional practice, or
conventional scale setting (CSS), this problem cannot be avoided
and is responsible for part of the theoretical errors, on the
other hand some strategies for the optimization of the truncated
expansion have been proposed, such as the Principle of Minimal
Sensitivity proposed by Stevenson~\cite{Stevenson:1981vj}, the
Fastest Apparent Convergence criterion introduced by
Grunberg~\cite{Grunberg:1980ja} and the Brodsky-Lepage-Mackenzie
(BLM) method~\cite{Brodsky:1982gc}. These are procedures commonly
in use for scale setting in perturbative QCD. In general, a
scale-setting procedure is considered reliable if it preserves
important self consistency requirements. All Renormalization Group
properties such as: {\it uniqueness, reflexivity, symmetry,} and
{\it transitivity} should be preserved also by the scale-setting
procedure in order to be generally applied~\cite{Brodsky:2012ms}.

In fact, once the optimal scale is set, by means of the RG
properties is possible to relate results in different schemes and
different observables. Other requirements are also suggested by
tested theories, by the convergence behavior of the series and
also for phenomenological reasons or scheme independence. We
discuss in this chapter the different optimization procedures and
their properties. An introduction to these methods can also be
found in Refs.~\cite{Wu:2013ei,Deur:2016tte}.


%


\section{Conventional scale setting - CSS}


According to common practice a first evaluation of the physical
observable is obtained by calculating perturbative corrections in
a given scheme (commonly used are $\rm MS$ or $\overline{\rm MS}$)
and at an initial renormalization scale $\mu_{r}=\mu_{r}^{\rm init
}$, obtaining the truncated expansion:

\begin{equation}
\rho_{n}=\mathcal{C}_{0}
\alpha_{s}^{p}\left(\mu_{r}\right)+\sum_{i=1}^{n}
\mathcal{C}_{i}\left(\mu_{r}\right)
\alpha_{s}^{p+i}\left(\mu_{r}\right), \quad(p \geq 0),
\label{observable-init0}
\end{equation}
where $\mathcal{C}_{0}$ is the tree-level term, while
$\mathcal{C}_{1}, \mathcal{C}_{2}, ...,\mathcal{C}_{n}$ are the
one-loop , two-loop, n-loop corrections respectively and $p$ is
the power of the coupling at tree-level.

In order to improve the pQCD estimate of the observable, after the
initial renormalization a change of scale using the RGE and a
chosen scale-setting method is performed in
Eq.~\ref{observable-init0}, which leads to: \be
\rho_{n}=\mathcal{C}_{0}
\alpha_{s}^{p}\left(\tilde{\mu}_{r}^{0}\right)+\sum_{i=1}^{n}
\overline{\mathcal{C}}_{i}\left(\tilde{\mu}_{\mathrm{r}}^{i}\right)
\alpha_{s}^{p+i}\left(\bar{\mu}_{r}^{i}\right), \quad(p \geq 0)
\label{observable-ren1} \ee where the new leading-order (LO) and
higher-order scales $\tilde{\mu}_{r}^{0}$ and
$\tilde{\mu}_{r}^{i}$ are functions of the initial renormalization
scale $\mu_{r}^{\rm init }$, and they depend the particular choice
of the scale-setting method. At the same time, the new
coefficients
$\overline{\mathcal{C}}_{i}\left(\bar{\mu}_{r}^{i}\right)$ are
changed accordingly in order to obtain a consistent result.

The simple CSS procedure starts from the scale and scheme
invariance of a given observable, which translates into complete
freedom for the choice of the renormalization scale. In practice
in this approach, the initial scale is directly fixed to the
typical momentum transfer of the process , $Q$ , or to a value
which minimizes the contributions of the loop diagrams and the
errors are evaluated varying the value of $Q$ in the range of 2,
$[Q / 2,2 Q]$.


It is often claimed that this simple scale setting method
estimates contributions from higher-order terms and that, due to
the perturbative nature of the expansion,  the introduction of
higher order corrections would reduce the scheme and scale
ambiguities order by order.


No doubt that the higher the loop corrections are calculated, the
greater is the precision of the theoretical estimations in
comparison with the experimental data, but we cannot know {\it a
priori} the level of accuracy necessary for the CSS to achieve the
desired precision and at present in the majority of cases only the
NNLO corrections are available. Besides this, the divergent nature
of the asymptotic perturbative series and the presence of
factorial growing terms (i.e. renormalons) severely compromise the
theoretical predictions.

However, even though this procedure may give an indication on the
level of conformality and convergence reached by the truncated
expansion, it leads to a numerical evaluation of theoretical
errors that is quite unsatisfactory and dependent strictly on the
value of the chosen scale. Different choices of the
renormalization scale may lead to very different results when
including higher order corrections. For example, the NLO
correction in W+3jets with the BlackHat code~\cite{Berger:2009ep}
can be either negligible or extremely severe depending on the
choice of the particular renormalization scale. One may argue that
the proper renormalization scale for a fixed-order prediction can
be judged by comparing theoretical results with experimental data,
but this method would be strictly process dependent and would
compromise the predictivity of the pQCD approach.

Besides the complexity of the higher order calculations and the
slow convergence of the perturbative series, there are many
critical points in the CSS method:

 \begin{itemize}
 \item
In general, no one knows the proper renormalization scale value,
$Q$, and the correct range where the scale and scheme parameters
should be varied in order to have the correct error estimate. In
fact, in some processes there can be more than one typical
momentum scale that can be taken as the renormalization scale
according to the CSS procedure, for example in processes involving
heavy quarks, typical scales are either the center-of-mass energy
$\sqrt{s}$ or also the heavy quark mass. Moreover, the idea of the
typical momentum transfer as the renormalization scale only sets
the order of magnitude of the scale, but does not indicate the
optimal scale;

 \item No distinction is made
among different sources of errors and their relative
contributions, e.g. in addition to the errors due to scale-scheme
uncertainties there are also the errors from missing higher-order
uncalculated terms. In such an approach, theoretical uncertainties
can become quite arbitrary and unreliable;

\item The convergence of the perturbative series in QCD is
affected by uncancelled large logarithms as well as by
``renormalon" terms that diverge as $\left(n ! \beta_{i}^{n}
\alpha_{s}^{n+1}\right)$ at higher
orders~\cite{Gross:1974jv,Lautrup:1977hs}, this is known as the
{\it renormalon problem}~\cite{Beneke:1998ui}. These renormalon
terms can give sizable contributions to the theoretical estimates,
as shown in $e^{+} e^{-}$annihilation, $\tau$ decay, deep
inelastic scattering, hard processes involving heavy quarks. These
terms are responsible for important corrections at higher orders
also in the perturbative region, leading to different predictions
according to different choices of the scale (as shown in
Ref.~\cite{Berger:2009ep}). Large logarithms on the other hand can
be resummed using the resummation
technique~\cite{Catani:1991kz,Catani:1992ua,Catani:1996yz,Aglietti:2006wh,DiGiustino:2011jn,Banfi:2014sua,Abbate:2010xh}
and results are IR renormalon free. This does not help for the
renormalization scale and scheme ambiguities, which still affect
theoretical predictions with or without resummed large logarithms.
In fact, as recently shown in Ref.~\cite{Catani:2020kkl} for the
$b\bar{b}$ production cross section at NNLO order of accuracy at
hadron colliders, the CSS scale setting leads to theoretical
uncertainties that are of the same order of the NNLO corrections
$\sim 20-30\%$ taking as the typical momentum scale the b-quark
mass $m_b\sim 4.92$GeV.

\item In the Abelian limit $N_{c} \rightarrow 0$ at fixed
$\alpha_{e m}=C_{F} \alpha_{s}$ with
$C_{F}=\left(N_{c}^{2}-1\right) / 2 N_{c}$, a QCD case approaches
effectively the QED analogous
case~\cite{Brodsky:1997jk,Kataev:2010tm}. Thus, in order to be
self-consistent any QCD scale-setting method should be also
extendable to $\rm QED$ and results should be in agreement with
the Gell-Mann and Low (GM-L) scheme. This is an important
requirement also for the perspective of a grand unified theory
(GUT), where only one method for setting the renormalization scale
can be applied and then it can be considered as a good criterion
for verifying if a scale setting is correct or not. CSS leads to
incorrect results when applied to QED processes. In the GM-L
scheme, the renormalization scale is set with no ambiguity to the
virtuality of the exchanged photon/photons, that naturally sums an
infinite set of vacuum polarization contributions into the running
coupling. Thus the CSS approach of varying the scale of a factor
of 2 does not apply to QED since the scale is already optimized.

\item The forthcoming large amount of high-precision experimental
data arising especially from the running of the high collision
energy and high luminosity Large Hadronic Collider (LHC), will
require more accurate and refined theoretical estimates. The CSS
appears to be more a ``lucky guess"; its results are affected by
large errors and the perturbative series poorly converges with or
without large-logarithm resummation or renormalon contributions.
Moreover, within this background, it is nearly impossible to
distinguish among SM and BSM signals and in many cases, improved
higher-order calculations are not expected to be available in the
short term.

\end{itemize}
To sum up, the conventional scale-setting method assigns an
arbitrary range and an arbitrary systematic error to fixed-order
perturbative calculations that greatly affects the predictions of
pQCD.




\section{The Principle of Minimal Sensitivity: PMS Scale-Setting}

The Principle of Minimal
Sensitivity~\cite{Stevenson:1981vj,Stevenson:1982qw} derives from
the assumption that, since observables should be independent of
the particular RS and scale, their optimal perturbative
approximations should be stable under small RS variations. The RS
scheme parameters $\beta_2,\beta_3,...$ and the scale parameter
$\Lambda$ (or the
 subtraction point $\mu_r$), are considered as ``unphysical" and
independent variables, and then their values are set in order to
minimize the sensitivity of the estimate to their small
variations. This is essentially the core of the Optimized
Perturbation Theory (OPT)~\cite{Stevenson:1981vj}, based on the
PMS procedure. The convergence of the perturbative expansion,
Eq.~\ref{observable-init0}, truncated to a given order $\rho_n$,
is improved by requiring its independence from the choice of RS
and $\mu$. The optimization is implemented by identifying the
RS-dependent parameters in the $\rho_n$-truncated series (the
$\beta_{i}$ for $2 \leq i \leq n$ and $\Lambda$ ), with the
request that the partial derivative of the perturbative expansion
of the observable with respect to the RS-dependent and scale
parameters vanishes. In practice the PMS scale setting is designed
to eliminate the remaining renormalization and scheme dependence
in the truncated expansions of the perturbative series.

More explicitly, the PMS requires the truncated series, i.e. the
approximant of a physical observable defined in
Eq.~\ref{observable-init0}, to satisfy the RG invariance given by
the \ref{eqn:pms}, with the substitution of the proper
$\beta_{(n)}$ function: \be \frac{\partial
\alpha_{\mathrm{s}}}{\partial
\beta_{j}}=-\beta\left(\alpha_{\mathrm{s}}\right)
\int_{0}^{\alpha_{\mathrm{s}}} \mathrm{d} \alpha^{\prime}
\frac{\alpha^{\prime
j+2}}{\left[\beta\left(\alpha^{\prime}\right)\right]^{2}}=\frac{\alpha_{\mathrm{s}}^{j+1}}{\beta_{0}}\left(\frac{1}{j-1}-\frac{\beta_{1}}{\beta_{0}}
\frac{j-2}{j(j-1)} \alpha_{\mathrm{s}}+\ldots\right) , \ee it
follows that:

 \ba
\frac{\partial \rho_{n}}{\partial
\tau}&=&\left(\left.\frac{\partial}{\partial
\tau}\right.+\beta\left(\alpha_{s}\right) \frac{\partial}{\partial
\alpha_{s}}\right) \rho_{n}  \equiv 0
\\ \frac{\partial \rho_{n}}{\partial
\beta_{j}}&=&\left(\left.\frac{\partial}{\partial
\beta_{j}}\right.-\beta\left(\alpha_{\mathrm{s}}\right)
\int_{0}^{\alpha_{\mathrm{s}}} \mathrm{d} \alpha^{\prime}
\frac{\alpha^{\prime
j+2}}{\left[\beta\left(\alpha^{\prime}\right)\right]^{2}}
\frac{\partial}{\partial \alpha_{\mathrm{s}}}\right) \rho_{n}
\equiv 0 \ea where $\tau=\ln \left(\mu_{r}^{2} / \Lambda_{\rm
QCD}^{2}\right)$ and $j \geq 2$. Scheme labels have been omitted.
The request of RS-independence modifies the series coefficients
$\mathcal{C}_{i}\left(1 \leq i \leq n\right)$ and the coupling
$\alpha_{s}$ to the PMS ``optimized" values
$\widetilde{\mathcal{C}_{i}}$ and $\widetilde{\alpha_{s}}$. At
first order the requirement of RS independence implies that
Eq.~\ref{observable-init0} satisfies the Callan-Symanzik equation
\ref{eqn:betag} with $\beta$ truncated at order $n$. This implies
that the scale dependence is not removed from the perturbative
coefficients: $\mathcal{C}_{n}\left(Q^{2}, \mu^{2}\right)$. We can
argue that this approach is more based on convergence rather than
physical criteria. In particular, the PMS is a procedure that can
be extended to higher order and it can be generally applied to
calculations obtained in arbitrary initial renormalization
schemes. Though this procedure leads to results that are suggested
to be unique and scheme independent, unfortunately it violates
important properties of the renormalization group as shown in
Ref.~\cite{Wu:2013ei}, such as reflexivity, symmetry, transitivity
and also the {\it existence and uniqueness} of the optimal PMS
renormalization scheme are not guaranteed since they are strictly
related to the presence of maxima and minima. Other
phenomenological implications will be shown in
section~\ref{sec:comparison}.


\section{The Fastest Apparent Convergence principle - FAC scale setting}


The Fastest Apparent Convergence (FAC) principle is based on the
idea of {\it effective charges}. As pointed out by
Grunberg~\cite{Grunberg:1980ja,Grunberg:1982fw,Grunberg:1989xf},
any perturbatively calculable physical quantity can be used to
define an effective coupling, or ``effective charge", by entirely
incorporating the radiative corrections into its definition.
Effective charges can be defined from an observable starting from
the assumption that the infinite series of a given quantity is
scheme and scale invariant. Given the perturbative series
$R=\mathcal{C}_0 \alpha_s^{p}+\cdots,$ the relative effective
charge $\alpha_{R}$ is given by \be \alpha_{R} \equiv
\left(\frac{R}{\mathcal{C}_{0}}\right)^{1/p}. \ee Since $R,
\mathcal{C}_{0}$ and $p$ are all renormalization scale and scheme
invariant, the effective charge $\alpha_{R}$ is scale and scheme
invariant.

The effective charge satisfies the same renormalization group
equations as the usual coupling. Thus, the running behavior for
both the effective coupling and the usual coupling are the same if
their RG equations are calculated in the same renormalization
scheme. This idea has been discussed in more detail in
Refs.~\cite{Dhar:1983py,Gupta:1990jq}.

Using the effective charge $\alpha_{R}$, the ratio $R_{\rm
e^+e^-}$ becomes~\cite{Gorishnii:1990vf}:
\begin{equation}
R_{e^{+} e^{-}}\left(Q^{2}\right) \equiv R_{e^{+}
e^{-}}^{0}\left(Q^{2}\right)\left[1+\frac{\alpha_{R}(Q)}{\pi}\right]
\end{equation}
where $R_{e^{+} e^{-}}^{0}\left(Q^{2}\right)$ is the Born result
and $s=Q^{2}$ is the center-of-mass energy squared.

An important suggestion is that all effective couplings defined in
the same scheme satisfy the same RG equations. While different
schemes or effective couplings, will differ through the third and
higher coefficients of the
$\left\{\beta_{i}^{\mathcal{R}}\right\}$-functions, which are
scheme $\mathcal{R}$ dependent. Hence, any effective coupling can
be used as a reference to define the renormalization procedure.

Given that expansions of the effective charges are known only up
to a certain order, $\alpha_{R} \simeq
\left(\frac{R_n}{\mathcal{C}_{0}}\right)^{1/p}$, an optimization
procedure is used to improve the perturbative calculations, namely
the FAC scale setting. The basic idea of the FAC scale setting
method is to set to zero all the higher order perturbative
coefficients,  i.e. $\mathcal{C}_{i(\geq 1)}\left(\mu_{r}^{\rm
FAC}\right) \equiv 0$, including all fixed order corrections into
the FAC renormalization scale of the leading term by means of the
RG equations in order to provide a reliable
estimate~\cite{Krasnikov:1981rp}.

In practice given a physical observable in an arbitrary
renormalization scheme written as:
$$
\sigma=A+B\left[\alpha_{s}\left(\mu_{r}\right)\right]^{d}\left[1+\sigma_{1}\left(\mu_{r}\right)
\alpha_{s}\left(\mu_{r}\right)+\mathcal{O}\left(\alpha_{s}^{2}\right)\right],
$$
the effective coupling $\bar{\alpha}_{s}\left(\mu_{r}\right)$ is
defined by the identity
$$
\sigma=A+B\left[\bar{\alpha}_{s}\left(\mu_{r}\right)\right]^{d},
$$
where $A$ and $B$ are general perturbative or non-perturbative
quantities predicted in principle by QCD, $d$ is the
$\alpha_{s}$-order at the Born level and
$\sigma_{1}\left(\mu_{r}\right)$ is the NLO coefficient.
Consequently, $\bar{\alpha}_{s}\left(\mu_{r}\right)$ is the object
effectively extracted from a LO analysis of the experimental data
on $\sigma$.

In general this method can be applied to any observable calculated
in any RS also in processes with large higher order corrections.
The FAC scale setting, as has been shown in Ref.~\cite{Wu:2013ei},
preserves the RG self-consistency requirements, although the FAC
method can be considered more an optimization approach rather than
a proper scale-setting procedure to extend order by order. FAC
results depend sensitively on the quantity to which the method is
applied. In general, when the NLO correction is large, the FAC
results to be a resummation of the most important higher order
corrections and then a RG improved perturbation theory is
achieved. Unfortunately phenomenological studies (see
Fig.~\ref{comparison}) with FAC show that this strategy is
acceptable only in a particular range of values of a given
variable related to the virtuality of the physical observable.


%

\section{The BLM Scale-Setting \label{sec:blm}}

The Brodsky-Lepage-Mackenzie (BLM)~\cite{Brodsky:1982gc} method
was designed to improve the pQCD estimate by absorbing the
$\left\{\beta_{i}\right\}$-terms arising in the perturbative
calculation into the running coupling using the RG equations. In
the BLM approach and subsequently in its generalization and
extension to all orders the Principle of Maximum Conformality
(PMC), the BLM/PMC-scales are identified with the running
$N_f$-terms related to UV divergent loops. In order to improve the
discussion, we will use the following notation: $n_f$ for the
general number of active flavors, $N_f$ for the number of active
flavors related to the UV divergent diagrams and $N_F$ for the
number of active flavors related to UV finite diagrams. We
underline that only terms related to the $\beta$-function, or
equivalently to $N_f$, must be included into the BLM/PMC scales.
The $N_F$ terms may arise in a calculation but are not responsible
for the running of the effective coupling $\alpha_s$ and thus they
do not determine the BLM/PMC scales.

The BLM shows a way to resolve the renormalization scheme-scale
ambiguities, by identifying the coefficients of the $\beta$ terms.
Once these coefficients are reabsorbed into the scale the
perturbative expansion no longer suffers of the renormalon growth
associated with the $n!\beta^n_0\alpha^{n+1}_s$, which are
eliminated. This improves the convergence of the perturbative
expansions in QED/QCD. More importantly, the renormalization scale
can be determined without computing all higher-order corrections
and in a unambiguous way. Thus, the lower-order or even the LO
analysis can be meaningfully compared with experiments. BLM scale
setting is greatly inspired by QED. The standard Gell-Mann-Low
scheme determines the correct renormalization scale identifying
the scale with the virtuality of the exchanged
photon~\cite{GellMann:1954fq}. For example, in electron-muon
elastic scattering, the renormalization scale is given by the
virtuality of the exchanged photon, i.e. the spacelike momentum
transfer squared $\mu_R^2 = q^2 = t$. Thus
\begin{equation} \alpha(t) = {\alpha(t_0) \over 1 - \Pi(t,t_0)}
\label{qed1}
\end{equation} where
$$ \Pi(t,t_0) = {\Pi(t) -\Pi(t_0)\over 1-\Pi(t_0) } $$ is the
vacuum polarization (VP) function. From Eq.~\ref{qed1} it follows
that the renormalization scale $\mu^2_R=t$ can be determined by
the $\beta_0$-term at the lowest order. This scale is sufficient
to sum all the vacuum polarization contributions into the dressed
photon propagator, both proper and improper to all orders.

Following the GM-L scheme in QED, the BLM scales can be determined
at LO order in perturbation theory by writing explicit
contributions coming from the different $N_f$ terms of the NLO
coefficient in a physical observable as~\cite{Brodsky:1982gc}:

 \ba
\rho &=& C_{0} \alpha_{s,\overline{ \rm MS}}^{p}\left(\mu_{r}\right)\left[1+\left(A N_{f}+B\right) \frac{\alpha_{s,\overline{ \rm MS}}\left(\mu_{r}\right)}{\pi}\right] \nonumber \\
&=& C_{0} \alpha_{s, \overline{ \rm
MS}}^{p}\left(\mu_{r}\right)\left[1+\left(-\frac{3}{2} A
\beta_{0}+\frac{33}{2} A+B\right) \frac{\alpha_{s, \overline{\rm
MS}}\left(\mu_{r}\right)}{\pi}\right] \ea

where $\mu_{r}=\mu_{r}^{\rm  init }$ stands for an initial
renormalization scale, which practically can be taken as the
typical momentum transfer of the process. The $N_{f}$ term is due
to the quark vacuum polarization. Calculations are in the
$\overline{\rm MS}$-scheme.

At the NLO level, all $N_{f}$ terms should be resummed into the
coupling. Using the NLO $\alpha_{s}$-running formula: \be
\alpha_{s, \overline{ \rm
MS}}\left(\mu_{r}^{*}\right)=\frac{\alpha_{s,\overline{ \rm
MS}}\left(\mu_{r}\right)}{1+\frac{\beta_{0}}{4 \pi} \alpha_{s,
\overline{ \rm MS}}\left(\mu_{r}\right) \ln\left(
\frac{\mu_r^*}{\mu_r}\right)}, \ee we obtain \be \rho=C_{0}
\alpha_{s,\overline{\rm
MS}}^{p}\left(\mu_{r}^{*}\right)\left[1+C_{1}^{*} \frac{\alpha_{s,
\overline{ \rm
MS}}\left(\mu_{r}^{*}\right)}{\pi}\right]\label{eqn:conformalform}
\ee where
$$
\mu_{r}^{*}=\mu_{r} \exp \left(\frac{3 A}{p}\right) $$ is the BLM
scale and
$$ C_{1}^{*}=\frac{33}{2} A+B ,
$$
is the {\it conformal} coefficient, i.e. the NLO coefficient not
depending on the RS and scale $\mu$. Both the effective BLM scale
$\mu_{r}^{*}$ and the coefficient $C_{1}^{*}$ are $N_{f}$
independent and conformal at LO.  By including the term $33 A / 2$
into the scale we eliminate the $\beta_0$ term of the NLO
coefficient $C_{1}$ which is responsible for the running of the
coupling constant, and the observable in the final results can be
written in its {\it maximal conformal form.}
Eq.\ref{eqn:conformalform}.

The BLM method can be extended to higher orders in a systematic
way by including the $n_f$ terms arising at higher order into the
BLM scales consistently. In order to extend the BLM beyond the
NLO, the following points are considered essential:
\begin{enumerate}
\item  All $n_f$-terms associated with the $\beta$-function (i.e.
$N_f$ terms) and then with the renormalization of the coupling
constant, must be absorbed into the effective coupling, while
those $n_f$ -terms that have no relation with UV divergent
diagrams (i.e. $N_F$-terms) should be identified and considered as
part of the conformal coefficients. After BLM scale setting, the
perturbative series for the physical observable becomes a
conformal series, all non-conformal terms should be absorbed into
the effective coupling in a consistent manner;

\item New $N_f$-terms (corresponding to new $\beta_0$
coefficients) arise at each perturbative order, thus a new BLM
scale that sums these terms consistently into the running
coupling, should be introduced at each calculated perturbative
order. In fact there is no reason to use a unified effective scale
for the whole perturbative series as shown in
Refs.~\cite{Grunberg:1991ac,Grunberg:1992mp}.

\item The BLM scales themselves should be a RG-improved
perturbative series~\cite{Brodsky:2011ta}. The length of the
perturbative series for each BLM scale depends on how many new
$N_f$-terms (or $\beta_i$-terms) we have from the higher-order
calculation and to what perturbative order we have performed.

\end{enumerate}
Actually the last point is not mandatory and needs clarification.
In order to apply the BLM/PMC using perturbative scales, the
argument of the coupling in the expansion of the BLM/PMC scale
should be the physical scale of the process $Q$, that can be
either the center-of-mass energy $\sqrt{s}$ or also another
variable such as $\sqrt{t},\sqrt{u},M_H,...,$ depending on the
process. Setting the initial scale to the physical scale would
greatly simplify the BLM/PMC procedure, preserving the original
scale invariance of the observable and eliminating the initial
scale dependence from the BLM/PMC scales. In case the BLM/PMC
scales are not perturbatively calculated, as it will be shown in
section~\ref{sec:icf}, the initial scale can be treated as an
arbitrary parameter.

In agreement with these indications, it is possible to achieve a
scale setting method extendible iteratively to all orders, which
leads to the correct coefficients ${\mathcal{C}_i}(\mu^*_{BLM})$
for the final ``maximally conformal" series:

\begin{equation}
\rho_{n}=\mathcal{C}_{0} \alpha_{s}^{p}\left(\mu_{\rm
BLM}^{*}\right)+{\mathcal{C}}_{1}\left(\mu_{\rm BLM}^{* *}\right)
\alpha_{s}^{p+1}\left(\mu_{\rm BLM}^{*
*}\right)+{\mathcal{C}}_{2}\left(\mu_{\rm BLM}^{* * *}\right)
\alpha_{s}^{p+2}\left(\mu_{\rm BLM}^{* * *}\right)+\cdots
\label{eqn:conf}
\end{equation}
where the BLM scales $\mu^*_{\rm BLM},\mu^{**}_{\rm BLM},...$ are
set by a recursive use of the RG equations in order to cancel all
the $N_f$ terms from the series. We remark that since the
coefficients ${\mathcal{C}_i}(\mu^*_{BLM})$ have been obtained
cancelling all $\beta$ terms related to running of the coupling
they actually are free from any scale and scheme dependence left.
In other words the ${\mathcal{C}_i}(\mu^*_{BLM}) \equiv
\tilde{\mathcal{C}_i} $ where the $\tilde{\mathcal{C}_i}$ are
conformal coefficients not depending on the renormalization scale.
Hence the BLM approach leads to an observable maximally conformal,
i.e. where all the renormalization scale and scheme dependence has
been confined to the effective coupling and to its renormalization
scale $\alpha_s(\mu_{\rm BLM})$.

Fundamental features of the BLM method:

\begin{enumerate}
\item [A)] BLM scales at LO , are set simply by identifying the
coefficient $A$ of the $N_f$ term;

\item [B)] since all $n_f$-terms related to the running of the
coupling are reabsorbed, scheme differences do not affect the
results and the perturbative expansions in $\alpha_s(\mu_r^*)$ in
two different schemes, e.g. $\rm MS$ and $\overline{\rm MS}$, are
identical. We notice that $n_f$-terms related to the UV finite
diagrams, may arise at every order in perturbation theory. These
terms might be related either to the particular kinematics of the
initial state or even to finite loop diagrams arising at higher
orders, thus in both cases are insensitive to the UV cutoff or to
the RS and cannot be considered as $\beta$-terms. We label these
terms as $N_F$-terms and they do not give contributions to the BLM
scales;

\item [C)] Using BLM scale setting, the perturbative expansion
does not change across quark threshold, given that all
vacuum-polarization effects due to a new quark are automatically
absorbed into the effective coupling. This implies that in a
process with fixed kinematic variables (e.g. a total cross
section), we can use a naive LO/NLO $\alpha_s(\mu_r^{\rm
BLM})$-running with the number of active flavor $N_f$ fixed to the
value determined by the BLM scale, to perform the
calculation~\cite{Brodsky:1998mf};

\item [D)]The BLM method preserves all the RG properties of {\it
existence and uniqueness, reflexivity, symmetry,} and {\it
transitivity}. As shown in
Refs.~\cite{Lu:1991yu,Lu:1991qr,Brodsky:1995ds,Brodsky:1994eh},
the RG invariance of the BLM leads to scheme independent
transformations that relate couplings in different schemes. These
are known as {\it commensurate scale relations} (CSRs), and it has
been shown that even though the expansion coefficients under
different renormalization schemes can be different, after a proper
scale setting, one can determine a relation between the effective
couplings leading to an invariant result for the calculated
quantity. Using this approach it is also possible to extend
conformal properties to renormalizable gauge theories, such as the
generalized Crewther
relation~\cite{Crewther:1972kn,Broadhurst:1993ru,Baikov:2010je,Brodsky:1995tb};

\item [E)] The BLM approach reduces to the GM-L scheme for QED in
the Abelian limit $N_c\rightarrow 0$~\cite{Brodsky:1997jk}, the
results are in perfect agreement;

\item [F)] The elimination of the $N_f$ term related with the
$\beta_0$ coefficient, from the perturbative series eliminates the
renormalon terms $n!\beta^n_0\alpha^{n+1}_s$ over the entire range
of the accessible physical energies and not only in the low-energy
domain. The convergence of the resulting series is then greatly
improved.
\end{enumerate}


\section{Phenomenological comparison}

\label{sec:comparison}

\begin{figure}[htb]
\centering
\includegraphics[width=10cm]{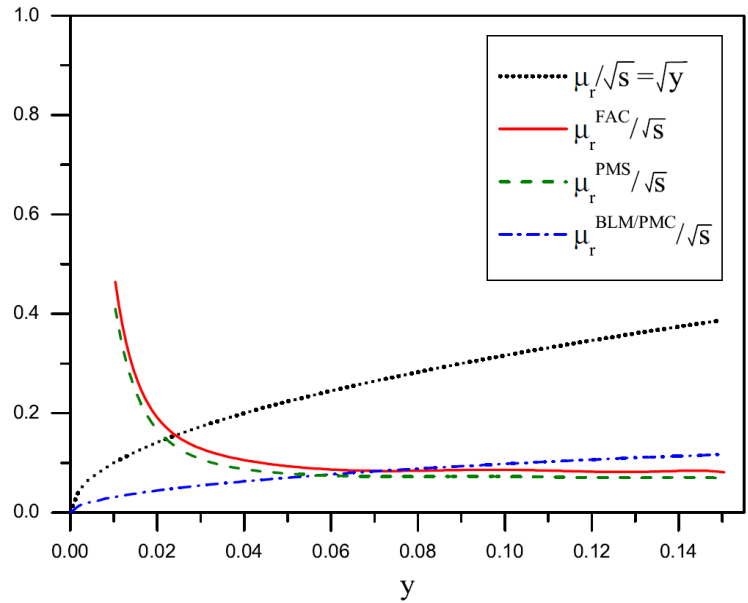}
\caption{Comparison of the results for the renormalization scale
$\mu_{r} / \sqrt{s}$ for the three-jet rate in $e^{+} e^{-}$
annihilation, under the BLM/PMC, PMS, FAC scale settings,
according to the usual $\sqrt{y}$, i.e. the jet virtuality, as
computed by Kramer and Lampe~\cite{Kramer:1990zt}. (Figure from
Ref.~\cite{Wu:2013ei}) } \label{comparison}
\end{figure}

Though results should be invariant with respect to the particular
method used for scale-setting, at a low level of accuracy
(NLO,NNLO,..) they can be different and even incorrect or
unphysical. Since the scale setting methods, such as FAC, PMS and
BLM/PMC, are designed from different fundamental principles, they
can give strikingly different results in practical applications.
In fact, as shown by Kramer and Lampe~\cite{Kramer:1990zt}, the
resulting PMS and FAC scales can be unphysical,
Fig.\ref{comparison};this is the case of the prediction of jet
production fractions in $e^{+} e^{-} \rightarrow q \bar{q} g$
annihilation, the PMS and the FAC scale grow without bound when
the gluon virtuality becomes soft.

Usually, jets are defined by clustering particles in narrow cones
with invariant mass less than $\sqrt{y s}$, where $y$ stands for
the resolution parameter and $\sqrt{s}$ is the total
center-of-mass energy. Physically, one expects the renormalization
scale $\mu_r$ to reflect the invariant mass of the jet. For
example, in the analogous problem in QED, the maximum virtuality
of the photon jet which sets the argument of the running coupling
$\alpha_s$ cannot be larger than $\sqrt{ys}$. Thus one expects
$\mu_r$ to decrease as $y \rightarrow 0$. However, as shown in the
Fig.~\ref{comparison}, the scales chosen by the FAC and PMS
methods do not reproduce this physical behavior: The predicted
scales $\mu_r^{\rm PMS}$ and $\mu_r^{\rm FAC}$ rise at small
values of the invariant mass $y$. This shows that the FAC and PMS
cannot get the right physical behavior in this limit, since they
have included also physics into the running coupling not
associated with renormalization. On the other hand, the BLM/PMC
scale has the correct physical behavior approaching low values, $y
\rightarrow 0$, which indicates the breaking of standard
pQCD~\cite{BURROWS1996157}, while in contrast, the scales chosen
by PMS and FAC show no indications of non-perturbative effects.



\chapter{The Principle of Maximum Conformality - PMC scale setting}

The Principle of Maximum Conformality
(PMC)~\cite{Brodsky:2011ig,Brodsky:2011ta, Brodsky:2012rj,
Mojaza:2012mf, Brodsky:2013vpa} is the principle underlying BLM
and it generalizes the BLM method to all possible applications and
to all orders.

Several extended versions of the BLM approach beyond the NLO have
been proposed in the literature such as the dressed skeleton
expansion, the large $\beta_0$-expansion, the BLM expansion with
an overall renormalization scale, the sequential BLM (seBLM), an
extension to the sequential BLM (xBLM) in
Refs.~\cite{Beneke:1994qe,Ball:1995ni,Brodsky:1994eh,Brodsky:1995tb,Grunberg:1991ac,Neubert:1994vb,Mikhailov:2004iq,Kataev:2014jba}.
These different extensions of the BLM are mostly partial or {\it
ad hoc} improvements of the first LO-BLM~\cite{Brodsky:1982gc} in
some cases up to NNLO, in other cases using a rather effective
approach, i.e. by introducing an overall effective BLM scale for
the entire perturbative expansion. Results obtained with these
approaches did not respect also the basic points (1-3) explained
in section~\ref{sec:blm}. Most important, these methods lead to
results that are still dependent on the initial renormalization
scale. The fundamental feature of the BLM is to obtain results
free from scale ambiguities and thus independent of the choice of
initial renormalization scale. The first aim of the BLM scale is
to eliminate the renormalization scale and scheme uncertainties;
thus any extension of the BLM not respecting this basic
requirement does not represent a real improvement of the standard
conventional scale setting CSS method.

The reasons for the different extensions of the BLM method to
higher order were mainly two: first it was not clear how to
generalize this approach to all possible quantities, which
translates into the question: what is the principle underlying the
BLM method? And second, what is the correct procedure to identify
and reabsorb the $N_f$-terms unambiguously order-by-order? A
practical reason that makes the extension to higher orders not
straightforward is the presence of UV finite corrections given by
the three and four-gluon vertices of the additional $N_{F}$-terms
that are unrelated to the running of $\alpha_{s}$.

In its first formulation in Ref.~\cite{Brodsky:2011ig} it was
suggested to use a unique $\rm PMC$ scale at LO to reabsorb all
$\beta$ contributions related to different skeleton graphs scale
by properly weighting the two contributions, such as that of the
t-channel and s-channel. This approach was more oriented towards a
single PMC scale that reabsorbs all $\beta$ terms related to the
running coupling. A multi-scale approach was later developed
considering different scales arising at each order of accuracy
including different $\beta$ coefficients according to the
perturbative expansion. We remark that the PMC method has all the
(A,B,C,D,E,F) properties of the BLM procedure (Sec.~\ref{sec:blm})
and it extends these properties to all orders eliminating the
renormalization scale and scheme ambiguities. The PMC also
generalizes this approach to all gauge theories. First this is
crucial in order to apply the same method to all SM sectors.
Secondly, in the perspective of a {\it grand unified theory}
(GUT), only one scale setting method can be applied for
consistency, and this method must agree with the GM-L scheme and
with the QED results.

In order to apply the PMC is convenient to follow the flowchart
shown in Fig.~\ref{flowchart} and to write the observable of
Eq.~\ref{observable-init0} with the explicit contributions of the
$n_f$ terms in the coefficients calculated at each order of
accuracy:

\begin{figure}[htb]
\begin{center}
  \includegraphics[width=10cm]{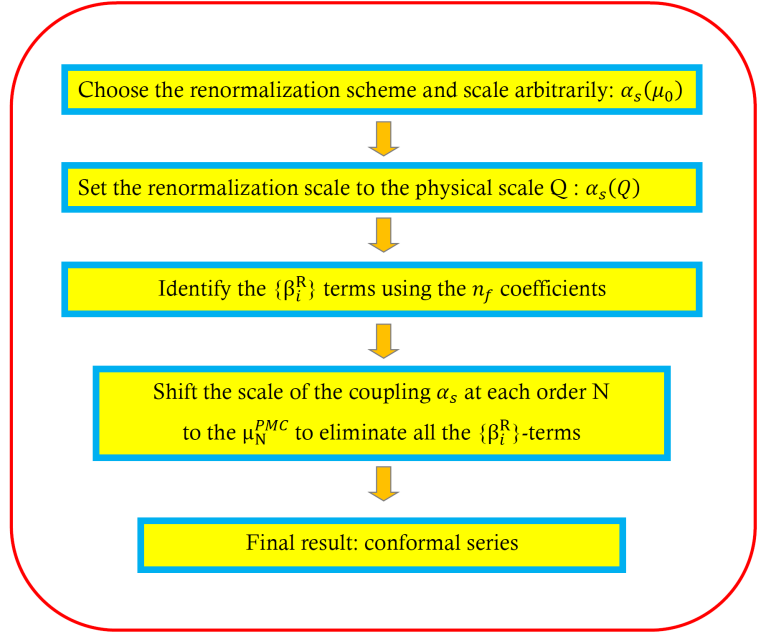}\\
  \caption{Flowchart for the PMC procedure.}\label{flowchart}
\end{center}
\end{figure}
\begin{equation}
\rho(Q)= \sum^{n}_{i=1} \left(\sum^{i-1}_{j=0} c_{i,j}(\mu_r, Q)
n_f^{j}\right) a_s^{p+i-1}(\mu_r), \label{nf}
\end{equation}
where $Q$ represents the kinematic scale or the physical scale of
the measured observable and $p$ is the power of $\alpha_s$
associated with the tree-level terms. In general, this procedure
is always possible either for analytic or numerical (e.g.
MonteCarlo) calculations, given that both strategies keep track of
terms related to different color factors. The core of the PMC
method, as was for BLM, is that all $N_f$ terms related to the
$\beta$-function arising in a perturbative calculation must be
summed, by proper definition of the renormalization scale
$\mu_{\rm PMC}$, into the effective coupling $\alpha_s$ by
recursive use of the RGE. Essentially the difference between the
two procedures is that while in BLM the scales are set iteratively
order by order to remove all $n_f$ terms, in PMC the $n_f$ terms
are written first as $\beta$ terms and then reabsorbed into the
effective coupling. The two procedures are related by the {\it
correspondence principle}~\cite{Brodsky:2011ta}.

\section{The multi-scale Principle of Maximum Conformality: PMCm \label{sec:pmcm}}

The PMCm method is based on a multi-scale application of the PMC.
We show in this section how to implement this method at any order
of accuracy. First, as shown in the flowchart in
Fig.~\ref{flowchart}, for the pQCD approximant (\ref{nf}), it is
convenient to transform the $\{n_f\}$ series at each order into
the $\{\beta_i\}$ series. The QCD degeneracy
relations~\cite{Bi:2015wea} ensure the realizability of such
transformation. For example, Eq.~\ref{nf} can be rewritten
as~\cite{Mojaza:2012mf, Brodsky:2013vpa}

\begin{eqnarray}
\rho(Q)&=&r_{1,0}a_s(\mu_r) + \bigg(r_{2,0}+\beta_{0}r_{2,1}\bigg)a_{s}^{2}(\mu_r)\nonumber\\
&&+\bigg(r_{3,0}+\beta_{1}r_{2,1}+ 2\beta_{0}r_{3,1}+ \beta_{0}^{2}r_{3,2}\bigg)a_{s}^{3}(\mu_r)\nonumber\\
&& +\bigg(r_{4,0}+\beta_{2}r_{2,1}+ 2\beta_{1}r_{3,1} + \frac{5}{2}\beta_{1}\beta_{0}r_{3,2} \nonumber\\
&&
+3\beta_{0}r_{4,1}+3\beta_{0}^{2}r_{4,2}+\beta_{0}^{3}r_{4,3}\bigg)
a_{s}^{4}(\mu_r)+\cdots, ~\label{rij1}
\end{eqnarray}
where $r_{i,j}$ can be derived from $c_{i,j}$; $r_{i,0}$ are
conformal coefficients, and $r_{i,j}$ $(j\neq0)$ are nonconformal.
For definiteness and without loss of generality, we have set $p=1$
and $n=4$ for illustrating the PMC procedures. Different types of
$\{\beta_i\}$-terms can be absorbed into $\alpha_s$ in an
order-by-order manner by using the RGE, which leads to distinct
PMC scales at each order:
\begin{eqnarray}
a^k_s(Q_k) &&\leftarrow a^k_s(\mu_r)\bigg\{1 + k \beta_0 \frac{r_{k+1,1}}{r_{k,0}} a_s(\mu_r) \nonumber \\
&&+k\left(\beta_1 \frac{r_{k+1,1}}{r_{k,0}} +\frac{k+1}{2} \beta_0^2 \frac{r_{k+2,2}}{r_{k,0}} \right) a^{2}_s(\mu_r)\nonumber \\
&&+k\bigg(\beta_2 \frac{r_{k+1,1}}{r_{k,0}}+\frac{2k+3}{2}\beta_0\beta_1 \frac{r_{k+2,2}}{r_{k,0}}\nonumber \\
&&+\frac{(k+1)(k+2)}{3!}\beta_0^3 \frac{r_{k+3,3}}{r_{k,0}}\bigg)
a^{3}_s(\mu_r)+\cdots \bigg\}. \label{scaledis1}
\end{eqnarray}
The coefficients $r_{i,j}$ are generally functions of $\mu_r$,
which can be redefined as
\begin{eqnarray}
r_{i,j}=\sum^j_{k=0}C^k_j{\hat r}_{i-k,j-k}{\rm
ln}^k(\mu_r^2/Q^2),~\label{rijrelation}
\end{eqnarray}
where the reduced coefficients ${\hat r}_{i,j}=r_{i,j}|_{\mu_r=Q}$
(specifically, we have ${\hat r}_{i,0}=r_{i,0}$), and the
combinatorial coefficients $C^k_j=j!/(k!(j-k)!)$. As discussed in
section~\ref{sec:blm}, we set the renormalization scale $\mu_r$ to
the physical scale of the process $Q$:
\begin{eqnarray}
\ln\frac{Q^2}{Q_1^2}&=& \frac{{\hat r}_{2,1}}{{\hat r}_{1,0}}+\beta_0\left(\frac{{\hat r}_{1,0}{\hat r}_{3,2}-{\hat r}_{2,1}^2}{{\hat r}_{1,0}^2}\right)a_s(Q)\nonumber \\
&&+\bigg[\beta_1\left(\frac{3{\hat r}_{3,2}}{2r_{1,0}}-\frac{3{\hat r}_{2,1}^2}{2{\hat r}_{1,0}^2}\right)+\beta_0^2\bigg(\frac{{\hat r}_{4,3}}{{\hat r}_{1,0}}-\frac{2{\hat r}_{3,2}{\hat r}_{2,1}}{{\hat r}_{1,0}^2}+\frac{{\hat r}_{2,1}^3}{{\hat r}_{1,0}^3} \bigg)\bigg]a^2_s(Q)+\cdots, \label{Q1} \\
\ln\frac{Q^2}{Q_2^2}&=& \frac{{\hat r}_{3,1}}{{\hat r}_{2,0}}+3\beta_0\frac{{\hat r}_{2,0}{\hat r}_{4,2}-{\hat r}_{3,1}^2}{2{\hat r}_{2,0}^2}a_s(Q)+\cdots, \label{Q2}\\
\ln\frac{Q^2}{Q_3^2}&=& \frac{{\hat r}_{4,1}}{{\hat
r}_{3,0}}+\cdots. \label{Q3}
\end{eqnarray}
Note that the PMC scales are of a perturbative nature, which is
also a kind of resummation and we need to know more loop terms to
achieve more accurate predictions. The PMC resums all the known
same type of $\{\beta_i\}$-terms to form precise PMC scales for
each order. Thus, the precision of the PMC scale for the
high-order terms decreases at the higher-and-higher orders due to
the less known $\{\beta_i\}$-terms in those higher-order terms.
For example, $Q_1$ is determined up to next-to-next-to-leading log
(N$^2$LL)-accuracy, $Q_2$ is determined up to NLL-accuracy and
$Q_3$ is determined at the LL-accuracy. Thus the PMC scales at
higher-orders are of less accuracy due to more of its perturbative
terms being unknown. This perturbative property of the PMC scale
causes the \textit{first kind of residual scale dependence}.

After fixing the magnitude of $a_s(Q_k)$, we achieve a conformal
series
\begin{eqnarray}
\rho(Q) &=& \sum_{i=1}^{4}{\hat r}_{i,0} a^i_s(Q_i) +\cdots.
\label{PMCmseries}
\end{eqnarray}
The PMC scale for the highest-order term, e.g. $Q_4$ for the
present case, is unfixed, since there is no $\{\beta_i\}$-terms to
determine its magnitude. This renders the last perturbative term
unfixed and causes the \textit{second kind of residual scale
dependence}. Usually, the PMCm suggests setting $Q_4$ as the last
determined scale $Q_{3}$, which ensures the scheme independence of
the prediction due to commensurate scale relations among the
predictions under different renormalization
schemes~\cite{Brodsky:1994eh, Huang:2020gic}. The pQCD series
(\ref{PMCmseries}) is renormalization scheme and scale
independent, and becomes more convergent due to the elimination of
the $\beta$ terms including those related to the renormalon
divergence. Thus a more accurate pQCD prediction can be achieved
by applying the PMCm. Two residual scale dependences are due to
perturbative nature of either the pQCD approximant or the PMC
scale, which is in principal different from the conventional
arbitrary scale dependence. In practice, we have found that those
two residual scale dependences are quite small even at low orders.
This is due to a generally faster pQCD convergence after applying
the PMCm. Some examples can be found in Ref.~\cite{Wu:2015rga}.

\section{The single-scale Principle of Maximum Conformality: PMCs \label{sec:pmcs} }


In some cases, the perturbative series might have a weak
convergence and the PMC scales might retain a comparatively larger
residual scale dependence. In order to overcome this a single
scale approach has been proposed, namely the PMCs, in order to
suppress the residual scale dependence by directly fixing a single
effective $\alpha_s$. Following the standard procedures of
PMCs~\cite{Shen:2017pdu}, the pQCD approximant (\ref{rij1})
changes to the following conformal series,
\begin{eqnarray}
\rho(Q)&=& \sum_{i=1}^{4} {\hat r}_{i,0}a^i_s(Q_*) + \cdots.
\label{conformal}
\end{eqnarray}
As in the previous section, we have set $p=1$ and $n=4$ for
illustrating the procedure. The PMC scale $Q_*$ can be determined
by requiring all the nonconformal terms vanish, which can be fixed
up to N$^2$LL-accuracy for $p=1$ and $n=4$, i.e. $\ln Q^2_* / Q^2$
can be expanded as a power series over $a_s(Q)$,
\begin{eqnarray}
\ln\frac{Q^2_*}{Q^2}=T_0+T_1 a_s(Q)+T_2 a^2_s(Q)+ \cdots,
\label{qstar}
\end{eqnarray}
where the coefficients $T_i~(i=0, 1, 2)$ are
\begin{eqnarray}
T_0&=&-\frac{{\hat r}_{2,1}}{{\hat r}_{1,0}}, \\
T_1&=&\frac{ \beta _0 ({\hat r}_{2,1}^2-{\hat r}_{1,0} {\hat
r}_{3,2})}{{\hat r}_{1,0}^2}+\frac{2 ({\hat r}_{2,0} {\hat
r}_{2,1}-{\hat r}_{1,0} {\hat r}_{3,1})}{{\hat r}_{1,0}^2},
\end{eqnarray}
and
\begin{eqnarray}
T_2&&=\frac{3 \beta _1 ({\hat r}_{2,1}^2-{\hat r}_{1,0} {\hat r}_{3,2})}{2 {\hat r}_{1,0}^2}\nonumber\\
&&+\frac{4({\hat r}_{1,0} {\hat r}_{2,0} {\hat r}_{3,1}-{\hat r}_{2,0}^2 {\hat r}_{2,1})+3({\hat r}_{1,0} {\hat r}_{2,1} {\hat r}_{3,0}-{\hat r}_{1,0}^2 {\hat r}_{4,1})}{ {\hat r}_{1,0}^3} \nonumber \\
&&+\frac{ \beta _0 (4 {\hat r}_{2,1} {\hat r}_{3,1} {\hat r}_{1,0}-3 {\hat r}_{4,2} {\hat r}_{1,0}^2+2 {\hat r}_{2,0} {\hat r}_{3,2} {\hat r}_{1,0}-3 {\hat r}_{2,0} {\hat r}_{2,1}^2)}{ {\hat r}_{1,0}^3}\nonumber\\
&&+\frac{ \beta _0^2 (2 {\hat r}_{1,0} {\hat r}_{3,2} {\hat
r}_{2,1}- {\hat r}_{2,1}^3- {\hat r}_{1,0}^2 {\hat r}_{4,3})}{
{\hat r}_{1,0}^3}.
\end{eqnarray}

Eq.~\ref{qstar} shows that the PMC scale $Q_*$ is also a power
series over $\alpha_s$, which resums all the known
$\{\beta_i\}$-terms and is explicitly independent of $\mu_r$ at
any fixed order, but depends only on the physical scale Q. It
represents the correct momentum flow of the process, and
determines an overall effective $\alpha_s$ value. Together with
the $\mu_r$-independent conformal coefficients, the resultant PMC
pQCD series is scheme and scale independent~\cite{Wu:2018cmb}. By
using a single PMC scale determined with the highest accuracy from
the known pQCD series, both the \textit{first} and the {\it second
kind of residual scale dependence} are suppressed.







\chapter{Infinite-Order Scale-Setting using the Principle of
Maximum Conformality: PMC$_\infty$ \label{pmcinfty}}

We introduce in this chapter a parametrization of the observables
that stems directly from the analysis of the perturbative QCD
corrections and which reveals interesting properties like scale
invariance independently of the process or of the kinematics. We
point out that this parametrization can be an intrinsic general
property of gauge theories and we define this property as
\textit{intrinsic conformality} (iCF{\footnote{Here the
conformality has to be understood as RG invariance only.}}). We
also show how this property directly indicates the correct
renormalization scale $\mu_R$ at each order of calculation and we
define this new method PMC$_\infty$: {\it Infinite-Order
Scale-Setting using the Principle of Maximum Conformality}. We
apply the iCF property and the PMC$_\infty$ to the case of the
thrust and C-parameter distributions in $e^+ e^-\rightarrow 3jets$
and we show the results.

\section{Intrinsic conformality (iCF) \label{sec:icf}}

In order to introduce intrinsic conformality (iCF), we consider
the case of a normalized IR-safe single-variable distribution and
write the explicit sum of pQCD contributions calculated up to NNLO
at the initial renormalization scale $\mu_0$:
\begin{eqnarray}
\frac{1}{\sigma_{0}} \! \frac{O d \sigma(\mu_{0})}{d O}\! &\!\! =
\!\! & \! \left\{ \! \frac{\alpha_{s}(\mu_0)}{2 \pi} \! \frac{ O d
A_{\mathit{O}}(\mu_0)}{d O}\! +\!\!
\left(\!\frac{\alpha_{s}(\mu_{0})}{2 \pi}\!\right)^{\!\!2} \!\!
\frac{ O d B_{\mathit{O}}(\mu_0)}{d O} \right. \nonumber  \\
 & & + \left. \left(\frac{\alpha_{s}(\mu_0)}{2
\pi}\right)^{\!\!3} \! \frac{O dC_{\mathit{O}}(\mu_0)}{d O}\!+
\!{\cal O}(\alpha_{s}^4) \right\},
 \label{observable1}
\end{eqnarray}
where the $\sigma_0$ is tree-level hadronic cross section, the
$A_O, B_O, C_O$ are respectively the LO, NLO and NNLO
coefficients, $O$ is the selected non-integrated variable. For the
sake of simplicity we will refer to the perturbatively calculated
differential coefficients as \textit{implicit coefficients} and we
drop the derivative symbol, i.e.
\begin{eqnarray}
A_O(\mu_0)~\!\!& \equiv &~~\frac{ O d A_{O}(\mu_0)}{d O},\,
B_O(\mu_0)~ \equiv~~\frac{ O d B_{O}(\mu_0)}{d O}, \nonumber \\
C_O(\mu_0)~\!\! & \equiv &~\frac{ O d C_{O}(\mu_0)}{d O}.
\label{implicitcoefficients}\end{eqnarray}
We define here the \textit{intrinsic conformality} as the property
of a renormalizable SU(N)/U(1) gauge theory, like QCD, which
yields a particular structure of the perturbative corrections that
can be made explicit representing the perturbative coefficients
using the following parametrization{\footnote{We are neglecting
here other running parameters such as the mass terms.}}:
\begin{eqnarray}
 A_{O}(\mu_0)\!\!\! &=& \!\!\! A_{\mathit{Conf}} , \nonumber \\
B_{O}(\mu_0) \!\!\! &=& \!\!\! B_{\mathit{Conf}}+\frac{1}{2} \beta_{0} \ln \left(\frac{\mu_0^{2}}{\mu_{\rm I}^{2}}\right) A_{\mathit{Conf}},  \nonumber \\
 C_{O}(\mu_0)\!\!\! &=& \!\!\! C_{\mathit{Conf}} +\beta_{0} \ln \left(\frac{\mu_{0}^{2}}{\mu_{\rm II}^{2}}\right)B_{\mathit{Conf}}+  \nonumber \\
 \!\!\! & & \!\!\! +\frac{1}{4}\left[\beta_{1}+\beta_{0}^{2}
 \ln \left(\frac{\mu_0^{2}}{\mu_{\rm I}^{2}}\right)\right] \ln \left(\frac{\mu_0^{2}}{\mu_{\rm I}^{2}}\right)
 A_{\mathit{Conf}},\nonumber \\
 \label{newevolution}
  \end{eqnarray}
where the $A_{\mathit{Conf}}, B_{\mathit{Conf}},
C_{\mathit{Conf}}$ are the scale invariant \textit{Conformal
Coefficients} (i.e. the coefficients of each perturbative order
not depending on the scale $\mu_0$) while we define the $\mu_{\rm
N}$ as \textit{Intrinsic Conformal Scales} and $\beta_0,\beta_1$
are the first two coefficients of the $\beta$-function. We recall
that the implicit coefficients are defined at the scale $\mu_0$
and that they change according to the standard RG equations under
a change of the renormalization scale according to :
\begin{eqnarray}
 A_{O}(\mu_R)\!\!\! &=& \!\!\! A_{O}(\mu_0) , \nonumber \\
B_{O}(\mu_R) \!\!\! &=& \!\!\! B_{O}(\mu_0)+\frac{1}{2} \beta_{0} \ln \left(\frac{\mu_R^{2}}{\mu_{0}^{2}}\right) A_{O}(\mu_0),  \nonumber \\
 C_{O}(\mu_R)\!\!\! &=& \!\!\! C_{O}(\mu_0) +\beta_{0} \ln \left(\frac{\mu_{R}^{2}}{\mu_{0}^{2}}\right)B_{O}(\mu_0)+  \nonumber \\
 \!\!\! & & \!\!\! +\frac{1}{4}\left[\beta_{1}+\beta_{0}^{2}
 \ln \left(\frac{\mu_R^{2}}{\mu_{0}^{2}}\right)\right] \ln \left(\frac{\mu_R^{2}}{\mu_{0}^{2}}\right)
 A_{O}(\mu_0)\nonumber \\
 \label{standardevolution}
  \end{eqnarray}
It can be shown that the form of Eq.~\ref{newevolution} is scale
invariant and it is preserved under a change of the
renormalization scale from $\mu_0$ to $\mu_R$ by standard RG
equations Eq.~\ref{standardevolution}, i.e.:
\begin{eqnarray}
 A_{O}(\mu_R)\!\!\! &=& \!\!\! A_{\mathit{Conf}} , \nonumber \\
B_{O}(\mu_R) \!\!\! &=& \!\!\! B_{\mathit{Conf}} +\frac{1}{2} \beta_{0} \ln \left(\frac{\mu_R^{2}}{\mu_{\rm I}^{2}}\right) A_{\mathit{Conf}},  \nonumber \\
 C_{O}(\mu_R)\!\!\! &=& \!\!\! C_{\mathit{Conf}} +\beta_{0} \ln \left(\frac{\mu_{R}^{2}}{\mu_{\rm II}^{2}}\right)B_{\mathit{Conf}}+  \nonumber \\
 \!\!\! & & \!\!\! +\frac{1}{4}\left[\beta_{1}+\beta_{0}^{2}
 \ln \left(\frac{\mu_R^{2}}{\mu_{\rm I}^{2}}\right)\right] \ln \left(\frac{\mu_R^{2}}{\mu_{\rm I}^{2}}\right) A_{\mathit{Conf}} \label{evolved}
  \end{eqnarray}
We note that the form of Eq.~\ref{newevolution} is invariant and
that the initial scale dependence is exactly removed by $\mu_R$.
Extending this parametrization to all orders we achieve a scale
invariant quantity: \textit{the iCF-parametrization is a
sufficient condition in order to obtain a scale invariant
observable}.

In order to show this property we collect together the terms
identified by the same \textit{conformal coefficient}, we name
each set as a \textit{conformal subset} and we extend the property
to order $n$:

\begin{eqnarray}
& \sigma_{\rm I} &  =\left\{ \left(\frac{\alpha_{s}(\mu_{0})}{2
\pi}\right)+\frac{1}{2}\beta_{0} \ln \left(
\frac{\mu_0^{2}}{\mu_{\rm
I}^{2}}\right)\left(\frac{\alpha_{s}(\mu_0)}{2 \pi}\right)^2
\right.  \nonumber \\
&& \qquad +\left. \frac{1}{4}\! \left[ \beta_{1} + \beta_{0}^{2}
\ln \left(\frac{\mu_0^{2}}{\mu_{\rm I}^{2}}\right) \right] \ln
\left(\frac{\mu_0^{2}}{\mu_{\rm I}^{2}}\right)
\left(\frac{\alpha_{s}(\mu_{0})}{2
\pi}\right)^3 +\ldots \right\} A_{\mathit{Conf}}   \nonumber \\
&  \sigma_{\rm II} & =  \left\{ \left(\frac{\alpha_{s}(\mu_{0})}{2
\pi}\right)^2  + \beta_0 \ln  \left(  \frac{\mu_0^2}{\mu_{\rm
II}^2}\right) \left(\frac{\alpha_{s}(\mu_{0})}{2 \pi}\right)^3
+\ldots \right\}
B_{\mathit{Conf}}   \nonumber \\
& \sigma_{\rm III} & = \left\{ \left(\frac{\alpha_{s}(\mu_{0})}{2
\pi}\right)^3 +\ldots \right\}C_{\mathit{Conf}}, \nonumber \\
& \vdots & \qquad
\qquad   .^{\huge .^{\huge \cdot}}  \nonumber \\
&  \sigma_{\rm n} & = \left\{ \left(\frac{\alpha_{s}(\mu_{0})}{2
\pi}\right)^n \right\}\mathcal{L}_{n \mathit{Conf}},
\label{confsubsets}
\end{eqnarray}
in each subset we have only one intrinsic scale and only one
conformal coefficient and the subsets are disjoint; thus no mixing
terms among the scales or the coefficients are introduced in this
parametrization. Moreover, the structure of the subsets remains
invariant under a global change of the renormalization scale, as
shown from Eq.~\ref{evolved}. The structure of each conformal set
$\sigma_{\rm I}, \sigma_{\rm II}, \sigma_{\rm III},...$ and
consequently the iCF are preserved also if we fix a different
renormalization scale for each conformal subset, i.e.
\begin{eqnarray}
\left(\mu^2 \frac{ \partial}{\partial \mu^2} +\beta
(\alpha_s)\frac{\partial}{\partial \alpha_s}\right) \sigma_{\rm
n}=0. \label{sigmainvariance}
\end{eqnarray}
We define here this property of Eq. ~\ref{confsubsets} of
separating an observable in the union of ordered scale invariant
disjoint subsets $\sigma_{\rm I}, \sigma_{\rm II}, \sigma_{\rm
III},...$ an \textit{ordered scale invariance}.

In order to extend the iCF to all orders, we perform the
$n\rightarrow \infty$ limit using the following strategy: we first
perform a partial limit $J_{/n}\rightarrow \infty$ including the
higher order corrections relative only to those
$\beta_0,\beta_1,\beta_2,...,\beta_{n-2}$ terms that have been
determined already at order $n$ for each subset, and then we
perform the complementary $\bar{n}$ limit, which consists in
including all the remaining higher order terms. For the $J_{/n}$
limit we have:

\begin{eqnarray}
 \lim_{J_{/n}\rightarrow\infty}&
\sigma_{\rm I} & \rightarrow
\left(\frac{\left. \alpha_{s}(\mu_{\rm I})\right|_{n-2}}{2 \pi}\right) A_{\mathit{Conf}} \nonumber \\
\lim_{J_{/n}\rightarrow\infty}& \sigma_{\rm II} & \rightarrow
\left(\frac{\left.\alpha_{s}(\mu_{\rm II})\right|_{n-3}}{2
\pi}\right)^2  B_{\mathit{Conf}}  \nonumber \\
 \lim_{J_{/n}\rightarrow\infty}& \sigma_{\rm III} & \rightarrow
\left(\frac{\left.\alpha_{s}(\mu_{\rm III})\right|_{n-4}}{2
\pi}\right)^3
C_{\mathit{Conf}} \nonumber \\
 & \vdots & \qquad
\qquad \vdots \nonumber \\
\lim_{J_{/n}\rightarrow\infty}& \sigma_{\rm n} & \equiv
 \left(\frac{\alpha_{s}(\mu_{0})}{2
\pi}\right)^n \mathcal{L}_{n \mathit{Conf}} \label{jnlimit}
\end{eqnarray}
where $\left. \alpha_{s}(\mu_{\rm I})\right|_{n-2}$ is the
coupling calculated up to $\beta_{n-2}$ at the intrinsic scale
$\mu_{\rm I}$. Given the particular ordering of the powers of the
coupling, in each conformal subset we have the coefficients of the
$\beta_0,...,\beta_{n-k-1}$ terms, where $k$ is the order of the
conformal subset and the $n$ is the order of the highest subset
with no $\beta$-terms. We note that the limit of each conformal
subset is finite and scale invariant up to $\sigma_{n-1}$. The
remaining scale dependence is confined in the coupling of the
$n^{th}$ term. Any combination of the $\sigma_{\rm
I},...,\sigma_{n-1}$ subsets is finite and scale invariant. We can
now extend the iCF to all orders performing the $\bar{n}$ limit.
In this limit we include all the remaining higher order
corrections. For the calculated conformal subsets this leads to
define the coupling at the same scales but including all the
missing $\beta$ terms. Thus each conformal subset remains scale
invariant. We point out that we are not making any assumption on
the convergence of the series for this limit. Thus, we have:

\begin{eqnarray}
\lim_{\bar{n}\rightarrow\infty} & \sigma_{\rm I} & \rightarrow
\left(\frac{ \alpha_{s}(\mu_{\rm I})}{2 \pi}\right) A_{\mathit{Conf}} \nonumber \\
\lim_{\bar{n}\rightarrow\infty}  & \sigma_{\rm II} &\rightarrow
\left(\frac{\alpha_{s}(\mu_{\rm II})}{2
\pi}\right)^2  B_{\mathit{Conf}}  \nonumber \\
\lim_{\bar{n}\rightarrow\infty} & \sigma_{\rm III} & \rightarrow
\left(\frac{\alpha_{s}(\mu_{\rm III})}{2 \pi}\right)^3
C_{\mathit{Conf}} \nonumber \\
 & \vdots & \qquad
\qquad \vdots \nonumber \\
\lim_{\bar{n}\rightarrow\infty} & \sigma_{\rm n} \!\! & \! \!
\equiv \!\! \lim_{n \rightarrow\infty}
\left(\frac{\alpha_{s}(\mu_{0})}{2
\pi}\right)^n \!\! \mathcal{L}_{n \mathit{Conf}} \!\! \rightarrow \hbox{Conformal Limit} \nonumber \\
 & & \label{nlimit}
\end{eqnarray}
where here now $ \alpha_{s}(\mu_{\rm I})$ is the complete coupling
determined at the same scale $\mu_{\rm I}$. Eq.~\ref{nlimit} shows
that the whole renormalization scale dependence has been
completely removed. In fact, neither the intrinsic scales
$\mu_{\rm N}$ nor the conformal coefficients
$A_{\mathit{Conf}},B_{\mathit{Conf}},C_{\mathit{Conf}},...,\mathcal{L}_{n
\mathit{Conf}},...$ depend on the particular choice of the initial
scale. The only term with a residual $\mu_0$ dependence is the
n-term, but this dependence cancels in the limit $n\rightarrow
\infty$. The scale dependence is totally confined to the coupling
$\alpha_s (\mu_0)$ and its behavior does not depend on the
particular choice of any scale $\mu_0$ in the perturbative region,
i.e. $\lim_{n\rightarrow\infty} \alpha_s(\mu_0)^n \sim a^n$ with
$a<1$. Hence the limit of $\lim_{n\rightarrow\infty} \sigma_{\rm
n}$ depends only on the properties of the theory and not on the
scale of the coupling in the perturbative regime.
 The proof given here shows that the iCF is {\it sufficient} to have a scale
invariant observable and it does not depend on the particular
convergence of the series. In order to show the {\it necessary}
condition we separate the two cases of a convergent series and an
asymptotic expansion. For the first case the {\it necessary}
condition stems directly from the uniqueness of the iCF form,
since given a finite limit and the scale invariance any other
parametrization can be reduced to the iCF by means of appropriate
transformations in agreement with the RG equations. For the second
case, we have that an asymptotic expansion though not convergent,
can be truncated at a certain order $n$, which is the case of
Eq.~\ref{confsubsets}. Given the particular structure of the iCF
we can perform the first partial limit $J_{/n}$ and we would
achieve a finite and scale invariant prediction ,
$\sigma_{N-1}=\Sigma_{i=1}^{n-1} \sigma_i$, for a truncated
asymptotic expansion, as shown in Eq.~\ref{jnlimit}. Given the
truncation of the series in the region of maximum of convergence
the n-th term would be reduced to the lowest value and so the
scale dependence of the observable would reach its minimum. Given
the finite and scale invariant limit $\sigma_{N-1}$ we conclude
that the iCF is unique and thus {\it necessary} for an {\it
ordered} scale invariant truncated asymptotic expansion up to the
$n^{th}$ order. We point out that in general the iCF form is the
most general and irreducible parametrization which leads to the
scale invariance, other parametrization are forbidden since if we
introduce more scales~\footnote{Here we refer to the form of
Eq.~\ref{newevolution}. In principle it is possible to write other
parametrizations preserving the scale invariance, but these can be
reduced to the iCF by means of appropriate transformations in
agreement with the RG equations.} into the logarithms of one
subset we would spoil the invariance under the RG transformation
and we could not achieve Eq.~\ref{evolved}, while on the other
hand no scale dependence can be introduced into the intrinsic
scales since it would remain in the observable already in the
first partial limit $J_{/n}$ and it could not be eliminated. The
conformal coefficients are conformal by definition at each order,
thus they do not depend on the renormalization scale and they do
not have a perturbative expansion. Hence {\it the iCF is a
necessary and sufficient condition for scale invariance.}

\section{Comments on the iCF and the ordered scale invariance \label{sec:osi}}

The iCF-parametrization can stem either from an inner property of
the theory, the iCF, or from direct parametrization of the
scale-invariant observable. In both cases the iCF-parametrization
makes the scale dependence of the observable explicit and it
exactly preserves the scale invariance. The iCF parametrization is
invariant with respect to the choice of initial scale $\mu_0$,
this implies that the same calculation performed choosing
different arbitrary initial scales, $\mu_0;\mu_0'$ leads to the
same result in the limit $J_{/n}$, a limit that is scale and
scheme independent. The iCF is also strongly motivated by the
renormalizability of QCD and by the uniqueness of the $\beta$
function in a given scheme, i.e. two different $\beta_i,\beta_i'$
do not occur in a perturbative calculation at any order in one RS
and the UV divergencies are cancelled by redefinition of the same
parameters at lowest and higher orders. We remark that the
conservation of the iCF form in one observable is strongly related
to the validity of the RG transformations, thus we expect the iCF
to be well preserved in the deep Euclidean region.

Once we have defined an observable in the iCF-form, we have not
only the scale invariance of the entire observable, but also the
\textit{ordered scale invariance} (i.e. the scale invariance of
each subset $\sigma_{\rm n}$ or $\sigma_{\rm N-1}$). The latter
property is crucial in order to obtain scale invariant observables
independently from the particular kinematic region and
independently from the starting order of the observable or the
order of the truncation of the series. Since in general, a theory
is blind with respect to the particular observable/process that we
might investigate, the theory should preserve the \textit{ordered}
scale invariance in order to define always scale invariant
observables. Hence if the iCF is an inner property of the theory,
it leads to implicit coefficients that are neither independent nor
conformal. This is made explicit in Eq.~\ref{newevolution}, but it
is hidden in the perturbative calculations in the case of the
implicit coefficients. For instance, the presence of the iCF
clearly reveals itself when a particular kinematic region is
approached and the $A_O$ becomes null. This would cause a breaking
of the scale invariance since a residual initial scale dependence
would remain in the observable in the higher order coefficients.
The presence of the iCF solves this issue by leading to the
correct redefinition of all the coefficients at each order
preserving the correct scale invariance exactly. Thus, in the case
of a scale-invariant observable $O$, defined according to the
implicit form (Eq.~\ref{observable1}), by the coefficients $
\{A_O, B_O, C_O,..,O_O,...\},$ it cannot simply undergo the change
$ \rightarrow \{0 , B_O, C_O,..,O_O,...\}$ , since this would
break the scale invariance. In order to preserve the scale
invariance, we must redefine the coefficients $\{\tilde{A}_O=0,
\tilde{B}_O, \tilde{C}_O,..,\tilde{O}_O,...\}$ cancelling out all
the initial scale dependence originating from the LO coefficient
$A_O$ at all orders. This is equivalent to subtracting out a whole
invariant conformal subset $\sigma_{\rm I}$ related to the
coefficient $A_{\mathit{Conf}}$ from the scale invariant
observable $O$. This mechanism is clear in the case of the
explicit form of the iCF, Eq.~\ref{newevolution}, where if
$A_{\mathit{Conf}}=0$ then the whole conformal subset is null and
the scale invariance is preserved. We underline that the conformal
coefficients can acquire all possible values without breaking the
scale invariance, they contain the essential information on the
physics of the process, while all the correlation factors can be
reabsorbed into the renormalization scales as shown by the PMC
method~\cite{Brodsky:2011ta,Brodsky:2011ta,Brodsky:2012rj,Mojaza:2012mf,Brodsky:2013vpa}.
Hence if a theory has the property of \textit{ordered scale
invariance}, it preserves exactly the scale invariance of
observables independently of the process, the kinematics and the
starting order of the observable. We underline that if a theory
has the intrinsic conformality all the renormalized quantities,
such as cross sections, can be parametrized with the iCF-form.
This property should be preserved by the renormalization scheme or
by the definition of IR safe quantities and it should be preserved
also in observables defined in effective theories. The iCF shows
that point (3) of the BLM/PMC approach (Section~\ref{sec:blm}) can
be improved by eliminating the perturbative expansion of the
BLM/PMC scales, leading to a scale and scheme invariant result.
Though we remark that the perturbative corrections in the BLM or
PMCm scales are suppressed in the perturbative region.


\section{The PMC$_\infty$ \label{sec:pmcinf}}

We introduce here a new method to eliminate the scale-setting
ambiguity in single variable scale invariant distributions, which
we call PMC$_\infty$. This method is based on the original PMC
principle and agrees with all the different PMC formulations for
the PMC-scales at the lowest order. Essentially the core of the
PMC$_\infty$ is the same for all the BLM-PMC prescriptions, i.e.
the effective running coupling value and hence its renormalization
scale at each order is determined by the $\beta_0$-term of the
next higher order, or equivalently by the \textit{intrinsic
conformal scale} $ \mu_{\rm N}$. The PMC$_\infty$ preserves the
iCF and thus the scale and scheme invariance, absorbing an
infinite set of $\beta$-terms to all orders. This method differs
from the other PMC prescriptions since, due to the presence of the
intrinsic conformality, no perturbative correction in $\alpha_s$
needs to be introduced at higher orders in the PMC-scales. Given
that all the $\beta$-terms of a single conformal subset are
included in the renormalization scale already with the definition
at lowest order, no initial scale or scheme dependence are left
due to the unknown $\beta$-terms in each subset. The
PMC$_\infty$-scale of each subset can be unambiguously determined
by $\beta_0$-term of each order, we underline that all logarithms
of each subset have the same argument and all the differences
arising at higher orders have to be included only in the conformal
coefficients. Reabsorbing all the $\beta$-terms into the scale
also the $n! \beta_0^n \alpha_s^n$ terms (related to
renormalons~\cite{Beneke:1998ui}) are eliminated, thus the
precision is improved and the perturbative QCD predictions can be
extended to a wider range of values. The initial scale dependence
is totally confined in the unknown PMC$_\infty$ scale of the last
order of accuracy (i.e. up to NNLO case in the
$\alpha_s(\mu_0)^3$).
 Thus if we fix the renormalization scale independently to
the proper intrinsic scale for each subset $\mu_{\rm N}$, we end
up with a perturbative sum of totally conformal contributions up
to the order of accuracy:
\begin{eqnarray}
\frac{1}{\sigma_{0}} \frac{O d \sigma(\mu_{\rm I},\mu_{\rm
II},\mu_{\rm III})}{d O}&=&\left\{ \frac{\alpha_{s}(\mu_{\rm
I})}{2 \pi} \frac{O dA_{\mathit{Conf}}}{d
O}+\left(\frac{\alpha_{s}(\mu_{\rm II})}{2 \pi}\right)^{2} \frac{O
d B_{\mathit{Conf}}}{d O} \right. \nonumber \\ & & \left. +
\left(\frac{\alpha_{s}(\mu_{\rm III})}{2 \pi}\right)^{3}\frac{O d
C_{\mathit{Conf}}}{d O} \right\}+{\cal O}(\alpha_{s}^4),
 \label{observable2}
\end{eqnarray}
at this order, the last scale is set to the physical scale Q, i.e.
$\mu_{\rm III}=\mu_0=Q$.

\section{iCF coefficients and scales: a new ``How-To" method}
\label{sec:icfcoeffandscales}

We describe here how all the coefficients of
Eq.~\ref{newevolution} can be identified from either a numerical
or analytical perturbative calculation. This method applies in
general to any perturbative calculation once results for the
different color factors are kept separate, however we refer to the
particular case of the NNLO thrust distribution results calculated
in Refs.~\cite{Weinzierl:2008iv,Weinzierl:2009ms} for the purpose.
Since the leading order is already ($A_{\mathit{Conf}}$) void of
$\beta$-terms we start with NLO coefficients. A general
numerical/analytical calculation keeps tracks of all the color
factors and the respective coefficients:
\begin{eqnarray}
B_{O}(N_f)=C_F \left[ C_A B_{O}^{N_c}+C_F B_{O}^{C_F}+ T_F N_f
B_{O}^{N_f}\right] \label{Bcoeff}
\end{eqnarray}
where $C_F=\frac{\left(N_{c}^{2}-1\right)}{2 N_{c}}$, $C_A=N_c$
and $T_F=1/2.$ The dependence on $N_f$ is made explicit here for
sake of clarity. We can determine the conformal coefficient
$B_{\mathit{Conf}}$ of the NLO order straightforwardly, by fixing
the number of flavors $N_f$ in order to kill the $\beta_0$ term:
\begin{eqnarray}
B_{\mathit{Conf}}&=& B_{O} \left( N_f \equiv \frac{33}{2} \right),\nonumber \\
B_{\beta_0} \equiv  \log  \frac{\mu_0^2}{\mu_{\rm I}^2}  & = & 2
\frac{B_O-B_{\mathit{Conf}}}{\beta_0 A_{\mathit{Conf}}}
\label{Bconf}
\end{eqnarray}
 we would achieve the same results in the usual PMC
way, i.e. identifying the $N_f$ coefficient with the $\beta_0$
term and then determining the conformal coefficient. Both methods
are consistent and results for the intrinsic scales and the
coefficients are in perfect agreement. At the NNLO a general
coefficient is composed of the contribution of six different color
factors:
\begin{eqnarray}
C_{O}(N_f)&=& \frac{C_F}{4} \left\{ N_{c}^{2} C_{O}^{N_c^2
}+C_{O}^{N_c^0}+\frac{1}{N_{c}^{2}} C_{O}^{\frac{1}{N_c^2}}
 \right. \nonumber \\ & & \left.  +N_{f} N_{c}\cdot C_{O}^{N_f N_c}+\frac{N_{f}}{N_{c}}
C_{O}^{N_f/N_c}+N_{f}^{2} C_{O}^{N_f^2}\right\}. \label{Ccoeff}
\end{eqnarray}
In order to identify all the terms of Eq.~\ref{newevolution} we
notice first that the coefficients of the terms $\beta_0^2$ and
$\beta_1$ are already given by the NLO coefficient $B_{\beta_0}$,
thus we need to determine only the $\beta_0$ and the conformal
$C_{\mathit{Conf}}$ terms. In order to determine the latter
coefficients we use the same procedure we used for the NLO; i.e.
we set the number of flavors $N_f \equiv 33/2$ in order to drop
off all the $\beta_0$ terms. We have then:
\begin{eqnarray}
C_{\mathit{Conf}}&=& C_{O} \left( N_f \equiv \frac{33}{2} \right)- \frac{1}{4}\overline{\beta}_1 B_{\beta_0} A_{\mathit{Conf}},\nonumber \\
C_{\beta_0} \equiv \log\left(\frac{\mu_0^2}{\mu_{\rm II}^2}\right)
& = & \frac{1}{\beta_0 B_{\mathit{Conf}}} {\bigg (}
C_O-C_{\mathit{Conf}}  \nonumber \\
&\!\!\!\!  &\!\!\!\! -\left.\frac{1}{4} \beta_0^2 B_{\beta_0}^2
A_{\mathit{Conf}}- \frac{1}{4} \beta_1 B_{\beta_0}
A_{\mathit{Conf}} \!\right)\!\!,\nonumber \\
 \label{Cconf}
\end{eqnarray}
with $\overline{\beta}_1\equiv \beta_1(N_f=33/2)=-107$. This
procedure can be extended to every order and one may decide
whether to cancel $\beta_0$, $\beta_1$ or $\beta_2$ by fixing the
appropriate number of flavors. The results can be compared leading
to determine exactly all the coefficients. We point out that
extending the intrinsic conformality to all orders we can predict
at this stage the coefficients of all the color factors of the
higher orders related to the $\beta$-terms except those related to
the higher order conformal coefficients and $\beta_0$-terms (e.g.
at NNNLO the $D_{\mathit{Conf}}$ and $D_{\beta_0}$). The
$\beta$-terms are coefficients that stem from UV-divergent
diagrams connected with the running of the coupling constant and
not from UV-finite diagrams. UV-finite $N_F$ terms may arise but
would not contribute to the $\beta$-terms. These terms can be
easily identified by the kinematic constraint at lowest order or
by checking deviations of the $n_f$ coefficients from the iCF
form. In fact, only the $N_f$ terms coming from UV-divergent
diagrams, depending dynamically on the virtuality of the
underlying quark and gluon subprocesses have to be considered as
$\beta$-terms and they would determine the intrinsic conformal
scales.  In general, each $\mu_{\rm N}$ is an independent function
of the physical scale of the process $\sqrt{s}$ (or
$\sqrt{t},\sqrt{u},...$), of the selected variable $O$ and it
varies with the number of colors $N_c$ mainly due to $ggg$ and
$gggg$ vertices. The latter terms arise at higher orders only in a
non-Abelian theory but they are not expected to spoil the
iCF-form. We underline that iCF applies to scale-invariant
single-variable distributions, in case one is interested in the
renormalization of a particular diagram, e.g. the $ggg$ vertex,
contributions from different $\beta$-terms should be singled out
in order to identify the respective intrinsic conformal scale
consistently with the renormalization of the non-Abelian $ggg$
vertex, as shown in~\cite{Binger:2006sj}.

\section{PMC$_\infty$ results for thrust and C-parameter \label{sec:thrustandc}}

The thrust distribution and the event shape variables are a
fundamental tool in order to probe the geometrical structure of a
given process at colliders. Being observables that are exclusive
enough with respect to the final state, they allow for a deeper
geometrical analysis of the process and they are also particularly
suitable for the measurement of the strong coupling $\alpha_s$~\cite{Kluth:2006bw}.\\
Given the high precision data collected at LEP and
SLAC~\cite{ALEPH:2003obs,DELPHI:2003yqh,OPAL:2004wof,L3:2004cdh,SLD:1994idb},
refined calculations are crucial in order to extract information
to the highest possible precision. Though extensive studies on
these observables have been released during the last decades
including higher order corrections from next-to-leading order
(NLO) calculations~\cite{Ellis:1980wv, Kunszt:1980vt,
Vermaseren:1980qz, Fabricius:1981sx, Giele:1991vf, Catani:1996jh}
to the next-to-next-to-leading
order(NNLO)~\cite{Gehrmann-DeRidder:2014hxk,Gehrmann-DeRidder:2007nzq,
GehrmannDeRidder:2007hr, Weinzierl:2008iv, Weinzierl:2009ms} and
including resummation of the large logarithms~\cite{Abbate:2010xh,
Banfi:2014sua}, the theoretical predictions are still affected by
significant theoretical uncertainties that are related to large
renormalization scale ambiguities.

In the particular case of the three-jet event-shape distributions
the conventional practice of CSS leads to results that do not
match the experimental data and the extracted values of $\alpha_s$
deviate from the world average~\cite{ParticleDataGroup:2020ssz}.


The thrust ($T$) and C-parameter ($C$) are defined as
\begin{eqnarray}
T=\max\limits_{\vec{n}}\left(\frac{\sum_{i}|\vec{p}_i\cdot\vec{n}|}{\sum_{i}|\vec{p}_i|}\right),
\end{eqnarray}
\begin{eqnarray}
C=\frac{3}{2}\frac{\sum_{i,j}|\vec{p_i}||\vec{p_j}|\sin^2\theta_{ij}}{\left(\sum_i|\vec{p_i}|\right)^2},
\end{eqnarray}
where the sum runs over all particles in the hadronic final state,
and $\vec{p}_i$ denotes the three-momentum of particle $i$. The
unit vector $\vec{n}$ is varied to maximize thrust $T$, and the
corresponding $\vec{n}$ is called the thrust axis and denoted by
$\vec{n}_T$. The variable $(1-T)$ is often used, which for the LO
of 3-jet production is restricted to the range $(0<1-T<1/3)$. We
have a back-to-back or a spherically symmetric event respectively
at $T=1$ and at $T=2/3$ respectively.
 For the C-parameter, $\theta_{ij}$ is the angle between
$\vec{p_i}$ and $\vec{p_j}$. At LO for the 3 jet production the
C-parameter is restricted to the range: $0\leq C\leq 0.75$ by
kinematics.

In general a normalized IR-safe single-variable observable, such
as the thrust distribution for the $e^+ e^-\rightarrow
3jets$~\cite{DelDuca:2016ily,DelDuca:2016csb}, is the sum of pQCD
contributions calculated up to NNLO at the initial renormalization
scale $\mu_0=\sqrt{s}=M_{Z}$:
\begin{eqnarray}
\frac{1}{\sigma_{tot}} \! \frac{O d \sigma(\mu_{0})}{d O}\! & = &
\left\{ x_0 \cdot \frac{ O d \bar{A}_{\mathit{O}}(\mu_0)}{d O} +
x_0^2 \cdot \frac{ O d \bar{B}_{\mathit{O}}(\mu_0)}{d O} \right. \nonumber  \\
 & & + \left. x_0^{3} \cdot \frac{O d\bar{C}_{\mathit{O}}(\mu_0)}{d
O}+ {\cal O}(\alpha_{s}^4) \right\},
 \label{observable1-thrust}
\end{eqnarray}
where $x(\mu)\equiv \alpha_s(\mu)/(2\pi)$, $O$ is the selected
event shape variable, $\sigma$ the cross section of the process,\\
$$\sigma_{tot}=\sigma_{0} \left( 1+x_0
A_{t o t}+ x_0^{2} B_{t o t}+ {\cal
O}\left(\alpha_{s}^{3}\right)\right)$$ \\ is the total hadronic
cross section and $\bar{A}_O, \bar{B}_O, \bar{C}_O$ are
respectively the normalized LO, NLO and NNLO coefficients:
\begin{eqnarray}
\bar{A}_{O} &=&A_{O} \nonumber \\
\bar{B}_{O} &=&B_{O}-A_{t o t} A_{O} \\
\bar{C}_{O} &=&C_{O}-A_{t o t} B_{O}-\left(B_{t o t}-A_{t o
t}^{2}\right) A_{O}, \nonumber
\end{eqnarray}
where $A_O, B_O, C_O$ are the coefficients normalized to the
tree-level cross section $\sigma_0$ calculated by MonteCarlo (see
e.g. the EERAD and Event2 codes~\cite{Gehrmann-DeRidder:2014hxk,
Gehrmann-DeRidder:2007nzq, GehrmannDeRidder:2007hr,
Weinzierl:2008iv, Weinzierl:2009ms}) and $A_{\mathit{tot}},
B_{\mathit{tot}}$ are:
\begin{eqnarray}
A_{\mathit{tot}} &= & \frac{3}{2} C_F ; \nonumber \\
B_{\mathit{tot}} &= & \frac{C_F}{4}N_c +\frac{3}{4}C_F
\frac{\beta_0}{2} (11-8\zeta(3)) -\frac{3}{8} C_F^2,
 \label{norm}
\end{eqnarray}
where $\zeta$ is the Riemann zeta function.

In general according to CSS the renormalization scale is set to
$\mu_0=\sqrt{s}=M_Z$ and theoretical uncertainties are evaluated
using standard criteria. In this case, we have used the definition
given in Ref.~\cite{Gehrmann-DeRidder:2007nzq} of the parameter
$\delta$, we define the average error for the event shape variable
distributions as:
 \begin{equation}
\bar{\delta}=\frac{1}{N} \sum_i^N \frac{ {\rm
max}_{\mu}(\sigma_i(\mu))- {\rm min}_{\mu} (\sigma_i(\mu))}{2
\sigma_i(\mu=M_Z)} \label{delta}
\end{equation} where $i$ is the index of the bin and $N$ is the
total number of bins, the renormalization scale is varied in the
range: $\mu \in [M_Z/2; 2 M_Z]$.

\section{The PMC$_\infty$ scales at LO and NLO
\label{sec:lonloscales}}

According to the PMC$_\infty$ prescription we fix the
renormalization scale to $\mu_{\rm N}$ at each order absorbing all
the $\beta$ terms into the coupling. We notice a small mismatch
between the zeroes of the conformal coefficient
$B_{\mathit{Conf}}$ and those of the remaining $\beta_0$ term in
the numerator (the formula is shown in Eq.~\ref{Cconf}). Due to
our limited knowledge of the strong coupling at low energies, in
order to avoid singularities in the NLO-scale $\mu_{\rm II}$, we
introduce a regularization that leads to a finite scale
$\tilde{\mu}_{\rm II}$ in the whole range of values of the
variable $(1-T)$. These singularities might be due either to the
presence of UV finite $N_F$ terms or to the logarithmic behavior
of the conformal coefficients when low values of the variable
$1-T$ are approached. Large logarithms arise from the
IR-divergence cancellation procedure and they can be resummed in
order to restore a predictive perturbative
regime~\cite{Catani:1991kz,Catani:1992ua,Catani:1996yz,Aglietti:2006wh,Aglietti:2007bp,Banfi:2014sua,Abbate:2010xh}.
We point out that IR cancellation should not spoil the iCF
property. Whether this is an actual deviation from the iCF-form
has to be further investigated. However, since the discrepancies
between the coefficients are rather small, we introduce a
regularization method based on redefinition of the norm of the
coefficient $B_{\mathit{Conf}}$ in order to cancel out these
singularities in the $\mu_{\rm II}$-scale. This regularization is
consistent with the PMC principle and up to the accuracy of the
calculation it does not introduce any bias effect in the results
or any ambiguity in the NLO-PMC$_\infty$ scale. All the
differences introduced by the regularization would start at the
$\rm N^3LO$ order of accuracy and they can be reabsorbed after in
the higher order PMC$_\infty$ scales. Thus we obtain for the
PMC$_\infty$ scales, $\mu_{\rm N}$ :


\begin{eqnarray}
\large{\mu}_{\rm I} & = & \sqrt{s} \cdot e^{f_{sc}-\frac{1}{2} B_{\beta_0}},\hspace{1.9cm}{ \scriptstyle (1-T)<0.33}  \label{icfscale1}  \\
\large{\tilde{\mu}}_{\rm II} & =& \left\{
\begin{array}{lr} \sqrt{s} \cdot e^{f_{sc}-\frac{1}{2} C_{\beta_0}
\cdot \frac{B_{\mathit{Conf}}}{B_{\mathit{Conf}}+\eta \cdot
A_{\mathit{tot}} A_{\mathit{Conf}} }},\\ \hspace{3.9cm} {\scriptstyle (1-T)<0.33}  \\
  \sqrt{s}\cdot e^{f_{sc}-\frac{1}{2} C_{\beta_0}},\\ \hspace{3.9cm}{ \scriptstyle (1-T)>0.33}
  \label{icfscale2}\\
 \end{array} \right.
 \label{PMC12}
\end{eqnarray}
where $\sqrt{s}=M_{Z}$ and the third scale is set to $\mu_{\rm
III}=\mu_0=\sqrt{s}$. The renormalization scheme factor for the
QCD results is set to $f_{sc}\equiv 0$. This scheme factor
reabsorbs also the scheme difference into the renormalization
scale and is related to the particular choice of the scale
parameter $\Lambda$ as discussed in section~\ref{sec:lambdapar}.
The coefficients $B_{\beta_0},C_{\beta_0}$ are the coefficients
related to the $\beta_0$-terms of the NLO and NNLO perturbative
order of the thrust distribution respectively. They are determined
from the calculated $A_O, B_O, C_O$ coefficients.

The $\eta$ parameter is a regularization term to cancel the
singularities of the NLO scale, $\mu_{\rm II}$, in the range
$(1-T)<0.33$, depending on non-matching zeroes between numerator
and denominator in the $C_{\beta_0}$. In general this term is not
mandatory for applying the PMC$_{\infty}$, it is necessary only in
case one is interested in applying the method all over the entire
range covered by the thrust, or any other observable. Its value
has been determined as $\eta=3.51$ for the thrust distribution and
it introduces no bias effects up to the accuracy of the
calculations and the related errors are totally negligible up to
this stage.

\begin{figure}[htb]
\centering
\includegraphics[width=12cm]{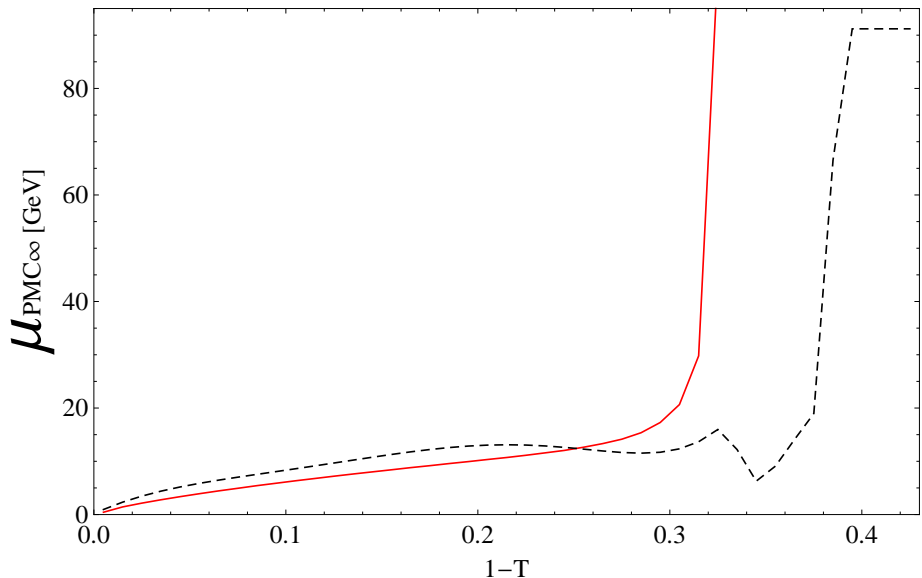}
\caption{The LO-PMC$_\infty$ (solid red) and the NLO-PMC$_\infty$
(dashed black) scales for thrust.~\cite{DiGiustino:2020fbk}}
\label{Tscales}
\end{figure}

 We point out that in the region $(1-T)>0.33$ we
have a clear example of intrinsic conformality-iCF where the
kinematic constraints set the $A_{\mathit{Conf}}=0$. According to
Eq.~\ref{evolved} setting the $A_{\mathit{Conf}}=0$ the whole
conformal subset $\sigma_{\rm I}$ becomes null. In this case all
the $\beta$ terms at NLO and NNLO disappear except the
$\beta_0$-term at NNLO which determines the $\mu_{\rm II}$ scale.
The surviving $n_f$ terms at NLO or the $n_f^2$ at NNLO are
related to the finite $N_F$-term at NLO and to the mixed $N_f
\cdot N_F$ term arising from $B_O \cdot \beta_0$ at NNLO. Using
the parametrization with explicit $n_f$ terms we have for
$(1-T)>0.33$:
\begin{eqnarray}
 A_{O} &=&  0, \nonumber \\
B_{O}  &=& B_0+ B_1 \cdot N_F, \nonumber \\
 C_{O} &=& C_0+C_1 \cdot n_f +C_2 \cdot N_f\cdot N_F.
  \label{Cparpar}
  \end{eqnarray}
 we can determine the $\tilde{\mu}_{\rm II}$ for the region $(1-T)>0.33$ as shown in Eq.~\ref{icfscale2}:

  \begin{equation}
 C_{\beta_0}=\left( \frac{C_1}{\frac{11}{3} C_A B_1-\frac{2}{3}
 B_0}\right)
  \label{mu2}
  \end{equation}
by identifying the $\beta_0$-term at NNLO. The LO and NLO
PMC$_\infty$ scales are shown in Fig.~\ref{Tscales}.
We notice that the two PMC$_\infty$ scale have similar behaviors
in the range $(1-T)<0.33$ and the LO-PMC$_\infty$ scale agrees
with the PMC scale used in Ref.~\cite{Wang:2019ljl}. This method
totally eliminates both the ambiguity in the choice of the
renormalization scale and the scheme dependence at all orders in
QCD.

\section{NNLO Thrust distribution results}

We use here the results of
Ref.~\cite{Weinzierl:2008iv,Weinzierl:2009ms} and for the running
coupling $\alpha_s(Q)$ we use the RunDec
program~\cite{Chetyrkin:2000yt}. In order to normalize
consistently the thrust distribution we expand the denominator in
$\alpha_0\equiv \alpha_s(\mu_0)$ while the numerator has the
couplings renormalized at different PMC$_\infty$ scales $\alpha_I
\equiv \alpha_s(\mu_{\rm I})$, $\alpha_{II}\equiv
\alpha_s(\tilde{\mu}_{II})$. We point out here that the proper
normalization would be given by the integration of the total cross
section after renormalization with the PMC$_\infty$ scales,
nonetheless since the PMC$_\infty$ prescription involves only
absorption of higher order terms into the scales, the difference
would be within the accuracy of the calculations, i.e. $\sim
\mathcal{O}(\alpha_s^4(\mu_0))$. Eq.~\ref{observable1-thrust}
becomes:
\begin{equation}
\frac{1}{\sigma_{tot}} \! \frac{O d \sigma(\mu_{\rm
I},\tilde{\mu}_{\rm II},\mu_{0})}{d O}=
\left\{\overline{\sigma}_{\rm I}+\overline{\sigma}_{\rm
II}+\overline{\sigma}_{\rm III}+ {\cal O}(\alpha_{s}^4) \right\},
 \label{observable3}
\end{equation}
where the $\overline{\sigma}_{N}$ are normalized subsets that are
given by:
\begin{eqnarray}
\overline{\sigma}_{\rm I} &=& A_{\mathit{Conf}} \cdot x_{\rm I} \nonumber  \\
\overline{\sigma}_{\rm II} &=& \left( B_{\mathit{Conf}}+\eta
A_{\mathit{tot}} A_{\mathit{Conf}} \right)\cdot x_{\rm II}^2
 - \eta A_{\textrm{tot}} A_{\mathit{Conf}} \cdot x_0^2 \nonumber \\
 & & -A_{\textrm{tot}} A_{\mathit{Conf}}\cdot x_0 x_{\rm I}  \nonumber \\
\overline{\sigma}_{\rm III} &=&\!\! \left( C_{\mathit{Conf}} \!-\!
A_{\textrm{tot}} \!B_{\mathit{Conf}}\!-\!(B_{\textrm
{tot}}-A_{\textrm{tot}}^{2}) A_{\mathit{Conf}}\right) \cdot x_0^3 ,\nonumber \\
\label{normalizedcoeff}
\end{eqnarray}
$A_{\mathit{Conf}}, B_{\mathit{Conf}}, C_{\mathit{Conf}}$ are the
scale-invariant conformal coefficients (i.e. the coefficients of
each perturbative order not depending on the scale $\mu_R$) while
$x_{\rm I},x_{\rm II},x_0$ are the couplings determined at the
$\mu_{\rm I},\tilde{\mu}_{\rm II},\mu_0$ scales respectively.


Normalized subsets for the region $(1-T)>0.33$ can be achieved
simply by setting $A_{\mathit{Conf}}\equiv 0$ in the
Eq.~\ref{normalizedcoeff}. Within the numerical precision of these
calculations there is no evidence of the presence of spurious
terms, such as any further UV-finite $N_F$ term up to
NNLO~\cite{Gehrmann:2014uva} besides the kinematic term at lowest
order in the multi-jet region. These terms, if there are any, must
remain rather small all over the range of the thrust variable in
comparison with the $\beta$ term or even be compatible with
numerical fluctuations. Moreover, we notice a small rather
constant difference between the iCF-predicted and the calculated
coefficient for the $N_f^2$-color factor of
Ref.~\cite{Weinzierl:2008iv} which might be due to a $n_f^2$
UV-finite coefficient or possibly to statistics. This small
difference must be included in the conformal coefficient but it
has a completely negligible impact on the total thrust
distribution.
\begin{figure}[htb]
\centering
\includegraphics[width=12cm]{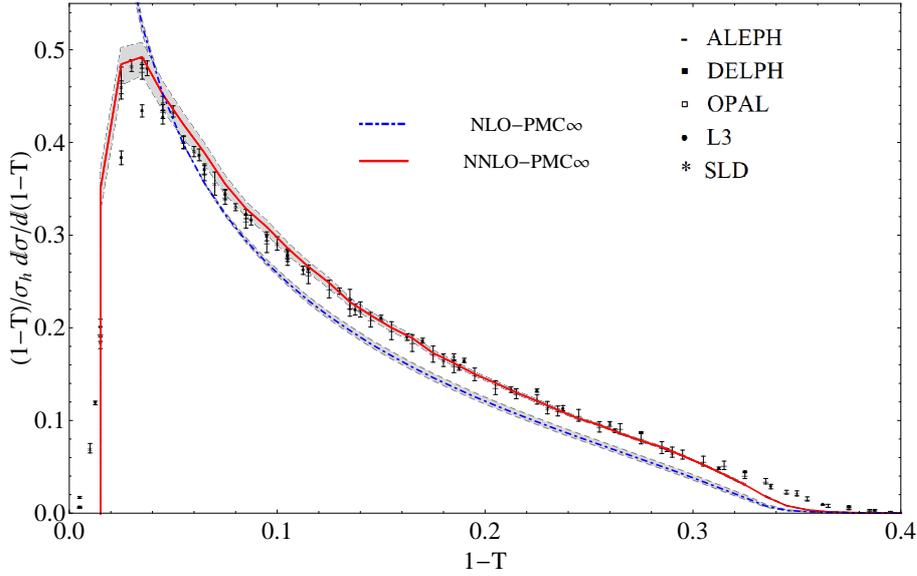}
\caption{The thrust distribution under the PMC$_\infty$ at NLO
(dotdashed blue) and at NNLO (solid
red)~\cite{DiGiustino:2020fbk}. The experimental data points are
taken from the ALEPH, DELPHI,OPAL, L3, SLD
experiments~\cite{ALEPH:2003obs,DELPHI:2003yqh,OPAL:2004wof,L3:2004cdh,SLD:1994idb}.
The shaded area shows theoretical errors for the PMC$_\infty$
predictions at NLO and at NNLO.} \label{thrust}
\end{figure}
In Fig.~\ref{thrust} we show the thrust distribution at NLO and at
NNLO with the use of the PMC$_\infty$ method. Theoretical errors
for the thrust distribution at NLO and at NNLO are also shown (the
shaded area). Conformal quantities are not affected by a change of
renormalization scale. Thus the errors shown give an evaluation of
the level of conformality achieved up to the order of accuracy and
they have been calculated using standard criteria, i.e. varying
the remaining initial scale value in the range $\sqrt{s}/2 \leq
\mu_0 \leq 2 \sqrt{s}$. Using the same definition of the parameter
$\bar{\delta}$ given in Eq.~\ref{delta}, we have in the interval
$0<(1-T)<0.33$ an average error of $\bar{\delta}\simeq 3.54\%$ and
$1.77\%$ for the thrust at NLO and at NNLO respectively. A larger
improvement has been obtained in the entire range of reliable
results for thrust distribution, i.e. $0<(1-T)<0.42$, from
$\bar{\delta}\simeq 7.36\%$ to $1.95\%$ from NLO to the NNLO
accuracy with the PMC$_\infty$.

\begin{figure}[htb]
\centering
\includegraphics[width=12cm]{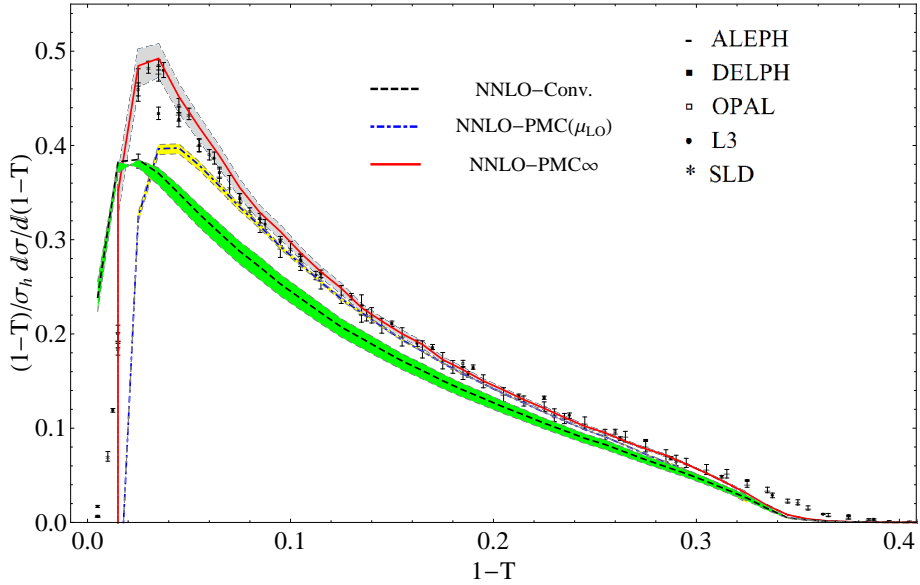}
\caption{The thrust distribution at NNLO under the Conventional
(dashed black), the PMC($\mu$\textsubscript{LO}) (dotdashed blue)
and the PMC$_\infty$ (solid red)~\cite{DiGiustino:2020fbk}. The
experimental data points are taken from the ALEPH, DELPHI,OPAL,
L3, SLD
experiments~\cite{ALEPH:2003obs,DELPHI:2003yqh,OPAL:2004wof,L3:2004cdh,SLD:1994idb}.
The shaded areas show theoretical error predictions at NNLO, which
have been calculated varying the remaining initial scale value in
the range $\sqrt{s}/2 \leq \mu_0 \leq 2 \sqrt{s}$.}
\label{thrust2}
\end{figure}

In Fig.~\ref{thrust2} a direct comparison of the PMC$_\infty$ with
the CSS results (obtained in ~\cite{Weinzierl:2008iv} and
~\cite{Gehrmann-DeRidder:2007nzq,GehrmannDeRidder:2007hr}) is
shown. In addition, we have shown also the results of the first
PMC approach used in~\cite{Wang:2019ljl}, which we indicate as
PMC($\mu$\textsubscript{LO}) extended to NNLO accuracy. In this
approach the last unknown PMC scale $\mu$\textsubscript{NLO} of
the NLO was set to the last known PMC scale
$\mu$\textsubscript{LO} of the LO, while the NNLO scale
$\mu$\textsubscript{NNLO}$\equiv \mu_0$ was left unset and varied
in the range $\sqrt{s}/2 \leq \mu_0 \leq 2 \sqrt{s}$. This
analysis was performed in order to show that the procedure of
setting the last unknown scale to the last known one leads to
stable and precise results and is consistent with the proper PMC
method in a wide range of values of the $(1-T)$ variable.

\begin{table}[h!]
\centering
 \begin{tabular}{||c|c|c|c||}
   \hline
 $\bar{\delta}[\%]$  &  Conv.  & PMC($\mu$\textsubscript{LO}) & PMC$_\infty$ \\
    \hline
   $0.10 < (1-T) < 0.33$ & 6.03 & 1.41 & 1.31 \\
    $0.21 < (1-T) < 0.33$ & 6.97 & 2.19 & 0.98 \\
    $0.33 < (1-T) < 0.42$ & 8.46 & -    & 2.61 \\
    $0.00 < (1-T) < 0.33$ &  5.34  & 1.33 & 1.77\\
   $0.00 < (1-T) <0.42$ & 6.00 & -  &  1.95 \\
   \hline
   \end{tabular}
 \caption{Average error, $\bar{\delta}$, for NNLO Thrust distribution under Conventional, PMC($\mu$\textsubscript{LO}) and PMC$_\infty$
 scale settings calculated in different ranges of values of the $(1-T)$ variable.}
 \label{tab:1}
\end{table}
Average errors calculated in different regions of the spectrum are
reported in Table~\ref{tab:1}. From the comparison with the CSS we
notice that the PMC$_\infty$ prescription significantly improves
the theoretical predictions. Moreover, results are in remarkable
agreement with the experimental data in a wider range of values
$(0.015 \leq 1-T \leq 0.33)$ and they show an improvement of the
PMC$(\mu_{LO})$ results when the two-jets and the multi-jets
regions are approached, i.e. the region of the peak and the region
$(1-T)>0.33$ respectively. The use of the PMC$_\infty$ approach in
perturbative thrust QCD calculations restores the correct behavior
of the thrust distribution in the region $(1-T)>0.33$ and this is
a clear effect of the iCF property. Comparison with experimental
data has been improved all over the spectrum and the introduction
of the $\rm N^3LO$ order correction would improve this comparison
especially in the multi-jet $(1-T) > 0.33$ region. In the
PMC$_\infty$ method theoretical errors are given by the unknown
intrinsic conformal scale of the last order of accuracy. We expect
this scale not to be significantly different from that of the
previous orders. In this particular case, as shown in
Eq.~\ref{normalizedcoeff}, we have also a dependence on the
initial scale $\alpha_s(\mu_0)$ left due to the normalization and
to the regularization terms. These errors represent 12.5\% and
1.5\% respectively of the whole theoretical errors in the range
$0<(1-T)<0.42$ and they could be improved by means of a correct
normalization.

\section{NNLO C-parameter distribution results}

\begin{figure}[htb]
\centering
\includegraphics[width=12cm]{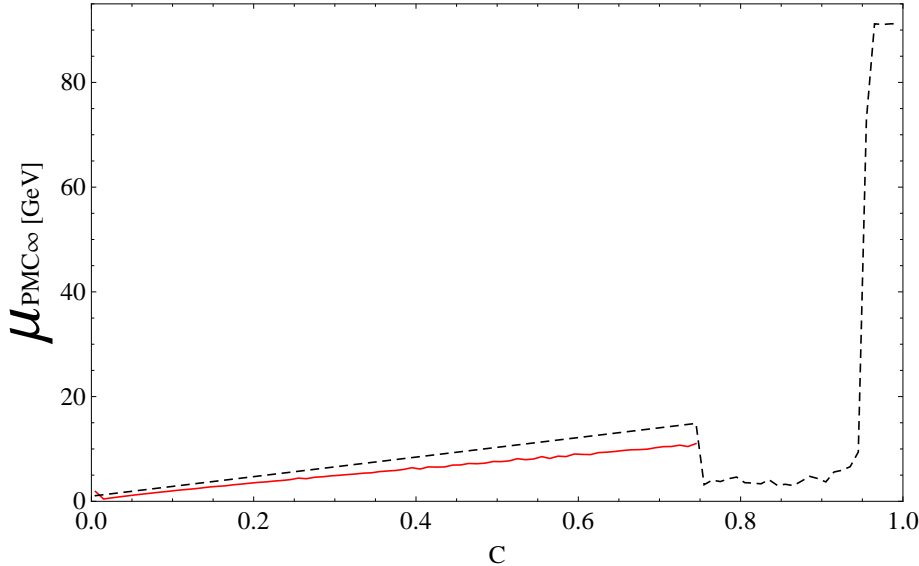}
\caption{The LO-PMC$_\infty$ (solid red) and the NLO-PMC$_\infty$
(dashed black) scales for C-parameter.~\cite{DiGiustino:2020fbk}}
\label{Cpar-scales}
\end{figure}

\begin{figure}[htb]
\centering
\includegraphics[width=12cm]{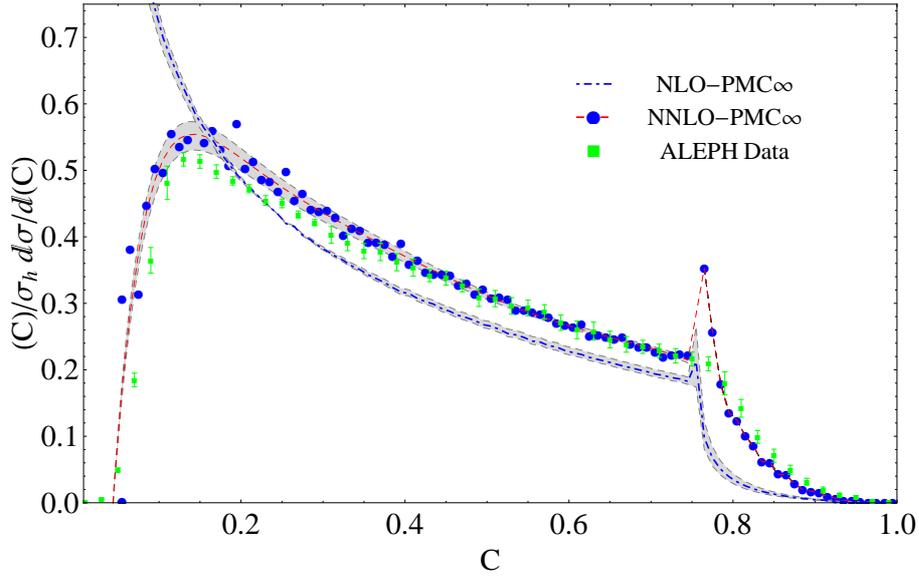}
\caption{The C-parameter distribution under the PMC$_\infty$ at
NLO (dotdashed blue) and at NNLO (dashed
red)~\cite{DiGiustino:2020fbk}. Blue points indicate the
NNLO-PMC$_\infty$ thrust distribution obtained with $\mu_{\rm
III}=\mu_0=M_Z$. The experimental data points (green) are taken
from the ALEPH experiment~\cite{ALEPH:2003obs}. The dashed lines
of the NNLO distribution show fits of the theoretical calculations
with interpolating functions for the values of the remaining
initial scale $\mu_0 = 2 M_Z$ and  $M_Z/2$. The shaded area shows
theoretical errors for the PMC$_\infty$ predictions at NLO and at
NNLO calculated varying the remaining initial scale value in the
range $\sqrt{s}/2 \leq \mu_0 \leq 2 \sqrt{s}$.} \label{Cpar}
\end{figure}

The same analysis applies straightforwardly to the C-parameter
distribution including the regularizing $\eta$ parameter, which
has been set to the same value $3.51$. The same scales of
Eq.~\ref{icfscale1} and Eq.~\ref{icfscale2} apply to the
C-parameter distribution in the region $0<C<0.75$ and in the
region $0.75<C<1$. In fact, due to kinematic constraints that set
the $A_{\mathit{Conf}}=0$, we have the same iCF effect also for
the C-parameter. Results for the C-parameter scales and
distributions are shown in Fig.~\ref{Cpar-scales} and
Fig.~\ref{Cpar} respectively.

Theoretical errors have been calculated, as in the previous case,
using standard criteria and results indicate an average error over
the whole spectrum $0<C<1$ of the C-parameter distribution at NLO
and at NNLO of $\bar{\delta}\simeq 7.26\%$ and $2.43\%$
respectively.

\begin{table}[htb]
\centering
 \begin{tabular}{ || c |c|c|c||  }
   \hline
 $\bar{\delta}$[\%]  &  Conv.  & PMC($\mu$\textsubscript{LO}) & PMC$_\infty$ \\
   \hline
   $0.00 < (C) < 0.75$ & 4.77 & 0.85 & 2.43 \\
    $0.75 < (C) < 1.00$ & 11.51 & 3.68 & 2.42 \\
    $0.00 < (C) < 1.00$ & 6.47 & 1.55 & 2.43 \\
  \hline
 \end{tabular}
 \caption{Average error  ,$\bar{\delta}$, for NNLO C-parameter distribution under Conventional, PMC($\mu$\textsubscript{LO}) and PMC$_\infty$
 scale settings calculated in different ranges of values of the $(C)$ variable.}
 \label{tab:2}
\end{table}
A comparison of average errors according to the different methods
is shown in Table~\ref{tab:2}. Results show that the PMC$_\infty$
improves the NNLO QCD predictions for the C-parameter distribution
over the entire spectrum.

\begin{figure}[htb]
\centering
\includegraphics[width=12cm]{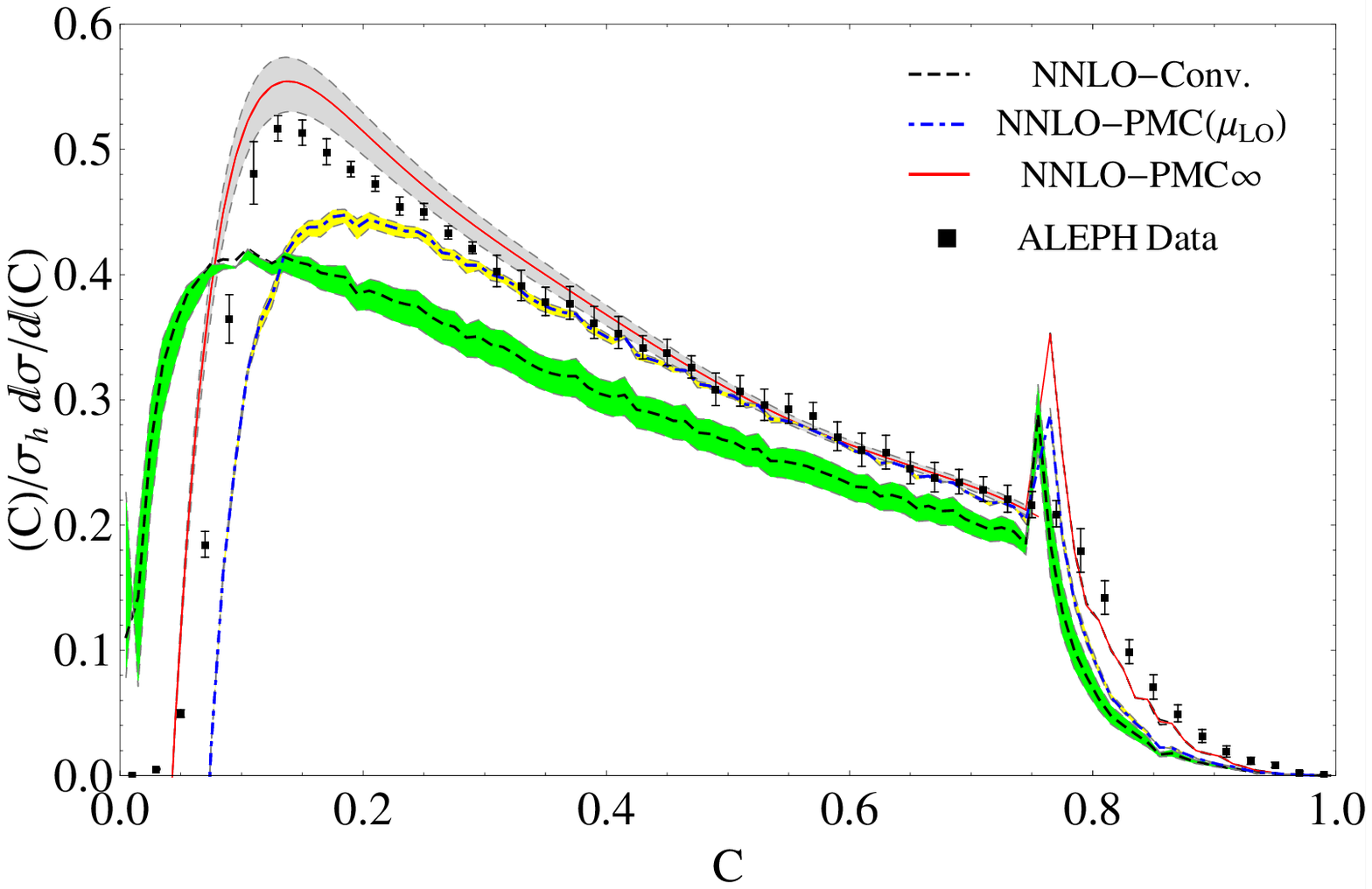}
\caption{The NNLO C-parameter distribution under CSS (dashed
black), the PMC($\mu$\textsubscript{LO}) (dotdashed blue) and the
PMC$_\infty$ (solid red)~\cite{DiGiustino:2020fbk}. The
experimental data points (Black) are taken from the ALEPH
experiment~\cite{ALEPH:2003obs}. The shaded area shows theoretical
error predictions at NNLO calculated varying the remaining initial
scale value in the range $\sqrt{s}/2 \leq \mu_0 \leq 2 \sqrt{s}$.}
\label{Cpar2}
\end{figure}

A comparison of the distributions calculated with the CSS, the
PMC($\mu$\textsubscript{LO})~\cite{Wang:2019isi} and the
PMC$_\infty$ is shown in Fig.~\ref{Cpar2}. Results for the
PMC$_\infty$ show remarkable agreement with the experimental data
away from the regions $C<0.05$ and $C\simeq 0.75$. The errors due
to the normalization and to the regularization terms
(Eq.~\ref{normalizedcoeff}) are respectively $8.8\%$ and $0.7\%$
of the whole theoretical errors. The perturbative calculations
could be further improved using a correct normalization and also
by introducing the large-logarithm resummation technique in order
to extend the perturbative regime.

\chapter{The thrust distribution in the QCD perturbative conformal window and in QED \label{sec:confwin}}

We employ, for the first time, the perturbative regime of the
quantum chromodynamics (pQCD) infrared conformal window as a
laboratory to investigate in a controllable manner (near)
conformal properties of physically relevant quantities such as the
thrust distribution in electron-positron annihilation
processes~\cite{DiGiustino:2021nep}.

The conformal window of pQCD has a long and noble history
conveniently summarized and generalized to arbitrary
representations in Ref.~\cite{Dietrich:2006cm}. Several lattice
gauge theory applications and results have been summarized in a
recent report on the subject in Ref.~\cite{Cacciapaglia:2020kgq}.

\section{The thrust distribution according to $N_f$}

It would be highly desirable to compare the PMC and CSS methods
along the entire renormalization group flow from the highest
energies down to zero energy. This is precluded in standard QCD
with a number of active flavors less than six because the theory
becomes strongly coupled at low energies. We therefore employ the
perturbative regime of the conformal window (Sec.~\ref{twoloops})
which allows us to arrive at arbitrary low energies and obtain the
corresponding results for the SU($3$) case at the cost of
increasing the number of active flavors. Here we are able to
deduce the full solution at NNLO in the strong coupling.
 In this chapter we will consider the region of flavors and
colors near the upper bound of the conformal window , i.e.
$N_f\sim 11/2 N_c$ , where the IR fixed point can be reliably
accessed in perturbation theory and we compare the two
renormalization scale setting methods, the CSS and the
PMC$_\infty$.

\begin{figure}[h]
 \centering
\hspace*{-0.5cm}
\includegraphics[width=12cm]{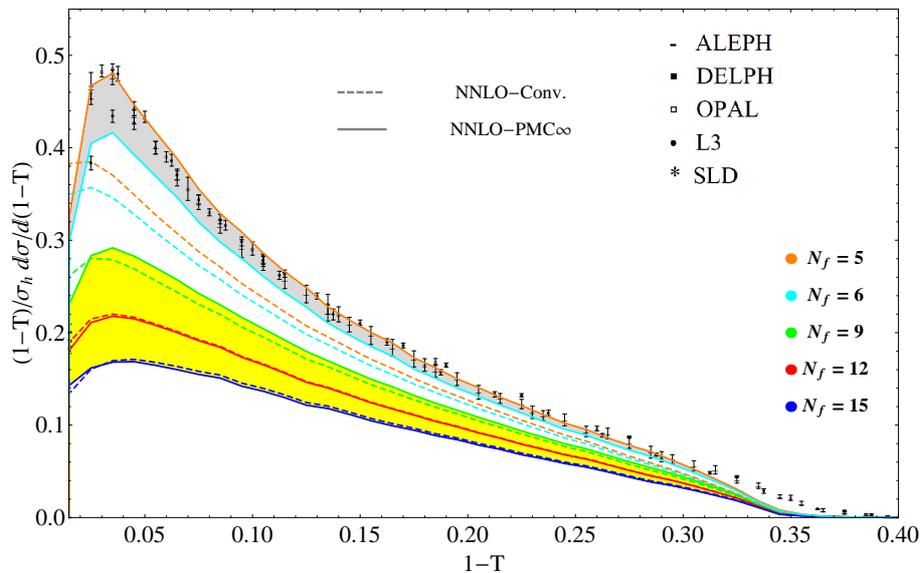}
\caption{Thrust distributions for different values of $N_f$, using
the PMC$_\infty$ (solid line) and the CSS (dashed
line)~\cite{DiGiustino:2021nep}. The Yellow shaded area is the
results for the values of $N_f$ taken in the conformal window. The
experimental data points are taken from the ALEPH, DELPHI,OPAL,
L3, SLD
experiments~\cite{ALEPH:2003obs,DELPHI:2003yqh,OPAL:2004wof,L3:2004cdh,SLD:1994idb}.}
\label{confthrust}
\end{figure}

Results for the thrust distribution calculated using the  NNLO
solution for the coupling $\alpha_s(\mu)$, at different values of
the number of flavors, $N_f$,  is shown in Fig.~\ref{confthrust}.

\begin{figure}[htb]
\centering
\includegraphics[width=10cm]{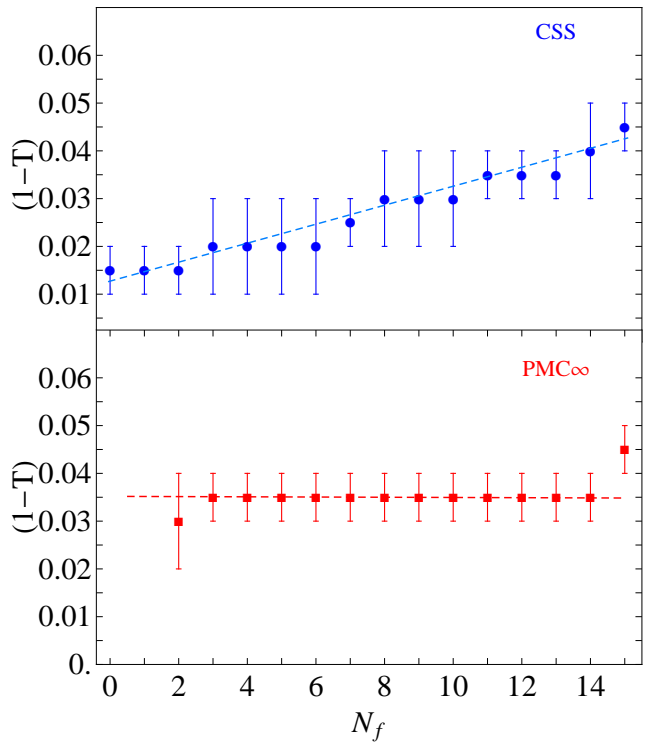}
\caption{Comparison of the position of the peak for the thrust
distribution using the CSS and the PMC$_\infty$ vs the number of
flavors, $N_f$. Dashed lines indicate the particular trend in each
graph~\cite{DiGiustino:2021nep}.} \label{peaks}
\end{figure}

A direct comparison between PMC$_\infty$ (solid line) and CSS
(dashed line) is shown at different values of the number of
flavors. We notice that, despite the phase transition (i.e. the
transition from an infrared finite coupling to an infrared
diverging coupling), the curves given by the PMC$_\infty$ at
different $N_f$, preserve with continuity the same characteristics
of the conformal distribution setting $N_f$ out of the conformal
window of pQCD. In fact, the position of the peak of the thrust
distribution is well preserved varying $N_f$ in and out of the
conformal window using the PMC$_\infty$, while there is constant
shift towards lower values using the CSS. These trends are shown
in Fig.~\ref{peaks}. We notice that in the central range,
$2<N_f<15$, the position of the peak is exactly preserved using
the PMC$_\infty$ and overlaps with the position of the peak shown
by the experimental data. According to our analysis for the case
PMC$_\infty$, in the range, $N_f<2$ the number of bins is not
enough to resolve the peak, though the behavior of the curve is
consistent with the presence of a peak in the same position, while
for $N_f \rightarrow 0$ the peak is no longer visible.
Theoretical uncertainties on the position of the peak have been
calculated using standard criteria, i.e. varying the remaining
initial scale value in the range $M_Z/2 \leq \mu_0 \leq 2 M_Z$,
and considering the lowest uncertainty given by the half of the
spacing between two adjacent bins.

\begin{figure}[htb]
\centering
\includegraphics[width=10cm]{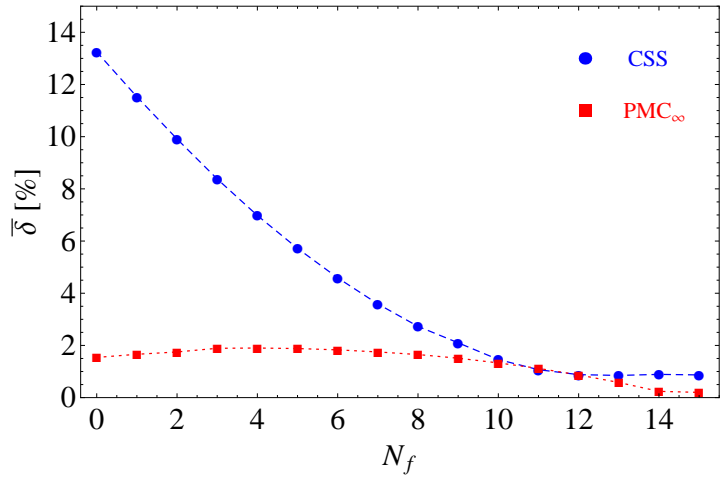}
\caption{Comparison of the average theoretical error,
$\bar{\delta}$, calculated using standard criteria in the range:
$0.005<(1-T)<0.4$, using the CSS and the PMC$_\infty$ for the
thrust distribution vs the number of flavors,
$N_f$~\cite{DiGiustino:2021nep}.} \label{err}
\end{figure}

Using the definition given in Eq.~\ref{delta}, we have determined
the average error, $\bar{\delta}$, calculated in the interval
$0.005<(1-T)<0.4$ of thrust and results for CSS and PMC$_\infty$
are shown in Fig.~\ref{err}. We notice that the PMC$_\infty$ in
the perturbative and IR conformal window, i.e. $12<N_f<\bar{N}_f$,
which is the region where $\alpha_s(\mu)<1$ in the whole range of
the renormalization scale values, from $0$ up to $\infty$, the
average error given by PMC$_\infty$ tends to zero ($\sim
0.23-0.26\%$) while the error given by the CSS tends to remain
constant ($0.85-0.89\%$). The comparison of the two methods shows
that, out of the conformal window, $N_f<\frac{34 N_c^3}{13
N_c^2-3}$, the PMC$_\infty$ leads to a higher precision.

\section{The thrust distribution in the Abelian limit
$N_c\rightarrow 0$ \label{sec:qedthrust}}

\begin{figure}[htb]
\centering
\includegraphics[width=12cm]{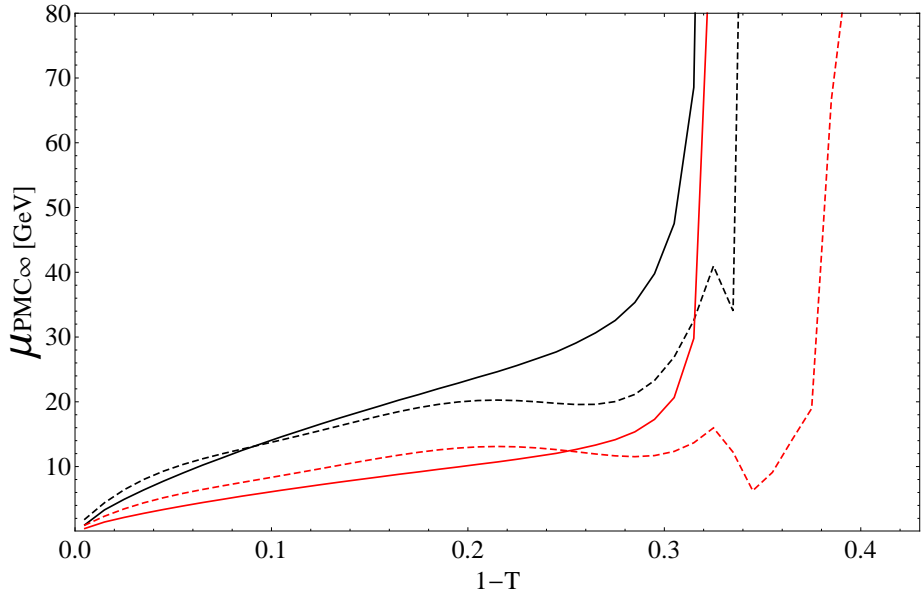}
\caption{PMC$_\infty$ scales for the thrust distribution: LO-QCD
scale (solid red); LO-QED scale (solid black);NLO-QCD scale
(dashed red); NLO-QED scale (dashed
black)~\cite{DiGiustino:2021nep}.} \label{scales}
\end{figure}

We consider now the thrust distribution in U(1) Abelian QED, which
rather than being infrared interacting is infrared free. We obtain
the QED thrust distribution performing the $N_c\rightarrow 0$
limit of the QCD thrust at NNLO according
to~\cite{Brodsky:1997jk,Kataev:2015yha}. In the zero number of
colors limit the gauge group color factors are fixed by $N_A=1,$
$C_F=1,$ $T_R=1,$ $C_A=0,$ $N_c=0,$ $N_f=N_l$, where $N_l$ is the
number of active leptons, while the $\beta$-terms and the coupling
rescale as $\beta_n/C_F^{n+1}$ and $\alpha_s \cdot C_F$
respectively. In particular $\beta_0=-\frac{4}{3}N_l$ and
$\beta_1=-4 N_l$ using the normalization of Eq.~\ref{betafun1}.
According to this rescaling of the color factors we have
determined the QED thrust and the QED PMC$_\infty$ scales. For the
QED coupling, we have used the analytic formula for the effective
fine structure constant in the $\overline{\textrm{MS}}$-scheme:
\begin{equation}
{\alpha(Q^2)} = {\alpha \over  { \left(1 -\Re e
\Pi^{\overline{\textrm{MS}}} (Q^2)\right)}},
\end{equation}
with $\alpha^{-1}\equiv \alpha(0)^{-1}= 137.036$ and the vacuum
polarization function ($\Pi$) calculated perturbatively at two
loops including contributions from leptons, quarks and $W$ boson.
The QED PMC$_\infty$ scales have the same form of
Eqs.~\ref{icfscale1} and \ref{icfscale2} with the factor for the
$\overline{\textrm{MS}}$-scheme set to $f_{sc}\equiv 5/6$ and the
$\eta$ regularization parameter introduced to cancel singularities
in the NLO PMC$_\infty$ scale $\mu_{\rm II}$ in the $N_c
\rightarrow 0$ limit tends to the same QCD value, $ \eta=3.51 $. A
direct comparison between QED and QCD PMC$_\infty$ scales is shown
in Fig.~\ref{scales}.

\begin{figure}[htb]
\centering \vspace{1cm}
\includegraphics[width=12cm]{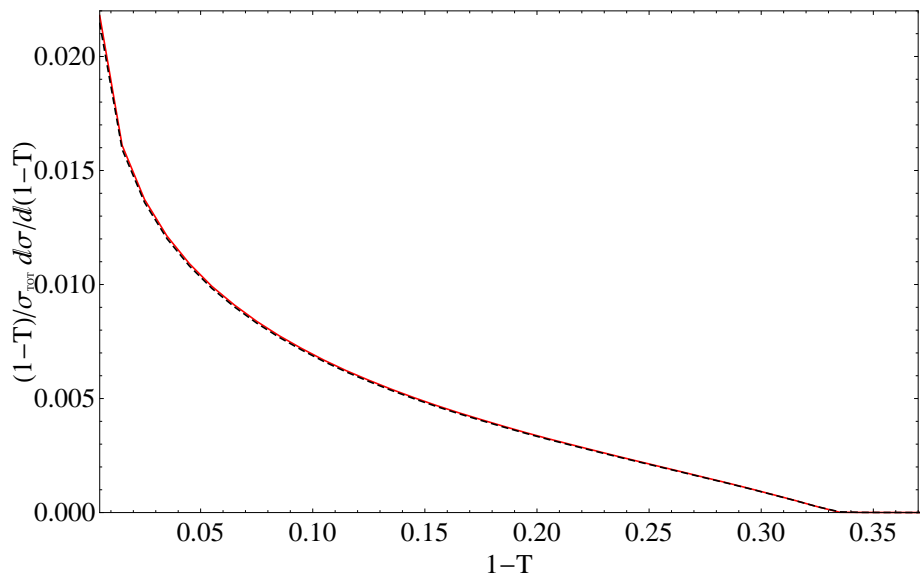}
\caption{Thrust distributions in the QED limit at NNLO using the
PMC$_\infty$ (solid red) and the CSS (dashed
black)~\cite{DiGiustino:2021nep}.} \label{qedthrust}
\end{figure}

We note that in the QED limit the PMC$_\infty$ scales have
analogous dynamical behavior as those calculated in QCD,
differences arise mainly owing to the $\overline{\textrm{MS}}$
scheme factor reabsorption, the effects of the $N_c$ number of
colors at NLO are very small. Thus we notice that perfect
consistency is shown from QCD to QED using the PMC$_\infty$
method. The normalized QED thrust distribution is shown in
Fig.~\ref{qedthrust}. We note that the curve is peaked at the
origin, $T=1$, which suggests that the three-jet event in QED
occurs with a rather back-to-back symmetry. Results for the CSS
and the PMC$_\infty$ methods in QED are of the order of
$O(\alpha)$ and show very small differences, given the good
convergence of the theory.

\chapter{Precise determination of the strong
coupling and its behavior from Event Shape variables using the PMC
\label{sec:novel}}

In this section we present a novel method for precisely
determining the running QCD coupling constant $\alpha_s(Q)$ over a
wide range of $Q$ from Event Shape variables for the
electron-positron annihilation process measured at a single
center-of-mass energy $\sqrt{s}$, based on PMC scale setting. In
particular, we show results obtained in
Refs.~\cite{Wang:2019isi,Wang:2019ljl} using the approach of a
single PMC scale at LO and at NLO , i.e. the
PMC($\mu$\textsubscript{LO}) of the previous chapter.


The precise determination of the strong coupling $\alpha_s(Q)$ is
one of the crucial tests of QCD. The dependence of $\alpha_s(Q)$
on the renormalization scale $Q$ obtained from many different
physical processes shows consistency with QCD predictions and
asymptotic freedom. The Particle Data Group (PDG) currently gives
the world average:
$\alpha_s(M_Z)=0.1181\pm0.0011$~\cite{Tanabashi:2018oca} in the
$\overline{\rm{MS}}$ renormalization scheme.

Particularly suitable to the determination of the strong coupling
is the process $e^+e^-\rightarrow 3jets$ since its lowest order is
$\mathcal{O}(\alpha_s)$~\cite{Kluth:2006bw}. Currently,
theoretical calculations for event shapes are based on CSS. By
using conventional scale setting, only one value of $\alpha_s$ at
the scale $\sqrt{s}$ can be extracted and the main source of the
uncertainty is given by the choice of the renormalization scale.
Several values for the strong coupling have been extracted from
several processes, e.g. $\alpha_s(M_Z)=0.1224\pm
0.0035$~\cite{Dissertori:2009ik} is obtained by using perturbative
corrections and resummation of the large logarithms at the
NNLO+NLL accuracy predictions. Other evaluations improving the
resummation calculations up to N$^3$LL give a result of
$\alpha_s(M^2_Z)=0.1135\pm0.0011$~\cite{Abbate:2010xh} from
thrust, and $\alpha_s(M_Z)=0.1123\pm0.0015$~\cite{Hoang:2015hka}
from the C-parameter. Non-perturbative corrections for
hadronization effects have also been included in
Ref.~\cite{Dasgupta:2003iq}, but as pointed out in
Ref.~\cite{Tanabashi:2018oca}, the systematics of the theoretical
uncertainties given by hadronization effects are not well
understood.

We show in this section that using the PMC is possible to
eliminate the renormalization scale ambiguities and obtain
consistent results for the strong coupling using the precise
experimental data of Event Shape variable distributions. We notice
that improved event shape distributions have been obtained in
Refs.~\cite{Kramer:1990zt, Gehrmann:2014uva,Hoang:2014wka} using
BLM and the soft and collinear effective theory (SCET) .

\section{Running behavior}
Given that the PMC scale (PMC($\mu$\textsubscript{LO})) is not a
single-valued but rather a monotonically increasing function of
$\sqrt{s}$ and of the selected observable,(as shown in
Figs.~\ref{Tscales} and \ref{Cpar-scales}), it is possible to
determine the strong coupling values at different scales from one
single experiment at one single center-of-mass energy. The
dependence of the scale on the observable reflects the dynamics of
the underlying gluon and quark subprocess. This dynamics varies
also the number of active flavors $N_f$.
Considering that PMC scales for QCD and QED show the same behavior
and their relation at LO is only given by a RS redefinition term:
$Q^2_{\rm QCD}/Q^2_{\rm QED}=e^{-5/3}$ this approach can also be
extended to QED.

\begin{figure}[htb]
\centering
\includegraphics[width=0.60\textwidth]{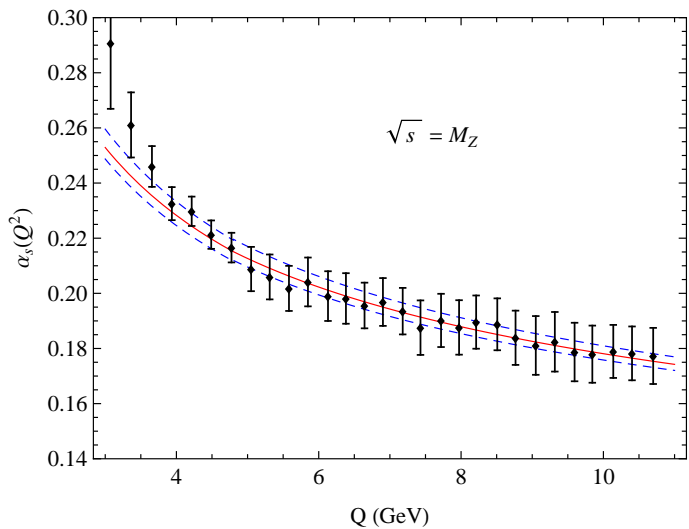}
\caption{The coupling constant $\alpha_s(Q)$ extracted by
comparing PMC predictions with the ALEPH data~\cite{ALEPH:2003obs}
at a single energy of $\sqrt{s}=M_Z$ from the C-parameter
distributions in the $\overline{\rm MS}$ scheme (from
Ref.~\cite{Wang:2019isi}). The error bars are the squared averages
of the experimental and theoretical errors. The three lines are
the world average evaluated from
$\alpha_s(M_Z)=0.1181\pm0.0011$~\cite{Tanabashi:2018oca}. }
\label{figasPMCC}
\end{figure}

\begin{figure}[htb]
\centering
\includegraphics[width=0.60\textwidth]{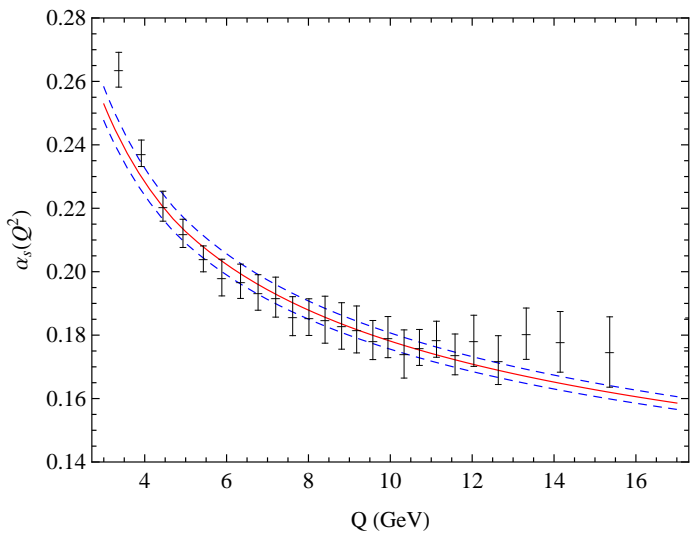}
\caption{The extracted $\alpha_s$ from the comparison of PMC
predictions with ALEPH data at $\sqrt{s}=M_Z$ (from
Ref.~\cite{Wang:2019ljl}). The error bars are from the
experimental data. The three lines are the world average evaluated
from $\alpha_s(M_Z)=0.1181\pm0.0011$~\cite{Tanabashi:2018oca}. }
\label{figasPMCT}
\end{figure}

We extract $\alpha_s$ at different scales bin-by-bin from the
comparison of PMC predictions for ($1-T$) and $C$ differential
distributions with measurements at $\sqrt{s}=M_Z$. The extracted
$\alpha_s$ values from the C-parameter distribution are shown in
Fig.~\ref{figasPMCC}. We note that the extracted $\alpha_s$ in the
scale range of $3$ GeV $<Q<11$ GeV are in excellent agreement with
the world average evaluated from
$\alpha_s(M_Z)$~\cite{Tanabashi:2018oca}. Given that the PMC scale
setting eliminates the scale uncertainties, the corresponding
extracted $\alpha_s$ values are not plagued by any ambiguity in
the choice of $\mu_r$. The extracted $\alpha_s$ values from the
thrust observable using PMC are shown in Fig.~\ref{figasPMCT}.
Also in this case there is good agreement in the range $3.5$ GeV
$<\mu_r<16$ GeV (corresponding to the ($1-T$) range:
$0.05<(1-T)<0.29$). These extracted values of $\alpha_s$ are in
good agreement also with the world average value of the
PDG~\cite{Tanabashi:2018oca}.

Thus, PMC scale setting provides a remarkable way to verify the
running of $\alpha_s(Q)$ from event shapes measured at a single
energy of $\sqrt{s}$. Analogously in QED, the running of QED
coupling $\alpha(Q)$ can be measured at a single energy of
$\sqrt{s}$ (see e.g.~\cite{OPAL:2005xqs}).

The differential distributions of event shapes are afflicted with
large logarithms especially in the two-jet region. Thus, the
comparison of QCD predictions with experimental data and thus the
extracted $\alpha_s$ values are restricted to the region where the
theory is able to describe the data well.

Choosing a different area of distributions leads to the different
values of $\alpha_s$.

\section{$\alpha_s(Q)$ from mean values}

The mean value of an event shape defined by:
\begin{eqnarray}
\langle y\rangle
&=&\int_0^{y_0}\frac{y}{\sigma_{h}}\frac{d\sigma}{dy}dy,
\end{eqnarray}
where $y_0$ is the kinematically allowed upper bound of the $y$
variable, involves an integration over the full phase space, it
thus provides an important platform to complement the differential
distributions and to determinate $\alpha_s$.

The mean values of event shapes have been extensively measured and
studied. At the present moment, pQCD predictions including NNLO
QCD corrections~\cite{Gehrmann-DeRidder:2009fgd, Weinzierl:2009yz}
based on CSS substantially underestimate experimental data.

PMC scales for the mean values of the event shapes are
respectively:
\begin{eqnarray}
\mu^{\rm{pmc}}_r|_{\langle1-T\rangle} = 0.0695\sqrt{s}, ~\rm{and}~
\mu^{\rm{pmc}}_r|_{\langle C\rangle} = 0.0656 \sqrt{s}, \nonumber
\end{eqnarray} for thrust and the C-parameter.
Both PMC scales are $\mu^{\rm pmc}_r\ll \sqrt{s}$.

When taking $\sqrt{s}=M_Z=91.1876$ GeV, the PMC scales are
$\mu^{\rm pmc}_r|_{\langle1-T\rangle}=6.3$ GeV and $\mu^{\rm
pmc}_r|_{\langle C\rangle}=6.0$ GeV for the thrust and the
C-parameter, respectively. The PMC scales of the differential
distributions for the thrust and the C-parameter are also very
small. The average of the PMC scales $\langle\mu^{\rm
pmc}_r\rangle$ of the differential distributions for the thrust
and the C-parameter are close to the PMC scales $\mu^{\rm
pmc}_r|_{\langle1-T\rangle}$ and $\mu^{\rm pmc}_r|_{\langle
C\rangle}$, respectively. This shows that PMC scale setting is
self-consistent from the differential distributions to the mean
values.

\begin{figure}[htb]
\centering
\includegraphics[width=0.60\textwidth]{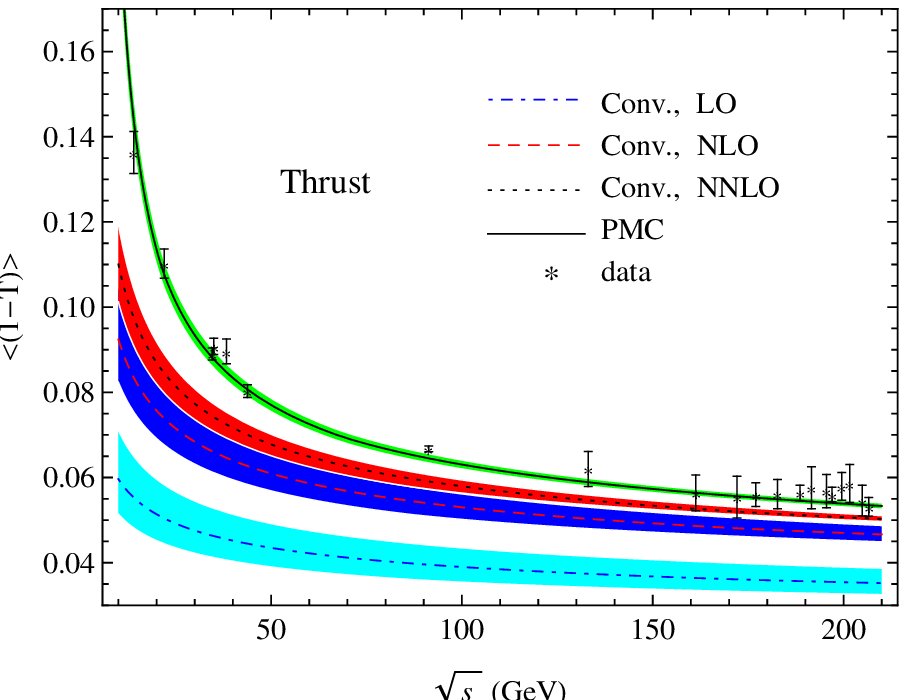}
\includegraphics[width=0.60\textwidth]{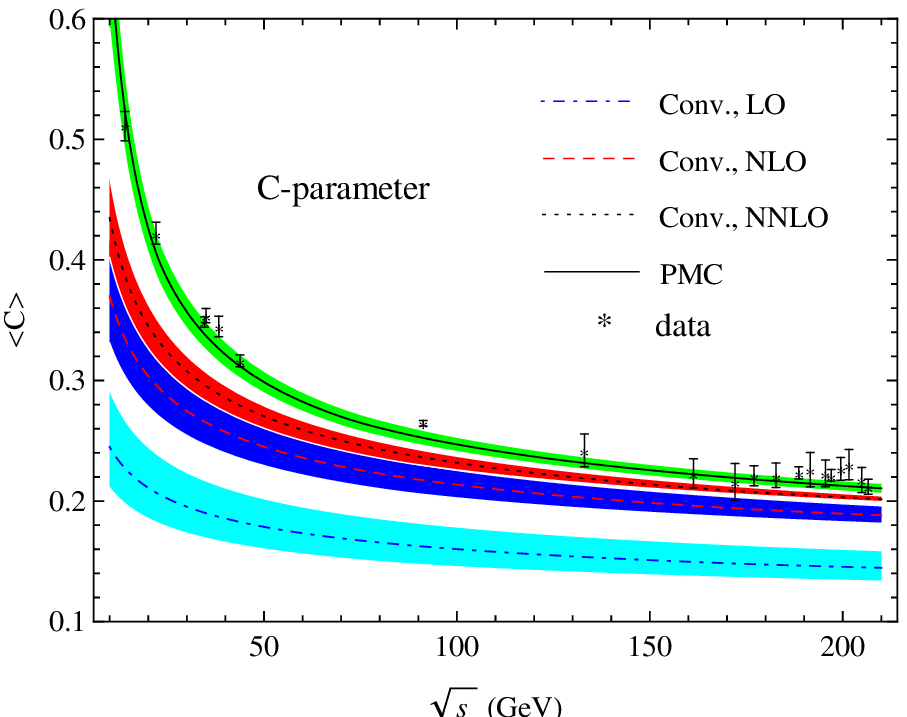}
\caption{The mean values for the thrust and C-parameter versus the
center-of-mass energy $\sqrt{s}$ using CSS (Conv.) and PMC scale
settings (from Ref.~\cite{Wang:2019isi}). The dotdashed, dashed
and dotted lines are the conventional results at LO, NLO and
NNLO~\cite{Gehrmann-DeRidder:2009fgd, Weinzierl:2009yz},
respectively, and the corresponding error bands are obtained by
varying $\mu_r\in[M_Z/2,2M_Z]$. The solid line is the PMC result
and its error band is the squared average of the errors for
$\alpha_s(M_Z)=0.1181\pm0.0011$~\cite{Tanabashi:2018oca} and for
estimated unknown higher-order terms $\pm0.2C_n$. The data are
from the JADE and OPAL experiments, taken
from~\cite{OPAL:2004wof,Pahl:2007zz}. } \label{convPMCmoment}
\end{figure}

The results for the mean values of thrust and the C-parameter
versus the center-of-mass energy $\sqrt{s}$ are shown in
Fig.~\ref{convPMCmoment}. We note that the results obtained with
conventional scale setting substantially deviate from experimental
data. In contrast, PMC scale setting leads to improved mean values
results for both thrust and the C-parameter. PMC predictions are
in excellent agreement with experimental data over a wide range of
energies.

We point out that PMC predictions eliminate the scale $\mu_r$
uncertainty thus errors estimated with the standard criteria are
almost negligible. A better estimate of the unknown higher-order
terms can be obtained by analyzing the magnitude of the
perturbative corrections and the convergence of the series.

The relative magnitude of corrections for the C-parameter
distribution is $C_{\rm LO}:C_{\rm NLO}:C_{\rm
NNLO}\sim1:0.5:0.2$~\cite{GehrmannDeRidder:2007hr} in the
intermediate region using CSS. Using the PMC, the relative
magnitude at NLO is improved to be $C_{\rm LO}:C_{\rm
NLO}\sim1:0.2$. The error estimate of an $n$th-order calculation
for the unknown $C_{n+1}$ term can be characterized by the last
known term, i.e., $\pm C_n$, where $n$ stands for LO, NLO, NNLO,
$\cdots$. The unknown $C_{n+1}$ term can be reasonably estimated
assuming the same rate of increment of the last known order,
$C_{n+1}/C_{n}=C_{n}/C_{n-1}$. Hence, in this case, we have that
error bars are given by: $\pm 0.2 C_n$ (as shown in
Fig.~\ref{convPMCmoment}).

\begin{figure}[htb]
\centering
\includegraphics[width=0.60\textwidth]{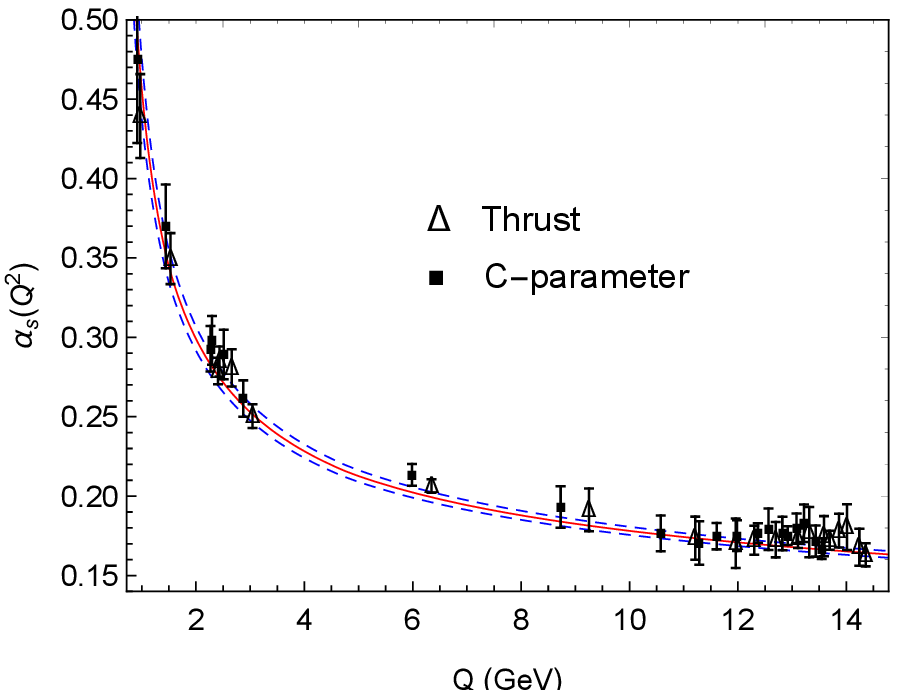}
\caption{The coupling constant $\alpha_s(Q)$ extracted from the
thrust and C-parameter mean values by comparing PMC predictions
with the JADE and OPAL data~\cite{OPAL:2004wof, Pahl:2007zz} in
the $\overline{\rm MS}$ scheme (from Ref.~\cite{Wang:2019isi}).
The error bars are the squared averages of the experimental and
theoretical errors. The three lines are the world average
evaluated from
$\alpha_s(M_Z)=0.1181\pm0.0011$~\cite{Tanabashi:2018oca}. }
\label{convPMCmomentas}
\end{figure}

Given the high level of agreement of the PMC predictions with the
measurements, we can extract $\alpha_s$ to a very high precision.

Results are shown in Fig.~\ref{convPMCmomentas}. We notice an
excellent agreement of the  $\alpha_s$ values with the world
average in the scale range of $1{\rm GeV}<Q<15{\rm GeV}$.

The extracted $\alpha_s$ are not plagued by the scale $\mu_r$
uncertainty. In addition, unlike the $\alpha_s$ extracted from the
differential distributions, the extracted $\alpha_s$ from the mean
values are not afflicted with the large logarithms nor the
non-perturbative effects.

\section{$\alpha_s(M_Z)$ from $\chi^2$-fit}

In order to obtain a reliable $\alpha_s$ at the scale of the $Z^0$
mass, we determine $\alpha_s(M_Z)$ from the fit of the PMC
predictions to measurements. In particular, we perform the fit by
minimizing the $\chi^2$ respect to the $\alpha_s(M_Z)$ parameter.
The variable  $\chi^2$ is defined as: $$\chi^2 =
\sum_{i}\left((\langle y\rangle^{\rm {expt}}_i - \langle
y\rangle^{\rm {th}}_i)/\sigma_i \right)^2,$$ where $\langle
y\rangle^{\rm {expt}}_i$ is the value of the experimental data,
$\sigma_i$ is the corresponding experimental uncertainty and
$\langle y\rangle^{\rm {th}}_i$ is the theoretical prediction. The
fit for thrust and the C-parameter leads to the results:
\begin{eqnarray}
\alpha_s(M^2_Z) &=& 0.1185\pm0.0011(\rm expt)\pm0.0005(\rm th) \nonumber \\
& =& 0.1185\pm0.0012,
\end{eqnarray}
with $\chi^2/$d.o.f.$=27.3/20$ for the thrust mean value, and
\begin{eqnarray}
\alpha_s(M^2_Z) &=& 0.1193^{+0.0009}_{-0.0010}(\rm expt)^{+0.0019}_{-0.0016}(\rm th) \nonumber \\
&& = 0.1193^{+0.0021}_{-0.0019},
\end{eqnarray}
with $\chi^2/$d.o.f.$=43.9/20$ for the C-parameter mean value,
where the first error is the experimental uncertainty and the
second error is the theoretical uncertainty. Both results are
consistent with the world average of
$\alpha_s(M^2_Z)=0.1181\pm0.0011$~\cite{Tanabashi:2018oca}.

The precision of the extracted $\alpha_s$ has been greatly
improved by using the PMC: the dominant scale $\mu_r$
uncertainties are eliminated and the convergence of pQCD series is
greatly improved. In particular, a strikingly much faster pQCD
convergence is obtained for the thrust mean
value~\cite{Wang:2019ljl}, theoretical uncertainties are even
smaller than the experimental uncertainties. We remark that these
results for $\alpha_s(M_Z)$ are one of the most precise
determinations of strong coupling at the $Z^0$ mass from event
shape variables.




\chapter{Heavy-quark pair production near the threshold region \label{sec:HQ}}


Heavy fermion pair production in $e^+e^-$ annihilation near the
threshold region is a fundamental process for the Standard Model
and of considerable interest for various phenomena. The precise
prediction of the production cross section for
$e^+e^-\rightarrow\tau^+\tau^-$ at the threshold region is
important in order to improve the measurement of the $\tau$-lepton
mass~\cite{Asner:2008nq}. Precise theoretical predictions for the
production cross section of $e^+e^-\rightarrow c\bar{c}/b\bar{b}$
at the thresholds are also crucial to determine the precise values
of the $c/b$ quark mass and of the coupling constant $\alpha_s$
determined by the sum rule method~\cite{Novikov:1976tn,
Novikov:1977dq, Voloshin:1995sf}.

 For the physics of the future high-energy electron-positron colliders, one of the most important
goals is the precise measurement of top-quark properties,
especially the top quark mass and width in the threshold
region~\cite{Seidel:2013sqa}; in order to reach the aimed
precision, precise predictions for the top-quark pair production
cross section are mandatory.

For heavy-fermion pair production in $e^+e^-$ annihilation, it is
well known that the Coulomb correction appears near the threshold
region, and the fixed-order perturbative calculations are spoilt
by the presence of singular terms. Physically, the renormalization
scale should become very soft in this region. However, the
renormalization scale is usually set to the mass of the heavy
fermion $\mu_r=m_f$ for the calculation of the production cross
sections using standard CSS. Results are then affected by
systematic errors, due to inherent scheme and scale uncertainties,
and predictions are quite unreliable in this kinematic region. In
this chapter, we apply the PMC to make a comprehensive analysis
for heavy-fermion pair production in $e^+e^-$ annihilation near
the threshold region. We show that two distinctly different scales
are determined for heavy fermion pair production near the
threshold region and we show consistency of PMC scale setting with
QED. For the purpose we show the results in both the modified
minimal subtraction scheme ($\overline{\rm MS}$ scheme) and the
V-scheme and we compare the results with the QED GM-L scheme.

\section{The QCD process of the quark pair production in the
$\overline{\rm MS}$ scheme } \label{sec:21}

The interaction between two heavy quarks can be described at
threshold by considering separate contributions arising from
non-relativistic and hard processes dynamics. Non-relativistic
contributions are determined by a non-relativistic potential that
is essentially the result of the effects of a static Coulomb-like
potential and of the linearly diverging confining term. While the
hard contributions encompass virtual perturbative QCD corrections
between heavy quarks in the final state. At the threshold,
kinematic boundaries enhance the effects of the Coulomb
non-relativistic terms that become large. Thus, we can distinguish
two typologies of perturbative corrections for the heavy quark
pair production cross section that we label as Coulomb and
non-Coulomb corrections. The cross section for
$e^{+}e^{-}\rightarrow\gamma^{*}\rightarrow Q\bar{Q}$ is known at
two-loop level (Ref.~\cite{Czarnecki:1997vz}) and it is given by:
\begin{eqnarray}
\sigma=\sigma^{(0)}\left[1 + \Delta^{(1)}\,a_s(\mu_r) +
\Delta^{(2)}(\mu_r)\,a^2_s(\mu_r) + {\cal O}(a^3_s)\right],
\label{sigma:1}
\end{eqnarray}
where $a_s(\mu_r)={\alpha_s(\mu_r)}/{\pi}$, $\mu_r$ is the
renormalization scale. The LO cross section is
\begin{eqnarray}
\sigma^{(0)}=\frac{4}{3}\frac{\pi\,\alpha^2}{s}N_c\,e^2_Q\frac{v\,(3-v^2)}{2},
\label{qqpairLOpmc}
\end{eqnarray}
and the quark velocity $v$ is
\begin{eqnarray}
v=\sqrt{1-\frac{4\,m_Q^2}{s}}.
\end{eqnarray}
Here, $N_c$ is the number of colors, $e_Q$ is the $Q$ quark
electric charge, $s$ is the center-of-mass energy squared and
$m_Q$ is the mass of the heavy quark $Q$. The one-loop correction
$\Delta^{(1)}$ near the threshold region can be written as
\begin{eqnarray}
\Delta^{(1)}=C_F\left(\frac{\pi^2}{2\,v}-4\right).
\end{eqnarray}
The two-loop correction $\Delta^{(2)}$ can be conveniently split
into terms proportional to various $\rm SU(3)$ color factors,
\begin{eqnarray}
\Delta^{(2)} &=& C_F^2\,\Delta^{(2)}_A + C_F\,C_A\,\Delta^{(2)}_{NA} \nonumber\\
&& + C_F\,T_R\,n_f\,\Delta^{(2)}_L + C_F\,T_R\,\Delta^{(2)}_H.
\end{eqnarray}
The terms $\Delta^{(2)}_{A}$, $\Delta^{(2)}_{L}$ and
$\Delta^{(2)}_{H}$ are the same in either Abelian or non-Abelian
theories; the term $\Delta^{(2)}_{NA}$ only arises in the
non-Abelian theory. This process provides the opportunity to
explore rigorously the scale-setting method in non-Abelian and
Abelian theories.

The Coulomb corrections play an important role in the threshold
region, since they are proportional to powers of $(\pi/v)^n$. This
implies that the renormalization scale can be relatively soft in
this region. Thus the PMC scales must be determined separately for
the non-Coulomb and Coulomb corrections~\cite{Brodsky:1995ds,
Brodsky:2012rj}. When the quark velocity $v\rightarrow0$, the
Coulomb correction dominates the contribution of the production
cross section, and the contribution of the non-Coulomb correction
will be suppressed.

From general grounds, one expects threshold physics to be governed
by the non-relativistic instantaneous potential.

The potential affects the cross section through final state
interactions when the scale is above threshold; it leads to bound
states when the scale is below threshold.

In order to apply the PMC, we write explicitly the $n_f$-dependent
and $n_f$-independent parts in the Eq.~\ref{sigma:1}, i.e.,

\begin{eqnarray}
\sigma &=&\sigma^{(0)}\Bigg[1+\Delta^{(1)}_h\,a_s(\mu_r) + \left(\Delta^{(2)}_{h,in}(\mu_r)+\Delta^{(2)}_{h,n_f}(\mu_r)\,n_f\right)\,a^2_s(\mu_r) \nonumber\\
&& + \left(\frac{\pi}{v}\right)\,\Delta^{(1)}_{v}\,a_s(\mu_r) +
\left(\frac{\pi}{v}\right)\,\left(\Delta^{(2)}_{v,in}(\mu_r)+\Delta^{(2)}_{v,n_f}(\mu_r)\,n_f\right)\,a^2_s(\mu_r) \nonumber\\
&&+\left(\frac{\pi}{v}\right)^2\,\Delta^{(2)}_{v^2}\,a^2_s(\mu_r)
+ {\cal O}(a^3_s)\Bigg]. \label{eqpairNNLO}
\end{eqnarray}

The coefficients $\Delta^{(1)}_h$ and $\Delta^{(2)}_h$ are for the
non-Coulomb corrections, and the coefficients $\Delta^{(1)}_v$,
$\Delta^{(2)}_v$ and $\Delta^{(2)}_{v^2}$ are for the Coulomb
corrections. These coefficients in the $\overline{\rm MS}$ scheme
are calculated in Refs.~\cite{Czarnecki:1997vz, Beneke:1997jm,
Bernreuther:2006vp} and at the scale $\mu_r=m_Q$ they can be
written as
\begin{eqnarray}
\Delta^{(1)}_h &=& -4\,C_F,\qquad
\Delta^{(1)}_{v}=\frac{C_F\,\pi}{2}, \\
\Delta^{(2)}_{h,in}&=&-{1\over72}\,C_F\,(C_A\,(302+468\zeta_3 + \pi^2(-179+192\ln2)) \nonumber\\
&& - 2\,(-16(-11+\pi^2)\,T_R + C_F\,(351+6\pi^4-36\zeta_3 \nonumber\\
&& + \pi^2\,(-70+48\ln2))) + 24(3\,C_A+2\,C_F)\,\pi^2\ln v), \nonumber\\
\Delta^{(2)}_{h,n_f}&=&{11\,C_F\,T_R\over9}, \nonumber\\
\Delta^{(2)}_{v,in}&=&-{1\over72}\,C_F\,\pi\,(-31\,C_A+144\,C_F+66\,C_A\ln(2v)), \nonumber\\
\Delta^{(2)}_{v,n_f}&=&{1\over18}\,C_F\,\pi\,T_R\,(-5+6\ln(2v)), \nonumber\\
\Delta^{(2)}_{v^2}&=&{C_F^2\,\pi^2\over12}.
\end{eqnarray}
After absorbing the nonconformal term
$\beta_0=11/3\,C_A-4/3\,T_R\,n_f$ into the coupling constant using
the PMC, we achieve:
\begin{eqnarray}
\sigma &=& \sigma^{(0)}\Bigg[1+\Delta^{(1)}_h\,a_s(Q_h) + \Delta^{(2)}_{h,\rm sc}(\mu_r)\,a^2_s(Q_h)  \nonumber\\
&& + \left(\frac{\pi}{v}\right)\,\Delta^{(1)}_v\,a_s(Q_v) + \left(\frac{\pi}{v}\right)\,\Delta^{(2)}_{v,\rm sc}(\mu_r)\,a^2_s(Q_v)\nonumber\\
&& + \left(\frac{\pi}{v}\right)^2\,\Delta^{(2)}_{v^2}\,a^2_s(Q_v)
+ {\cal O}(a^3_s)\Bigg]. \label{eqpairNNLOpmc}
\end{eqnarray}
The PMC scales $Q_i$ can be written as
\begin{eqnarray}
Q_i =
\mu_r\exp\left[\frac{3\,\Delta^{(2)}_{i,n_f}(\mu_r)}{2\,T_R\,\Delta^{(1)}_i}\right],
\end{eqnarray}
and the coefficients $\Delta^{(2)}_{i,\rm sc}(\mu_r)$ are
\begin{eqnarray}
\Delta^{(2)}_{i,\rm sc}(\mu_r) =
\frac{11\,C_A\,\Delta^{(2)}_{i,n_f}(\mu_r)}{4\,T_R}+\Delta^{(2)}_{i,in}(\mu_r),
\end{eqnarray}
where, $i=h$ and $v$ stand for the non-Coulomb and Coulomb
corrections, respectively. The nonconformal $\beta_0$ term is
eliminated, and the resulting pQCD series matches the conformal
series and thus only the conformal coefficients remain in the
cross section. The conformal coefficients are independent of the
renormalization scale $\mu_r$. At the present two-loop level, the
PMC scales are also independent of the renormalization scale
$\mu_r$. Thus, the resulting cross section in
Eq.~\ref{eqpairNNLOpmc} eliminates the renormalization scale
uncertainty.

Considering the color factors $C_A=3$, $C_F=4/3$ and $T_R=1/2$ for
QCD, the PMC scales in the $\overline{\rm MS}$ scheme are
\begin{eqnarray}
Q_h =e^{-11/24}\,m_Q \label{scaleQCDhms}
\end{eqnarray}
for the non-Coulomb correction, and
\begin{eqnarray}
Q_v =2\,e^{-5/6}\,v\,m_Q \label{scaleQCDbms}
\end{eqnarray}
for the Coulomb correction. The scale $Q_h$ stems from the hard
gluon-virtual corrections, and thus it is determined from the
short-distance process. The scale $Q_v$ originates from the static
Coulomb potential. As expected, the scale $Q_h$ is of order $m_Q$,
while the scale $Q_v$ is of order $v\,m_Q$.

\begin{figure}[htb]
\centering
\includegraphics[width=0.60\textwidth]{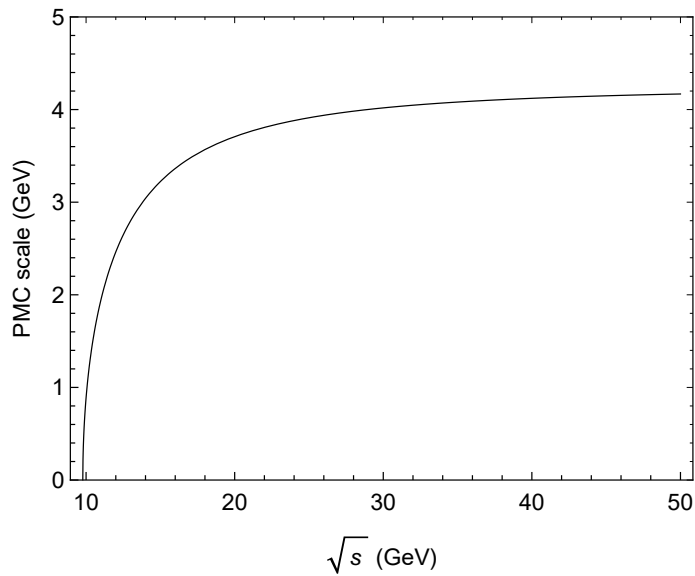}
\caption{The PMC scale $Q_v$ versus the center-of-mass energy
$\sqrt{s}$ for the $b$ quark pair production in the $\overline{\rm
MS}$ scheme~\cite{Wang:2020ckr}. $m_Q=4.89$ GeV. }
\label{figPMCscalebeta0}
\end{figure}

In the following, we take bottom-quark pair production as an
example to make a detailed analysis near the threshold region.
Taking $m_Q=4.89$ GeV~\cite{Bernreuther:2016ccf}, we obtain
\begin{eqnarray}
Q_h = 3.09~\rm GeV,
\end{eqnarray}
which is smaller than $m_Q$ in the $\overline{\rm MS}$ scheme. For
the Coulomb correction part, the scale $Q_v$ is shown in
Fig.~\ref{figPMCscalebeta0}. It shows that the scale $Q_v$ depends
continuously on the quark velocity $v$, and becomes soft for
$v\rightarrow0$, yielding the correct physical behavior of the
scale and reflecting the virtuality of the QCD dynamics. Also the
number of active flavors $n_f$ changes with the quark velocity $v$
according to the PMC scale.

When the quark velocity $v\rightarrow 0$, the small scale in the
coupling constant shows that the perturbative QCD theory becomes
unreliable and non-perturbative effects must be taken into
account. One can adopt light-front holographic
QCD~\cite{Brodsky:2014yha} to evaluate $\alpha_s$ in the low-scale
region.

In contrast, the renormalization scale is simply fixed to the
typical scale $\mu_r=m_Q$ using CSS. Our calculations show that in
the $\overline{\rm MS}$ scheme, the scale should be
$e^{-11/24}\,m_Q$, which is smaller than $m_Q$ for the non-Coulomb
correction. For the Coulomb correction, simply fixing the
renormalization scale $\mu_r=m_Q$ obviously violates the physical
behavior near the threshold region, since the scale becomes soft
for $v\rightarrow0$.

\begin{figure}[htb]
\centering
\includegraphics[width=0.60\textwidth]{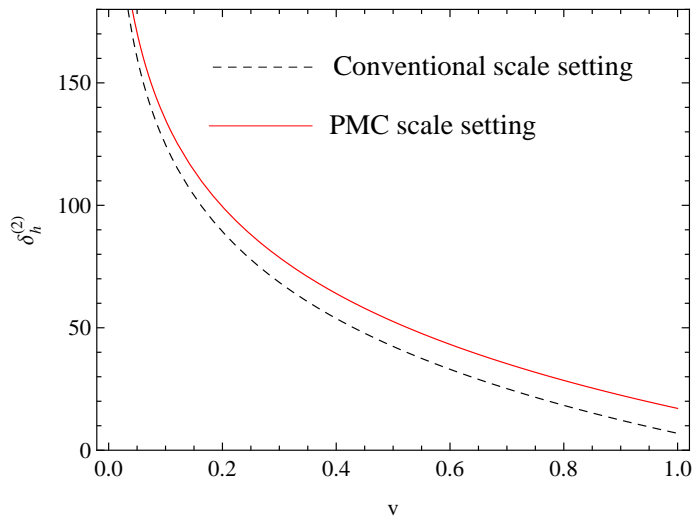}
\caption{The two-loop coefficients $\Delta^{(2)}_h$ of the
non-Coulomb correction in the $\overline{\rm MS}$ scheme for the
$b$ quark pair production, where
$\Delta^{(2)}_h=(\Delta^{(2)}_{h,in}+\Delta^{(2)}_{h,n_f}\,n_f)$
is for conventional scale setting while
$\Delta^{(2)}_h=\Delta^{(2)}_{h,\rm sc}$ is for PMC scale
setting~\cite{Wang:2020ckr}.} \label{figCoedetah2}
\end{figure}

The two-loop coefficients $\Delta^{(2)}_h$ of the non-Coulomb
correction in the $\overline{\rm MS}$ scheme using conventional
and PMC scale settings are shown in Fig.~\ref{figCoedetah2}. We
note that although the $v$-dependent behavior of the coefficients
$\Delta^{(2)}_h$ is the same, their magnitudes are different using
different scale settings. When the quark velocity $v\rightarrow0$,
both curves diverge $\Delta^{(2)}_h\rightarrow+\infty$ due to the
presence of the term $-\ln v$. As expected, after multiplying this
term by the $v$-factor in the LO cross section $\sigma^{(0)}$
given in Eq.~\ref{qqpairLOpmc}, the contribution of the
non-Coulomb corrections is finite and is suppressed near the
threshold region, i.e.
$(\sigma^{(0)}\,\Delta^{(2)}_h\,a^2_s)\rightarrow0$ for the quark
velocity $v\rightarrow0$.

\begin{figure}[htb]
\centering
\includegraphics[width=0.60\textwidth]{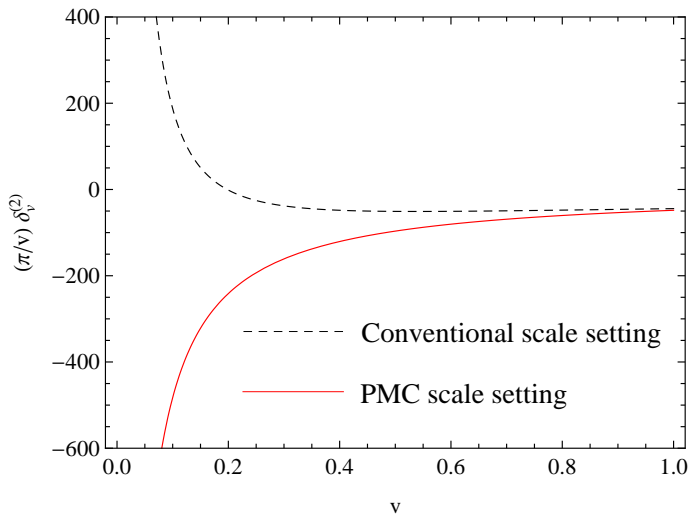}
\caption{The two-loop coefficients $(\pi/v)\Delta^{(2)}_v$ of the
Coulomb correction in the $\overline{\rm MS}$ scheme for the $b$
quark pair production, where
$\Delta^{(2)}_v=(\Delta^{(2)}_{v,in}+\Delta^{(2)}_{v,n_f}\,n_f)$
is for conventional scale setting and
$\Delta^{(2)}_v=\Delta^{(2)}_{v,\rm sc}$ is for PMC scale
setting~\cite{Wang:2020ckr}. } \label{figCoedetab2}
\end{figure}

In the small relative velocity domain, i.e. $(v \ll 1)$, the
effect of multiple-photon exchange between charged particles
becomes significant.

It was shown by Sommerfeld that the correction due to
re-scattering of charged particles in the final state is
proportional to the wave function at the origin squared,
$|\Psi(0)|^{2}$. Thus, the annihilation process acquires some
characteristics of the corresponding bound state. It was shown by
Gamow-Sommerfeld-Sakharov~\cite{sommerfeld,
Gamow:1928zz,Sakharov:1948plh} that the leading contributions
$\alpha_s^n/v^n$  can be resummed using the so-called SGS factor:
\begin{equation}
T \equiv |\Psi(0)|^{2} =\frac{z}{1-\exp (-z)},
\end{equation}
with $z=\frac{C_{F} \alpha_{s} \pi}{v}$. Thus the Coulomb term of
the form $(\pi/v)^2\,\Delta^{(2)}_{v^2}\,a^2_s$ can be reabsorbed
in the Sommerfeld rescattering
formula~\cite{Czarnecki:1997vz}.\footnote{It is also believed that
the subleading terms $\alpha_s^{n+1}/v$ can be resummed by adding
a hard coefficient factor, such as:
$$\sigma =\sigma^{(0)}\left(1-4 C_{F} \frac{\alpha_{s}}{\pi}\right)|\Psi(0)|^{2}.$$}

 For the Coulomb term of the form $(\pi/v)\,\Delta^{(2)}_{v}\,a^2_s$, we present
the coefficients $(\pi/v)\Delta^{(2)}_v$ in the $\overline{\rm
MS}$ scheme using conventional and PMC scale settings in
Fig.~\ref{figCoedetab2}. It shows that when the quark velocity
$v\rightarrow0$, the $v$-dependent behavior of the coefficients
$(\pi/v)\Delta^{(2)}_v$ is dramatically different using
conventional and PMC scale settings. In the case of conventional
scale setting, its behavior is
$(\pi/v)\Delta^{(2)}_v\rightarrow+\infty$ for $v\rightarrow0$ due
to the presence of the term $-\ln v/v$. After multiplying this
term by the $v$-factor in the LO cross section $\sigma^{(0)}$, the
contribution of the Coulomb correction using conventional scale
setting is not finite, i.e.,
$(\sigma^{(0)}\,(\pi/v)\,\Delta^{(2)}_v\,a^2_s)\rightarrow+\infty$
for $v\rightarrow0$. It should be stressed that the term $\ln v$
vanishes, and the term $-1/v$ remains in the conformal coefficient
after applying PMC scale setting. Thus the $v$-dependent behavior
is $(\pi/v)\Delta^{(2)}_{v}\rightarrow-\infty$ for
$v\rightarrow0$. This term $-1/v$ is cancelled by multiplying it
by the $v$-factor in the LO cross section $\sigma^{(0)}$, and thus
the contribution of the Coulomb correction using PMC scale setting
is finite for $v\rightarrow0$. For $v\rightarrow1$, contributions
of the Coulomb correction using conventional and PMC scale
settings are both suppressed.

\section{Quark-pair production in the V-scheme \label{sec:22} }

In order to investigate possible scheme dependence effects, we
perform the above analysis also in a different scheme.
Particularly suitable for the purpose is the scheme defined by the
effective charge $a^V_s=\alpha_V/\pi$ (V-scheme) given by the
static potential between two heavy quarks~\cite{Appelquist:1977tw,
Fischler:1977yf, Peter:1996ig, Schroder:1998vy, Smirnov:2008pn,
Smirnov:2009fh, Anzai:2009tm},
\begin{eqnarray}
V(Q^2) = -{4\,\pi^2\,C_F\,a^V_s(Q)\over Q^2},
\end{eqnarray}
which provides a physically-based alternative to the usual
$\overline{\rm MS}$ scheme. Analogously to QED and the GM-L
scheme, setting the scale to the virtuality of the exchanged
gluon, all vacuum polarization corrections are resummed into the
coupling $a^V_s$. By using the relation between $a_s$ and $a^V_s$
at the one-loop level, i.e.,
\begin{eqnarray}
a^V_s(Q) = a_s(Q) +
\left({31\over36}C_A-{5\over9}T_R\,n_f\right)a^2_s(Q) + {\cal
O}(a^3_s), \label{Vscheme:as}
\end{eqnarray}
we can perform the change of scheme for Eq.~\ref{eqpairNNLO}, from
the $\overline{\rm MS}$ scheme to the V-scheme. Thus, the
corresponding perturbative coefficients in the V-scheme are:

\begin{eqnarray}
\sigma^{(0)}|_V=\sigma^{(0)},
\end{eqnarray}
\begin{eqnarray}
\Delta^{(1)}_h|_V&=&\Delta^{(1)}_h, \quad \Delta^{(1)}_{v}|_V=\Delta^{(1)}_{v},\\
\Delta^{(2)}_{h,in}|_V&=&\Delta^{(2)}_{h,in} - {31\over36}\,C_A\,\Delta^{(1)}_h, \nonumber\\
\Delta^{(2)}_{h,n_f}|_V&=&\Delta^{(2)}_{h,n_f} + {5\over9}\,T_R\,\Delta^{(1)}_h, \nonumber\\
\Delta^{(2)}_{v,in}|_V&=&\Delta^{(2)}_{v,in} - {31\over36}\,C_A\,\Delta^{(1)}_{v}, \nonumber\\
\Delta^{(2)}_{v,n_f}|_V&=&\Delta^{(2)}_{v,n_f} + {5\over9}\,T_R\,\Delta^{(1)}_{v}, \nonumber\\
\Delta^{(2)}_{v^2}|_V&=&\Delta^{(2)}_{v^2}.
\end{eqnarray}
And the corresponding PMC scales in the V-scheme are:
\begin{eqnarray}
Q_h =e^{3/8}\,m_Q \label{scaleQCDh}
\end{eqnarray}
for the non-Coulomb correction, and
\begin{eqnarray}
Q_v =2\,v\,m_Q \label{scaleQCDb}
\end{eqnarray}
for the Coulomb correction. Again, in the V-scheme, $Q_h$ is of
order $m_Q$, while $Q_v$ is of order $v\,m_Q$, since the scale
$Q_h$ originates from the hard-gluon virtual corrections, and
$Q_v$ originates from Coulomb rescattering. The physical behavior
of the scales does not change using different renormalization
schemes. We notice that the difference between the PMC scales in
the $\overline{\rm MS}$ scheme and in the physically-based
V-scheme is given by $e^{\delta_C}$ scheme factor. This difference
is unphysical and is a consequence of the particular convention
adopted.

The PMC scale eliminates this conventional dependence (i.e. this
RS dependence). Scheme independent relations between two effective
charges can be achieved by  ``commensurate scale relations"
(CSR)~\cite{Brodsky:1994eh, Lu:1992nt} as discussed in
Sec.\ref{RSSQCD}.

\begin{figure}[htb]
\centering
\includegraphics[width=0.60\textwidth]{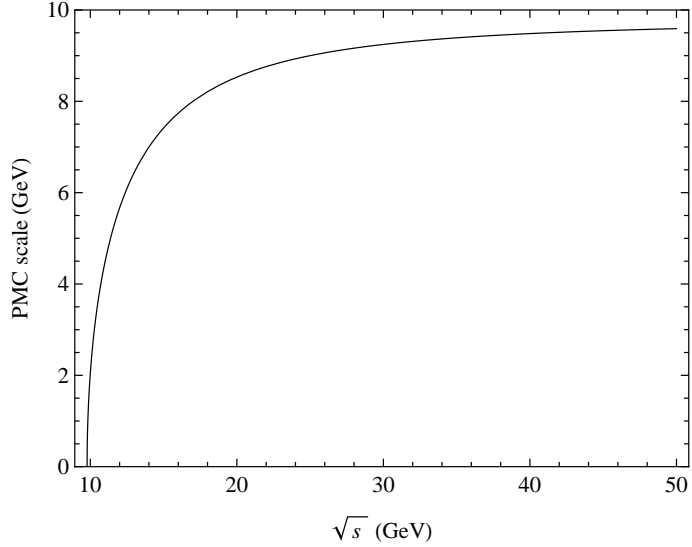}
\caption{The PMC scale $Q_v$ versus the center-of-mass energy
$\sqrt{s}$ for the $b$ quark pair production in the
V-scheme~\cite{Wang:2020ckr}. $m_Q=4.89$ GeV. }
\label{figPMCscalebeta}
\end{figure}

For $b$-quark production, setting the mass to $m_Q=4.89$ GeV, we
obtain the scale $Q_h=7.11$ GeV for the non-Coulomb correction,
and its value larger than the conventional choice $\mu_r=m_Q$. The
scale for the Coulomb correction depends on the velocity $v$, we
show the $Q_v$ scale versus the center-of-mass energy $\sqrt{s}$
in the V-scheme in Fig.~\ref{figPMCscalebeta}. The exponent
disappears in Eq.~\ref{scaleQCDb} compared to the scale in
Eq.~\ref{scaleQCDbms} in the $\overline{\rm MS}$ scheme. The scale
$Q_v$ becomes soft for $v\rightarrow0$, and $Q_v\rightarrow 2m_Q$
for $v\rightarrow1$, showing the correct physical behavior.

\begin{figure}[htb]
\centering
\includegraphics[width=0.60\textwidth]{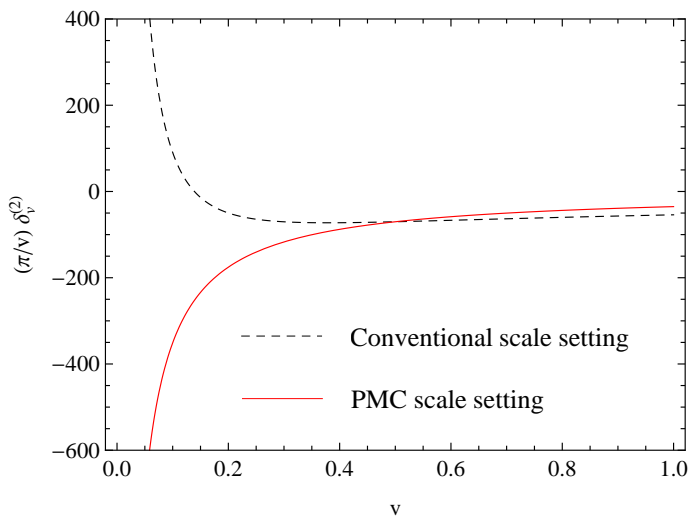}
\caption{The two-loop coefficients $(\pi/v)\Delta^{(2)}_v$ of the
Coulomb correction in the V-scheme for the $b$ quark pair
production, where
$\Delta^{(2)}_v=(\Delta^{(2)}_{v,in}|_V+\Delta^{(2)}_{v,n_f}|_V\,n_f)$
is for conventional scale setting and
$\Delta^{(2)}_v=\Delta^{(2)}_{v,\rm sc}|_V$ is for PMC scale
setting~\cite{Wang:2020ckr}. } \label{figCoedetabV2}
\end{figure}

As in the case of the $\overline{\rm MS}$ scheme, the
$v$-dependent behavior of the coefficients $\Delta^{(2)}_h$ of the
non-Coulomb correction in the V-scheme using conventional and PMC
scale settings is the same. The two-loop coefficients
$(\pi/v)\Delta^{(2)}_v$ of the Coulomb correction in the V-scheme
using conventional and PMC scale settings are presented in
Fig.~\ref{figCoedetabV2}. Figure~\ref{figCoedetabV2} shows that
close to the threshold region, the $v$-dependent behavior of the
coefficients $(\pi/v)\Delta^{(2)}_v$ using conventional and PMC
scale settings is dramatically different. When the quark velocity
$v\rightarrow0$, the coefficient is
$(\pi/v)\Delta^{(2)}_v\rightarrow+\infty$ due to the presence of
the term $-\ln v/v$ using conventional scale setting. After
applying PMC scale setting, the coefficient is
$(\pi/v)\Delta^{(2)}_v\rightarrow-\infty$ due to the term $\ln v$
vanishes and the term $-1/v$ remains in the conformal coefficient.
Thus, multiplying by the $v$-factor in the LO cross section
$\sigma^{(0)}$, the contribution of the Coulomb correction for
$v\rightarrow0$ is not finite using conventional scale setting and
is finite after using PMC scale setting. It is noted that
quark-pair and lepton-pair production in $e^+e^-$ annihilation
near the threshold region should show similar physical behavior.
This dramatically different behavior of the
$(\pi/v)\Delta^{(2)}_v$ between conventional and PMC scale
settings near the threshold region should be checked in QED.

\section{Lepton pair production \label{sec:3}}

Similar to quark-pair production, the lepton-pair production cross
section for the QED process
$e^{+}e^{-}\rightarrow\gamma^{*}\rightarrow l\bar{l}$ is expanded
in the QED fine structure constant $\alpha$. According to the
previous analysis, we can split the cross section into non-Coulomb
and Coulomb parts. Perturbative coefficients for the lepton-pair
production cross section can be obtained from Eq.~\ref{eqpairNNLO}
by replacing the QCD color factors: $C_A=3$, $C_F=4/3$ and
$T_R=1/2$ with the corresponding QED~\cite{Czarnecki:1997vz,
Hoang:1995ex, Hoang:1997sj}: $C_A=0$, $C_F=1$ and $T_R=1$.


Applying the PMC, vacuum-polarization corrections can be absorbed
into the QED running coupling, according to:
\begin{eqnarray}
\alpha(Q) = \alpha\left[1 +
\left({\alpha\over\pi}\right)\sum^{n_f}\limits_{i=1}{1\over3}\left(\ln\left({Q^2\over
m^2_i}\right)-{5\over3}\right)\right],
\end{eqnarray}
where $m_i$ is the mass of the light virtual lepton, and is far
smaller than the final-state lepton mass $m_l$.

The resulting PMC scales can be written as
\begin{eqnarray}
Q_i =
m_l\,\exp\left[{5\over6}+{3\over2}\,{\Delta^{(2)}_{i,n_f}\over\Delta^{(1)}_i}\right],
\end{eqnarray}
where, $i=h$ and $v$ stand for the non-Coulomb and Coulomb
corrections, respectively. After absorbing the vacuum polarization
corrections into the QED running coupling, the coefficients
$\Delta^{(2)}_{i,n_f}$ vanish. According to the replacements
$C_A=0$, $C_F=1$ and $T_R=1$ for QED, the PMC scales become:
\begin{eqnarray}
Q_h =e^{3/8}\,m_l \label{scaleQEDh}
\end{eqnarray}
for the non-Coulomb correction, and
\begin{eqnarray}
Q_v =2\,v\,m_l \label{scaleQEDb}
\end{eqnarray}
for the Coulomb correction.

PMC scales show the same physical behavior from QCD to QED. It is
noted that PMC scales in Eqs.~\ref{scaleQCDh} and \ref{scaleQCDb}
for QCD in the V-scheme coincide with the QED scales in
Eqs.~\ref{scaleQEDh}) and \ref{scaleQEDb}, respectively. This
result was expected for self-consistency, given that the PMC
method in QCD agrees with the standard Gell-Mann-Low
method~\cite{GellMann:1954fq} in QED. We also point out that the
V-scheme provides a natural scheme for the QCD process for
quark-pair production.

In the following, we take $\tau$-lepton pair production as an
example to make a detailed analysis near the threshold region.
Taking $m_\tau=1.777$ GeV~\cite{Tanabashi:2018oca}, we obtain the
scale $Q_h = 2.59$ GeV, which is larger than the scale $m_\tau$
for non-Coulomb correction. For the Coulomb correction, as in the
case of QCD, the scale becomes soft for $v\rightarrow0$ and
$Q_v\rightarrow2m_l$ for $v\rightarrow1$. The PMC scales thus
rigorously yield the correct physical behavior for the lepton pair
production near the threshold region.

\begin{figure}[htb]
\centering
\includegraphics[width=0.60\textwidth]{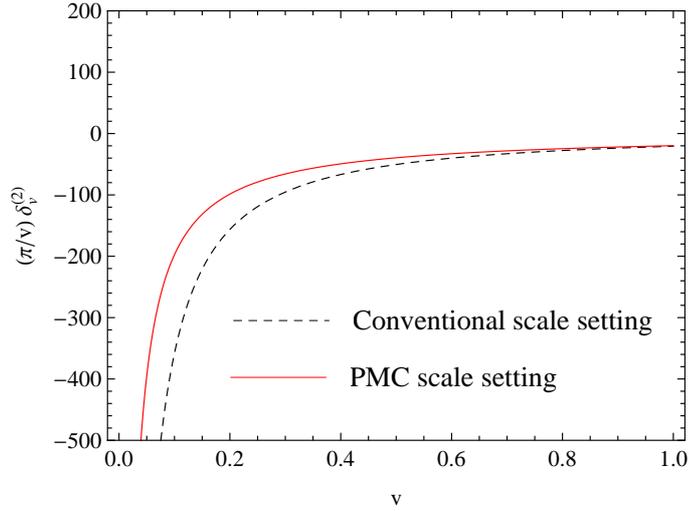}
\caption{The two-loop coefficients $(\pi/v)\Delta^{(2)}_v$ of the
Coulomb correction for the $\tau$ lepton pair production, where
$\Delta^{(2)}_v=(\Delta^{(2)}_{v,in}+\Delta^{(2)}_{v,n_f}\,n_f)$
is for conventional scale setting and
$\Delta^{(2)}_v=\Delta^{(2)}_{v,\rm in}$ is for PMC scale
setting(~\cite{Wang:2020ckr}). } \label{figCoeQEDb2}
\end{figure}

Non-Coulomb corrections have the same $v$-dependent behavior for
the coefficients $\Delta^{(2)}_h$ independently of the use of
conventional or PMC scale settings, as for the QCD case. For the
Coulomb correction, we show the coefficients
$(\pi/v)\Delta^{(2)}_v$ using conventional and PMC scale settings
in Fig.~\ref{figCoeQEDb2}.

It is noted that in contrast with the QCD case, using the CSS,
when the quark velocity $v\rightarrow0$, the coefficient
$(\pi/v)\Delta^{(2)}_v\rightarrow-\infty$ due to the presence of
the term $\ln v/v$; while using the PMC the coefficient
$(\pi/v)\Delta^{(2)}_v\rightarrow-\infty$ due to the presence of
the term $-1/v$ . Multiplying by the $v$-factor in the LO cross
section $\sigma^{(0)}$, the contribution of the Coulomb correction
is given by:
$(\sigma^{(0)}\,(\pi/v)\,\Delta^{(2)}_v\,(\alpha/\pi)^2)\rightarrow-\infty$
for $v\rightarrow0$ using conventional scale setting, and is a
finite for $v\rightarrow0$ using PMC scale setting. We note that
using different scale setting we achieve opposite behaviors, as
shown in Figs.~\ref{figCoedetab2}, \ref{figCoedetabV2}
and~\ref{figCoeQEDb2}, but only the PMC is in agreement with the
QED results of the GM-L scheme. Moreover, scale setting affects
results for the Coulomb corrections that are finite when using the
PMC scale setting, and in contrast diverge when using conventional
scale setting for both QCD and QED.



\chapter*{Summary}
\addcontentsline{toc}{chapter}{\numberline{}Summary}

We briefly summarize here the results.
\begin{itemize}
\item [I)] In section \ref{pmcinfty} we have illustrated the
PMC$_\infty$ method and how it applies to either numerical or
analytical calculated observables. The approach of the
PMC$_\infty$ is totally general and can be applied to any
perturbative calculation, such as fully integrated cross sections
and also differential distributions. This new approach preserves
intrinsic conformality (iCF), which introduces a new concept of
scale invariance: ordered scale invariance, i.e. the scale
invariance of any combination of conformal subsets. In this
approach scale invariance is preserved perturbatively, analogously
to gauge invariance.

In this thesis we have also presented a new procedure to identify
the conformal coefficients and the PMC$_\infty$-scales valid to
all orders, which can be applied to either numerical or analytical
calculations. The PMC$_\infty$ preserves the (A,B,C,D,E,F)
properties of the BLM/PMC method: it eliminates renormalon growth
$\alpha_s^{n+1}\beta_0^n n!$, scheme dependence and all
uncertainties related to the scale ambiguity up to the order of
accuracy. The iCF/PMC$_\infty$ scales are identified by the lowest
order logarithm related to the $\beta_0$-term at each order and
all the physics of the process is contained in the conformal
coefficients. This is in complete agreement with QED and with the
Gell-Mann and Low scheme~\cite{Brodsky:1982gc,Brodsky:1997jk}.

We have discussed why iCF should be considered a strict
requirement for a theory in order to preserve the scale invariance
of the observables and we have shown that iCF is consistent with
the single-variable thrust and C-parameter distributions. We point
out that other conformal aspects of QCD resulting from different
sectors such as Commensurate Scale Relations-CSR~\cite{Lu:1992nt}
or dual theories as AdS/CFT~\cite{ads1} might also be related to
the intrinsic conformality.

We underline that the iCF property in a theory would have two main
remarkable consequences: first, it indicates the correct coupling
constant at each order as a function of the intrinsic conformal
scale $\mu_N$, and second, since only the conformal and the
$\beta_0$ coefficients need to be identified in the observables at
each order, by means of the PMC$_\infty$ method the iCF would
reveal its predictive feature for the coefficients of the higher
order color factors. We point out that in many cases
implementation of the iCF in a multi-loop calculations procedure
would lead to a significant reduction of the color-factor
coefficients and it would speed up the calculations for higher
order corrections.



\item [II)] In section \ref{sec:thrustandc} we have shown results
for thrust and the C-parameter for $e^+e^-\rightarrow 3 jets$,
comparing the two methods for setting the renormalization scale in
pQCD: the Conventional Scale Setting (CSS) and the infinite-order
scale setting based on the Principle of Maximum Conformality
(PMC$_\infty$). The PMC$_\infty$ has been applied to the NNLO
thrust and C-parameter distributions and the results show perfect
agreement with the experimental data. The evaluation of
theoretical errors using standard criteria show that the
PMC$_\infty$ significantly improves the theoretical predictions
over the entire spectra of the shape variables. We have shown that
PMC$_\infty$ works particularly well for the case of the event
shape variables, where we found an extremely good agreement of the
$n_f$ coefficients with the iCF parametrization. We point out that
in fixed order calculations the PMC$_\infty$ last scale is set to
the kinematic scale of the process: in this case
$\mu_{III}=\sqrt{s}=M_{Z}$. As shown in Eq. \ref{confsubsets}, the
scale dependence on the initial scale is totally confined in the
last subset $\sigma_n$. Thus the the last term in the iCF
determines the level of {\it conformality} reached by the
expansion and is entangled with theoretical uncertainties given by
higher order uncalculated terms. Any variation of the last scale
has to be intended to evaluate theoretical uncertainties given by
higher order contributions and not as an ambiguity of the
PMC$_\infty$ method~\cite{Chawdhry:2019uuv}. Evaluation of the
theoretical errors using standard criteria shows that the
PMC$_\infty$ significantly improves the precision of the pQCD
calculations for thrust and C-parameter. We remark that an
improved analysis of theoretical errors might be obtained by
giving a prediction of the contributions of higher order terms
using a statistical approach as shown in
Refs.~\cite{Bonvini:2020xeo,Duhr:2021mfd}. This would lead to a
more rigorous method to evaluate errors and thus to restrict the
range of the last PMC$_\infty$ scale that, as we have shown here,
can also be fixed to the last known PMC$_\infty$ one leading to
precise and stable predictions.

\item [III)] In section \ref{sec:confwin} we have investigated,
for the first time, the thrust distribution in the conformal
window of pQCD and in the QED limit. Assuming, for
phenomenological reasons, the physical value of the strong
coupling to be that of the $Z^0$ mass scale, it restricts the
conformal window range, at two loops, to be within $\frac{34
N_c^3}{13 N_c^2-3}< N_f<\bar{N}_f$ with $ \bar{N}_f\simeq 15.22$.
The closer $N_f$ is to the higher value, the more perturbative and
conformal the system is. In this region, we have shown that the
PMC$_\infty$ leads to a higher precision with a theoretical error
that tends to zero. Moreover, results for the thrust distribution
in the conformal window have similar shapes to those of the
physical values of $N_f$ and the position of the peak is preserved
when one applies the PMC$_\infty$ method. Comparison with the
experimental data indicates also that PMC$_\infty$ agrees with the
expected number of flavors. A good fit with experimental data is
shown by the PMC$_\infty$ results for the range $5<N_f<6$, which
agrees with the active number of flavors of the Standard Model.
Outside the pQCD conformal window the PMC$_\infty$ leads to a
higher precision with respect to the CSS. In addition,
calculations for the QED thrust shown in section
\ref{sec:qedthrust} reveal a perfect consistency of the
PMC$_\infty$ with QED when taking the QED limit of QCD for both
the PMC$_\infty$ scale and for the regularization $\eta$ parameter
which tends to the same QCD value.

\item [IV)] In section \ref{sec:novel} we have presented a novel
method to determine the strong coupling constant from event-shape
variables. Though the event-shape variables, especially those for
the process $e^+e^-\rightarrow 3 jets$, are particularly suitable
for the determination of the strong coupling constant
$\alpha_s(M_Z)$, results obtained by using CSS do not match the
experimental average value and event-shape distributions
underestimate the experimental data at NNLO. Some extracted
$\alpha_s$ from event shapes by the CSS may also seem quite
unrealistic. Our analysis suggests that these conventional
difficulties can be overcome by using the PMC renormalization
scale setting leading to a comprehensive and self-consistent
analysis for both the differential distributions and the mean
values for event shapes. The physical observable should be
independent of the choice of renormalization scale. The PMC scale
yields the renormalization scale-independent physical behavior and
reflects the virtuality of the QCD dynamics. Thus the PMC provides
a remarkable way to verify the running of $\alpha_s(Q)$ from the
event shape differential measurement at a single energy of
$\sqrt{s}$. The values of $\alpha_s(M_Z)$ determined are
consistent with the world average and are more precise than the
values obtained from previous analysis performed on the same data.
Recently in Ref.~\cite{Wang:2021tak}, this method has also been
extended to the other event-shape variables $B_W,B_T,\rho$.

\item [V)] In section \ref{sec:HQ} we have investigated
heavy-fermion pair production in $e^+e^-$ annihilation as a
fundamental process of the SM. In the kinematic region close to
threshold, two major distinct contributions arise in perturbative
calculations given by the static Coulomb-like rescattering
potential and non-Coulomb hard-gluon virtual corrections. The
conventional procedure of simply setting the renormalization scale
as $\mu_r=m_{\rm f}$ violates the physical behavior and leads to
unreliable predictions near the threshold region. In contrast, the
PMC scale-setting method provides a self-consistent analysis, and
reveals perfect agreement for the physical behavior of the scale
near the threshold region, between QED and QCD. This fundamental
process shows a peculiar situation where two distinctly different
scales are applied using the PMC. The scale determined for the
hard virtual correction is of order the fermion mass $m_{f}$; the
scale determined for the Coulomb rescattering is of order
$v\,m_f$, which becomes soft at the threshold, given that
$v\rightarrow0$. This reflects the physical virtuality of the
exchanged gluons (or photons), given by the dynamics of the
underlying physical QCD (or QED) process.

\item While non-Coulomb corrections for fermion pair production
are suppressed near the threshold region, the Coulomb corrections
diverge in this kinematic region, using CSS. In contrast, the
$v$-dependent behavior of the scale changes after using PMC scale
setting, turning into a perfectly consistent behavior from QCD to
QED. The divergence is given only by the higher-order
$(\pi/v)^n\,a^n_s$-Coulomb terms, and the proper resummation using
the Sommerfeld (SGS) rescattering formula leads to a finite
result. A finite prediction can also be obtained using the PMC
scale setting in both QCD and in QED.

\item A straight comparison between QED and QCD is obtained using
the V-scheme, which provides a natural scheme for the QCD
calculation for the quark pair production process. Converting the
QCD calculations in the $\overline{\rm MS}$ scheme to the
V-scheme, the resulting PMC predictions in the QED limit are
consistent with the GM-L scheme. QCD and QED scales differ only by
a scheme factor: $Q^2_{\rm QCD}/Q^2_{\rm QED}=e^{-5/3}$; this
factor converts the underlying scale predictions in the
$\overline{\rm MS}$ scheme used in QCD to the scale of the
V-scheme conventionally used in QED~\cite{Brodsky:1994eh}. In
fact, the PMC scales are:  $Q_h =e^{3/8}\,m_f$ for the hard
virtual correction and $Q_v =2\,v\,m_f$ for Coulomb rescattering
for both QCD and QED. The predictions in the V-scheme based on the
conventional scale-setting method do not agree with the QED limit
where the renormalization scale can be set unambiguously by using
the GM-L method. The PMC scale-setting method in QCD can be
reduced in the Abelian limit $N_c\rightarrow0$ to the
Gell-Mann-Low method. This consistency provides strong support for
the PMC scale-setting method.
\end{itemize}

\hspace{1cm}

\chapter*{Conclusions}
\addcontentsline{toc}{chapter}{\numberline{}Conclusions}



In this thesis we have introduced a new scale setting procedure,
namely PMC$_\infty$, which stems from the general Principle of
Maximum Conformality (PMC) and preserves a particular property
that we have defined as \textit{intrinsic conformality} (iCF). The
iCF is a particular parametrization of the perturbative series
that preserves exactly the scale invariance of an observable
perturbatively. The PMC$_\infty$ solves the conventional
renormalization-scale ambiguity in QCD, it preserves not only the
iCF but also all the features of the PMC approach and leads to a
final conformal series at any order of the perturbative
calculation. In fact, the final series is given by perturbative
conformal coefficients with the couplings determined at conformal
renormalization scales. This method agrees with the BLM/PMCm
approach at LO and given the uniqueness of the iCF, any other
alternative method, e.g. the method proposed in
Ref.~\cite{Chawdhry:2019uuv}, which we indicate as ``PMCa", to fix
a scale in order to obtain a perturbative scale-invariant quantity
leads to the same results as the PMC$_\infty$, as shown in
Ref.~\cite{Huang:2021hzr}. Unlike the previous BLM/PMC approaches,
the PMC$_\infty$ scales are not perturbatively calculated, but are
conformal functions of the physical scale/scales of the process
and any other non-integrated momentum/variable, e.g. the
event-shape variable $\rm (1-T)$ or $C$.


The PMC$_\infty$ is totally independent of the initial scale
$\mu_0$ used for renormalization in perturbative calculations and
it preserves the scale invariance at any stage of calculation,
independently of the kinematic boundary conditions, independently
of the starting order of the observable or the order of the
truncated expansion.

The iCF improves the general BLM/PMC procedure and the point ``3"
of section~\ref{sec:blm}. In the same section, we have suggested
that an improvement and simplification of the perturbatively
calculated BLM/PMC scales, would be achieved by setting the
renormalization scale $\mu_R$ directly to the physical scale Q of
the process, before applying the BLM/PMC procedure. This would
remove the initial scale $\mu_0$ dependence from the
perturbatively calculated BLM/PMC scales.


We underline that in contrast with the other PMCm and PMCs
approach the PMC$_\infty$ preserves the iCF, thus scales are set
automatically in kinematic regions where kinematic constraints
cancel the effects of the lower order conformal coefficients.
These effects are particularly visible in the case of event-shape
variables in the multi-jet regions. The iCF effects in these
kinematic regions are neglected by the PMCm and PMCs approach,
unless an {\it ad hoc} prescription is introduced. As recently
shown in Refs.~\cite{Gao:2021wjn} and \cite{Huang:2021hzr},
application of the PMC$_\infty$ improves the theoretical
predictions for $\Gamma(H \to gg)$, $R_{e^+e^-}$, $R_{\tau}$,
$\Gamma(H \to b \bar{b})$ with respect to the CSS. It has been
noted in Ref.~\cite{Huang:2021hzr} that in general also the PMCs
approach works remarkably well especially for fully integrated
quantities.

We point out that this approach may have an effect of averaging
the differences of the PMCm/PMC$_\infty$ scales arising at each
order, which might be significant for the precision of a
prediction at a certain level of accuracy.

However, given the small differences that we have found in the
first two consecutive PMC$_\infty$ scales, i.e. $\mu_I,\mu_{II}$,
in the LO allowed kinematic region, i.e. $0<(1-T)<1/3$ and in
$0<C<0.75$, we may argue that in the same accessible kinematic
domain two consecutive PMC$_\infty$ scales have such small
differences that a single-scale approach, such as the PMCs, would
be justified leading to analogously precise predictions.


We recall that only the $n_f$ terms related to the UV diverging
diagrams (i.e. the $N_f$ terms) must be reabsorbed into the
PMC$_\infty$ scales. Thus PMC$_\infty$ perfectly agrees with the
PMCm when an observable has a manifest iCF form. We remark that
the iCF underlies scale invariance perturbatively, i.e. the
ordered scale invariance. We also remark that PMC$_\infty$ agrees
with the single-scale approach PMCs in the case of an observable
with a particular iCF form with all scales equal, i.e. $\mu_{\rm
I}=\mu_{\rm II}=\mu_{\rm III}....=\mu_{\rm N}$. In this sense the
PMCm and PMCs may be considered more as ``optimization procedures"
that follow the purpose of the maximal conformal series by
transforming the original perturbative series into an iCF-like
final series by using the PMC scales. In contrast, the
PMC$_\infty$ does not indicate any particular value of the
renormalization scale to be used, but indicates the final limit
obtained by each conformal subset and then by the perturbative
expansion once all the terms related to each conformal subset are
resummed. The PMC$_\infty$ is RG invariant at each order of
accuracy, which means we can perform a change of scale at any
stage and reobtain the initial perturbative quantity. In this
sense PMC$_\infty$ is not to be understood as an ``optimization
procedure", but as an explicit RG invariant form to parametrize a
perturbative quantity that leads faster to the conformal limit. By
setting the renormalization scale of each subset to the
corresponding PMC$_\infty$ scale, one simply cancels the infinite
series of $\beta$ terms, leading to the same conformal result as
the original series. Given that both scales and coefficients are
conformal in the PMC$_\infty$, also the scheme and scale
dependence is completely removed in the perturbative series up to
infinity. It has been pointed out in Ref.~\cite{Huang:2021hzr},
that the PMC$_\infty$ scale might become quite small at a certain
order and that PMC$_\infty$ retains the second kind of scale
dependence. We underline that the last scale in the PMC$_\infty$
controls the level of convergence and the conformality of the
perturbative series and is thus entangled with the theoretical
error of a given prediction. According to the PMC$_\infty$
procedure the last scale must be set to the invariant physical
scale of the process, given by $\sqrt{s},M_H...$. We have shown in
this thesis that the usual PMC practice of setting the last scale
equal to the last unknown scale is consistent also for the
PMC$_\infty$ and leads to precise and stable results. At the
moment, improvements to these points are under investigation.

We remark that the evaluation of the theoretical errors using
standard criteria, shows that the PMC$_\infty$ significantly
improves the precision of the pQCD calculations and eliminates the
scheme and scale ambiguities. An improved analysis of theoretical
errors might be obtained by using a statistical approach for
evaluating the contributions of the uncalculated higher order
terms as shown in Refs.~\cite{Bonvini:2020xeo,Duhr:2021mfd}. This
approach would lead to a more rigorous method to evaluate errors
giving also indications on the possible range of values for the
last unknown PMC$_\infty$ scale.




\addcontentsline{toc}{chapter}{\numberline{}Bibliography}
\bibliographystyle{unsrt}
\bibliography{bibliografia}

%

\end{document}